\documentclass[aps,prd,twocolumn,nofootinbib,showpacs,superscriptaddress,10pt]{revtex4-2}

\usepackage{textcomp}
\usepackage{amsmath}
\usepackage{amsfonts}
\usepackage[bbgreekl]{mathbbol}
\usepackage{calc}
\usepackage{mathrsfs}
\usepackage{amssymb}
\usepackage{array}
\usepackage{color}
\usepackage{bbm}
\usepackage{graphicx}
\usepackage{hyperref}
\usepackage{color}
\usepackage[usenames,dvipsnames,svgnames,table]{xcolor}
\usepackage{tabulary}
\usepackage{booktabs}
\usepackage{cancel}

\newcommand{\beq}{\begin{equation}}
\newcommand{\eeq}{\end{equation}}
\renewcommand{\P}{{\cal P}}
\newcommand{\res}{{\cal R}}
\renewcommand{\O}{{\cal O}}
\newcommand{\w}{s}
\newcommand{\R}{R}

\newcommand{\e}{\varepsilon}

\begin{document}
\title{Two-timescale evolution of extreme-mass-ratio inspirals: waveform generation scheme for quasicircular orbits in Schwarzschild spacetime} 
\author{Jeremy Miller}
\affiliation{Department of Physics, National Institute of Oceanography,
Israel Oceanographic and Limnological Research,
Haifa, Israel}
\affiliation{Jerusalem College of Technology,
21 Havaad Haleumi Street, Jerusalem 9372115, Israel}
\author{Adam Pound}
\affiliation{School of Mathematical Sciences and STAG Research Centre, University of 
Southampton, Southampton, SO17 1BJ, United Kingdom}
\date{\today}
\begin{abstract}
Extreme-mass-ratio inspirals, in which a stellar-mass compact object spirals into a supermassive black hole in a galactic core, are expected to be key sources for LISA. Modelling these systems with sufficient accuracy for LISA science requires going to second (or {\em post-adiabatic}) order in gravitational self-force theory. Here we present a practical two-timescale framework for achieving this and generating post-adiabatic waveforms. The framework comprises a set of frequency-domain field equations that apply on the fast, orbital timescale, together with a set of ordinary differential equations that determine the evolution on the slow, inspiral timescale. Our analysis is restricted to the special case of quasicircular orbits around a Schwarzschild black hole, but its general structure carries over to the realistic case of generic (inclined and eccentric) orbits in Kerr spacetime. In our restricted context, we also develop a tool that will be useful in all cases: a formulation of the frequency-domain field equations using hyperboloidal slicing, which significantly improves the behavior of the sources near the boundaries. We give special attention to the slow evolution of the central black hole, examining its impact on both the two-timescale evolution and the earlier self-consistent evolution scheme.
\end{abstract}
\maketitle


\section{Introduction and summary}
Four years after the first direct detection of gravitational waves~\cite{LIGO16a}, the LIGO-Virgo collaboration now announces new detections on a regular basis~\cite{LIGO-catalog,LIGO-alerts,LIGO-catalog2}. To date, all the signals have originated from compact binary inspirals, involving either black holes or neutron stars spiralling toward each other and eventually merging. Observations of these systems have provided a wealth of information about the population of black holes in the universe~\cite{LIGO-BH-Population}, the equation of state of neutron stars~\cite{LIGO-NS-EOS}, and the validity of general relativity in the strong-field regime~\cite{LIGO-GR-tests,LIGO-GR-tests2}.

However, these binaries have all occupied a restricted region of the parameter space, in which the two objects are of roughly equal size and their orbits are approximately quasicircular. When the space-based detector LISA is launched, one of its key sources will be a very different class of binaries called extreme-mass-ratio-inspirals (EMRIs), comprising a stellar-mass compact object of mass $\mu$ slowly spiraling into a black hole of mass $M\sim10^5$--$10^7 M_\odot$~\cite{eLISA-white-paper:13}. Because the inspiral is very slow, an EMRI can lie in the LISA band for the full duration of the mission. The small object can execute hundreds of thousands of intricate orbits in that time, generating a high-resolution map of the massive black hole's spacetime. This map, as represented by precise measurements of the black hole's multipole moments, for example, is encoded in the emitted gravitational radiation, along with other detailed information about the strong-field dynamics~\cite{Babak:2017tow}.

To extract this information from a detected EMRI waveform using matched filtering, we require a model that maintains phase coherence over potentially $\sim 10^5$ wave cycles. {\em Gravitational self-force theory} currently provides the only viable route to meeting this stringent accuracy goal~\cite{Barack-COST-Action:18}. Broadly speaking, the gravitational self-force describes a gravitating object's deviation from test-body motion due to the object's own gravitational field. In the context of an EMRI, the small object of mass $\mu$ slightly perturbs the spacetime of the large black hole of mass $M$, such that the total metric takes the form ${\sf g}_{\alpha\beta}=g_{\alpha\beta}+\e h^1_{\alpha\beta}+\e^2 h^2_{\alpha\beta}+\O(\e^3)$, where $g_{\alpha\beta}$ is the Kerr metric of the large black hole and $\e$ is a formal counting parameter that counts powers of the small mass ratio $\mu/M$. The perturbations $h^n_{\alpha\beta}$ accelerate the small object away from geodesic motion in $g_{\alpha\beta}$, driving the object's slow inspiral, and we interpret this to be the effect of a self-force. If we represent the object's trajectory with a worldline $z^\mu$, its covariant acceleration in $g_{\alpha\beta}$ becomes
\beq\label{EOM}
\frac{D^2 z^\alpha}{d\tau^2} = \e f^\alpha_{1} + \e^2 f^\alpha_{2} + \O(\e^3),
\eeq
where the proper time $\tau$ and covariant derivative $\frac{D}{d\tau}=\frac{dz^\alpha}{d\tau}\nabla_{\!\alpha}$ are defined with respect to $g_{\alpha\beta}$. The forces (per unit mass) $f^\alpha_{n}$ include both the gravitational self-force, due to $h^n_{\alpha\beta}$, and finite-size effects, due to the object's spin (in $f^\alpha_{1}$), quadrupole moments (in $f^\alpha_{2}$), and higher moments (in $f^\alpha_{n>2}$). 

\subsection{EMRI models: requirements and status}

It has been stressed for some time~\cite{Rosenthal:2006nh} that to accurately model EMRIs, one must include the second-order terms in the equation of motion~\eqref{EOM}. This follows from a simple scaling argument. Take ${\cal E}$ to be the energy of the small object and $\dot {\cal E}$ to be the gravitational-wave flux of energy out of the system. Given that ${\cal E}\sim \mu$ and $\dot {\cal E}\sim (\e h^1_{\alpha\beta})^2$, the inspiral will take place over the {\em radiation-reaction time} $t_{rr} ={\cal E}/\dot {\cal E}\sim M/\e$. On this time scale, the second-order force $f^\alpha_{2}$ causes a cumulative shift in $z^\alpha$ of order 
\beq
\delta z^\alpha \sim \e^2f^\alpha_{2} t^2_{rr}\sim \e^0.\label{dez}
\eeq
Since matched filtering will require errors in orbital phase to be much less than 1 radian, this suggests that the contribution of $f^\alpha_{2}$ cannot be neglected. However, the same reasoning shows that the effect of $f^\alpha_{3}$ can be safely neglected. Therefore, it is both necessary and sufficient to include second-order effects.

This scaling argument was made more precise by Hinderer and Flanagan~\cite{Hinderer-Flanagan:08}, who showed that on the radiation-reaction time scale, the phase of the gravitational waveform has an expansion of the form
\beq\label{phasing}
\varphi = \frac{1}{\e}\left[\varphi_0(\e t) + \e\varphi_1(\e t)+\O(\e^2)\right].
\eeq
The leading term  in this expansion, $\frac{1}{\e}\varphi_0$, is said to be of {\em adiabatic} order. Computing it requires only a certain time average of the dissipative piece of $f^\alpha_{1}$, $f^\alpha_{1,\rm diss}$. The first subleading term, $\varphi_1$, is said to be of {\em first post-adiabatic} order. Computing it requires the complete $f^\alpha_{1}$ (including the conservative piece, $f^\alpha_{1,\rm cons}$) and a certain time average of the dissipative piece of $f^\alpha_{2}$, $f^\alpha_{2,\rm diss}$.\footnote{This description is somewhat altered by the existence of transient resonances, but the main conclusions are unchanged~\cite{Flanagan-Hinderer:12}.} The {\em second} post-adiabatic correction can be neglected, as it only contributes $\sim \e$ to the accumulated phase.


In recent decades, there has been a significant effort to develop and implement adiabatic and post-adiabatic EMRI models, reviewed in Ref.~\cite{Barack-Pound:18}. Practical methods of calculating only the necessary input for adiabatic evolution have been formulated and implemented~\cite{Drasco-etal:05,Sago-etal:06,Fujita-etal:09,Isoyama-etal:19}. Due to the high-dimensional parameter space, significant work remains to actually generate adiabatic waveforms using this input~\cite{Hughes-Capra}, but it is expected that a template bank based on such waveforms will suffice to detect many (or even most) EMRI signals, though not to perform high-precision parameter estimation. In the absence of such adiabatic templates, existing ``kludge models'' may even suffice for detection~\cite{Chua:2017ujo}.

For post-adiabatic modeling, the full first-order self-force can now also be calculated along generic bound orbits in Kerr spacetime~\cite{vandeMeent:17b}, and there is ongoing work to incorporate first-order effects of the small object's spin~\cite{Warburton-Osburn-Evans:17,Witzany:2019dii,Akcay:2019bvk}. However, calculations of $f^\alpha_2$ are far less mature. The basic formalism of self-force theory at second order, including the fundamental analytical ingredients, was derived by one of us in Ref.~\cite{Pound:12a} (see also~\cite{Rosenthal:2006iy,Detweiler:12,Gralla:2012db,Pound:2012dk,Pound:2017psq}). Since then, there has been steady progress in developing this formalism into a practical numerical scheme~\cite{Pound:2014xva,Warburton-Wardell:14,Wardell:2015ada,Pound:2015wva,Miller:2016hjv,PhD_thesis,Pound-Wardell:21}. In Ref.~\cite{Pound-etal:19}, we reported the first implementation of that scheme, resulting in a calculation of the gravitational binding energy of quasicircular EMRIs around Schwarzschild black holes. But there remain many challenges in generating post-adiabatic waveforms, particularly in the astrophysically relevant case of generic orbits around Kerr black holes. 

\subsection{Orbital evolution and the two-timescale approximation}

One way that second-order calculations are more complex than first-order ones is that they must incorporate the system's evolution {\em ab initio}. At first order there is a sense in which one can delay the choice of evolution scheme. On short enough time scales, the orbit can be approximated as a geodesic. The metric perturbation $h^1_{\alpha\beta}$, and the self-force $f^\alpha_1$, can then be calculated as if generated by a point mass moving on that geodesic, with the freedom to later choose how to utilize those results to drive the evolution. At second order, this is no longer true: the evolution of the system acts as a source for the second-order field, $h^2_{\alpha\beta}$, meaning one must choose an evolution scheme before one can even write down the second-order field equation.

The simplest approach  to evolution, used in derivations by Gralla and Wald~\cite{Gralla:08,Gralla:2012db}, is to consider perturbative corrections to the trajectory, as in $z^\alpha(\tau,\e)=z^\alpha_0(\tau)+\e z^\alpha_1(\tau)+\O(\e^2)$. In this approach, the mass $\mu$ moving on the geodesic $z^\alpha_0$ creates the metric perturbation $h^1_{\alpha\beta}$, which drives the correction to the trajectory, $z^\alpha_1$.  $z^\alpha_1$ then contributes to the next-order perturbation $h^2_{\alpha\beta}$, which contributes to the next-order correction to the trajectory, $z^\alpha_2$, and so on. This is conceptually simple, but it is obvious from the outset that it can only be accurate on short time scales: Over the course of an inspiral, $\mu$ will move far from $z^\alpha_0$, such that the ``small correction'' $\e z^\alpha_1$ will grow large with time. This growth in $z^\alpha_1$ will create a commensurate growth in $h^2_{\alpha\beta}$. When the corrections become comparable to the leading terms, the expansion will no longer be valid.

Another approach is the {\em self-consistent} approximation, which treats $z^\alpha$ nonperturbatively and makes each $h^n_{\alpha\beta}$ a functional of that nonperturbative trajectory~\cite{Pound:10a,Pound:2017psq}. The trajectory and metric perturbation are then to be determined together as a coupled system. This approximation accurately accounts for the long-term evolution of the trajectory. It has also been used in many of the foundational derivations in self-force theory, and it will be our starting point in this paper. However, it has only been concretely implemented in a scalar toy model~\cite{Diener-etal:12}, and because it uses a trajectory that is always evolving, it abandons the usual advantages of having approximately geodesic motion at first order.

The alternative that has generally been used in practice is a method of {\em osculating geodesics}~\cite{Mino:05,Pound-Poisson:08a,Gair-etal:10,Warburton:2011fk,Osburn-Warburton-Evans:16,Warburton-Osburn-Evans:17,vandeMeent-Warburton:18}. In this approach, at each instant $\tau$ along the accelerated worldline, one computes the self-force as if, for its entire past history, the mass $\mu$ had been moving on the geodesic that is tangential to the worldline at that instant $\tau$. The orbit then effectively evolves smoothly from one geodesic to the next. In principle this method can be carried to second order (and beyond), as sketched in Ref.~\cite{Pound:2015tma}. However, it does not make maximal use of the properties of an EMRI---specifically, the system's near-periodicity. 

The second-order calculation reported in Ref.~\cite{Pound-etal:19} was instead based on another alternative: a {\em two-timescale expansion}~\cite{Kevorkian-Cole:96} (or multiscale expansion) of the Einstein field equations. This expansion, previously utilized within self-force analyses in Refs.~\cite{Pound-Poisson:08b,Mino:2008rr,Hinderer-Flanagan:08,Pound:2010pj,Pound:10c,Pound:2015wva,Yang:2019iqa} and within post-Newtonian analyses in Refs.~\cite{Damour:2004bz,Klein:2013qda,Will:2016pgm,Will:2019lfe}, {\em is} tailored to the properties of an EMRI. An EMRI has two disparate time scales: the orbital period $\sim M$ and the much longer radiation-reaction time $\sim M/\e$. On the orbital time scale, the system is triperiodic, with frequencies $\Omega_r$, $\Omega_\theta$, and $\Omega_\phi$ associated with the radial, polar, and azimuthal motions. This leads to a leading-order metric perturbation with a discrete frequency spectrum,
\beq\label{h FD}
h^1_{\alpha\beta} =\sum_{mpq}h^{1,\omega_{mpq}}_{\alpha\beta}(r,\theta,\phi)e^{-i \left(p\Omega_r  + q\Omega_\theta + m\Omega_\phi \right)t},
\eeq
where the frequency label refers to $\omega_{mpq}=p\Omega_r + q\Omega_\theta + m\Omega_\phi$, and $(t,r,\theta,\phi)$ denote the Boyer-Lindquist coordinates of the background Kerr geometry. On the radiation-reaction time scale $\sim M/\e$, the system's amplitudes and frequencies slowly evolve. To efficiently and accurately capture the behavior on both time scales, a two-timescale expansion introduces multiple time variables: {\em fast times} $\varphi_r=\int \Omega_r dt$, $\varphi_\theta=\int \Omega_\theta dt$, and $\varphi_\phi=\int \Omega_\phi dt$, which vary on the orbital time scale; and a {\em slow time} $\tilde t \sim \e t$, which varies on the radiation-reaction time. The analog of Eq.~\eqref{h FD} is then
\beq\label{h multiscale}
h^n_{\alpha\beta} = \sum_{mpq}\tilde h^{n,\omega_{mpq}}_{\alpha\beta}(\tilde t,r,\theta,\phi)e^{- i \left(p\varphi_r + q\varphi_\theta + m\varphi_\phi\right)}.
\eeq
On short time scales, the amplitudes $\tilde h^{n,\omega_{mpq}}_{\alpha\beta}$ are approximately constant, while the phases are approximately $\varphi_\alpha\sim \Omega_\alpha t$, such that the two-timescale expansion reduces to the ordinary Fourier expansion~\eqref{h FD} at leading order. However, the form~\eqref{h multiscale} remains approximately triperiodic for {\em all} orders $n$ over the entire inspiral, while an expansion of the form~\eqref{h FD}, which freezes the frequencies and amplitudes, can only be accurate at leading order and for short intervals of time.

We will see below that the ansatz~\eqref{h multiscale} splits the Einstein equations into two distinct sets: frequency-domain equations that govern the amplitudes $\tilde h^{n,\omega_{mpq}}_{\alpha\beta}$ at each fixed value of $\tilde t$, and evolution equations that determine the amplitudes and frequencies as functions of $\tilde t$. At first order, the equations for $\tilde h^{1,\omega_{mpq}}_{\alpha\beta}$ are identical to those for the ordinary Fourier coefficients $h^{1,\omega_{mpq}}_{\alpha\beta}$, meaning a two-timescale computation can naturally build on existing frequency-domain codes~\cite{Akcay:2013wfa,Osburn-etal:14,vandeMeent:17b}.

\subsection{Overview of this paper and outline of wave-generation framework}\label{this_paper}

Although two-timescale expansions have been utilized in the past to explore features of EMRIs, they have largely centred on the expansion of the equation of motion~\eqref{EOM}. The work of Hinderer and Flanagan~\cite{Hinderer-Flanagan:08} has been particularly influential in this regard, having provided a complete treatment of that expansion, and having led to Eq.~\eqref{phasing}. However, because the mass $\mu$'s trajectory $z^\alpha$ is coupled to the metric perturbation, an expansion of the equation of motion is by itself incomplete; it must be combined with an expansion of the Einstein field equations. 

Here we focus on that coupled problem, building on our previous work on a scalar toy model~\cite{Pound:2015wva}. We restrict our analysis to the simplest case of quasicircular orbits in Schwarzschild spacetime. The orbital dynamics in this scenario is comparatively trivial, but the field equations have most of the essential features of the full problem. In this restricted context, we present the explicit form of the expanded Einstein equations and the framework they provide for generating post-adiabatic waveforms. A forthcoming series of papers~\cite{two-timescale-0,two-timescale-1,two-timescale-2,two-timescale-3} will extend our analysis to generic orbits.\footnote{Some months after this paper was submitted for publication, one of us provided an overview of the method for generic orbits~\cite{Pound-Wardell:21}. Future papers will provide the complete details.}

In the remainder of this section, we provide a complete outline of the paper and of the wave-generation framework. The body of the paper then fills in the technical details for interested readers.

We begin in Sec.~\ref{sec_field_equations} with a review of self-force theory in the self-consistent framework, keeping the discussion general enough to describe inspirals into a Kerr black hole. Through second order in $\e$, the coupled field equations and equation of motion take the form
\begin{align}
E_{\alpha\beta}[\bar h^{1\res}] &= - E_{\alpha\beta}[\bar h^{1\P}],\label{EFE1-schematic}\\
E_{\alpha\beta}[\bar h^{2\res}] &= 2\delta^2 G_{\alpha\beta} - E_{\alpha\beta}[\bar h^{2\P}],\label{EFE2-schematic}\\
\frac{D^2z^\alpha}{d\tau^2} &= \e f_1^\alpha[h^{1\res}] +\e^2 f_2^\alpha[h^{1\res},h^{2\res}],\label{EOM-schematic}
\end{align}
given below in Eqs.~\eqref{SFhres} and \eqref{puncture_scheme_1}--\eqref{puncture_scheme_2} with \eqref{d2G}, \eqref{Ealphabeta}, \eqref{h1P form}, and \eqref{h2P form}. Here $E_{\alpha\beta}$ is  the linearized Einstein tensor in the Lorenz gauge (up to a factor of $-1/2$), a bar denotes a trace-reversed field $\bar h^n_{\alpha\beta}:= h^n_{\alpha\beta}-\frac{1}{2}g_{\alpha\beta}g^{\mu\nu}h^n_{\mu\nu}$, and $\delta^2 G_{\alpha\beta}$ is the piece of the full Einstein tensor that is quadratic in $h^1_{\alpha\beta}$. Rather than solving for the physical fields $h^n_{\alpha\beta}$, we solve for the {\em residual} fields $h^{n\res}_{\alpha\beta}:=h^n_{\alpha\beta}-h^{n\P}_{\alpha\beta}$, where $h^{n\P}_{\alpha\beta}$ are analytically known {\em punctures}, which diverge on $z^\alpha$ but guarantee that the full fields $h^n_{\alpha\beta}=h^{n\P}_{\alpha\beta}+h^{n\res}_{\alpha\beta}$ agree with the physical fields outside the small compact object. At first order, this is equivalent to approximating the small object as a point mass $\mu$ on $z^\alpha$.

Typically, $h^1_{\alpha\beta}$ is taken to {\em only} include the linear field of the mass $\mu$. However, because the large black hole absorbs gravitational radiation, its mass and spin slowly change with time. To accurately account for this, we modify previous descriptions of the self-consistent approximation such that the solution to Eq.~\eqref{EFE1-schematic} includes perturbations proportional to $\delta M$ and $\delta S$, the small, evolving corrections to the black hole's mass and spin. (The background metric, on the other hand, remains stationary.)

In Sec.~\ref{sec_twotimescale_expansion}, we review the two-timescale method, and in Sec.~\ref{sec_expanded_eom} we apply it to the equation of motion~\eqref{EOM-schematic}, specializing to quasicircuclar orbits in a Schwarzschild background in the process. For a quasicircular orbit, rather than three orbital frequencies and associated phases, there is only one frequency, $\Omega(t,\e):=\frac{d\phi_p}{dt}$, associated with the azimuthal angle $\phi_p$ of the particle's orbit. (Here and below we adopt standard Schwarzschild coordinates.) Both $\Omega$ and the orbital radius $r_p$ are slowly evolving functions of $\tilde t = \e t$, given by the expansions
\begin{align}
\Omega &= \Omega_0(\tilde t) + \e\Omega_1(\tilde t) + \O(\e^2),\\
r_p &= r_0(\tilde t) + \e r_1(\tilde t) + \O(\e^2).
\end{align}
The terms $r_n$ satisfy ordinary differential equations with the schematic form 
\begin{align}
\frac{dr_0}{d\tilde t} &= \dot r_0(r_0,f^\alpha_{1,\rm diss}),\label{r0dot-schematic}\\
\frac{dr_1}{d\tilde t} &= \dot r_1(r_0,r_1, f^\alpha_{2,\rm diss}),\label{r1dot-schematic}
\end{align}
given explicitly by Eqs.~\eqref{r0dot} and \eqref{r1dot}, and the terms $\Omega_n$ are determined by relations
\begin{align}
\Omega_0 &= \Omega_0(r_0),\label{Omega0-schematic}\\
\Omega_1 &= \Omega_1(r_0,r_1,f^\alpha_{1,\rm cons}),\label{Omega1-schematic}
\end{align}
given in Eqs.~\eqref{Omega0} and \eqref{Omega1}. Here and throughout this paper, an overdot denotes a derivative with respect to slow time.

\begin{figure}[t]
\centering
\includegraphics[width=.45\columnwidth]{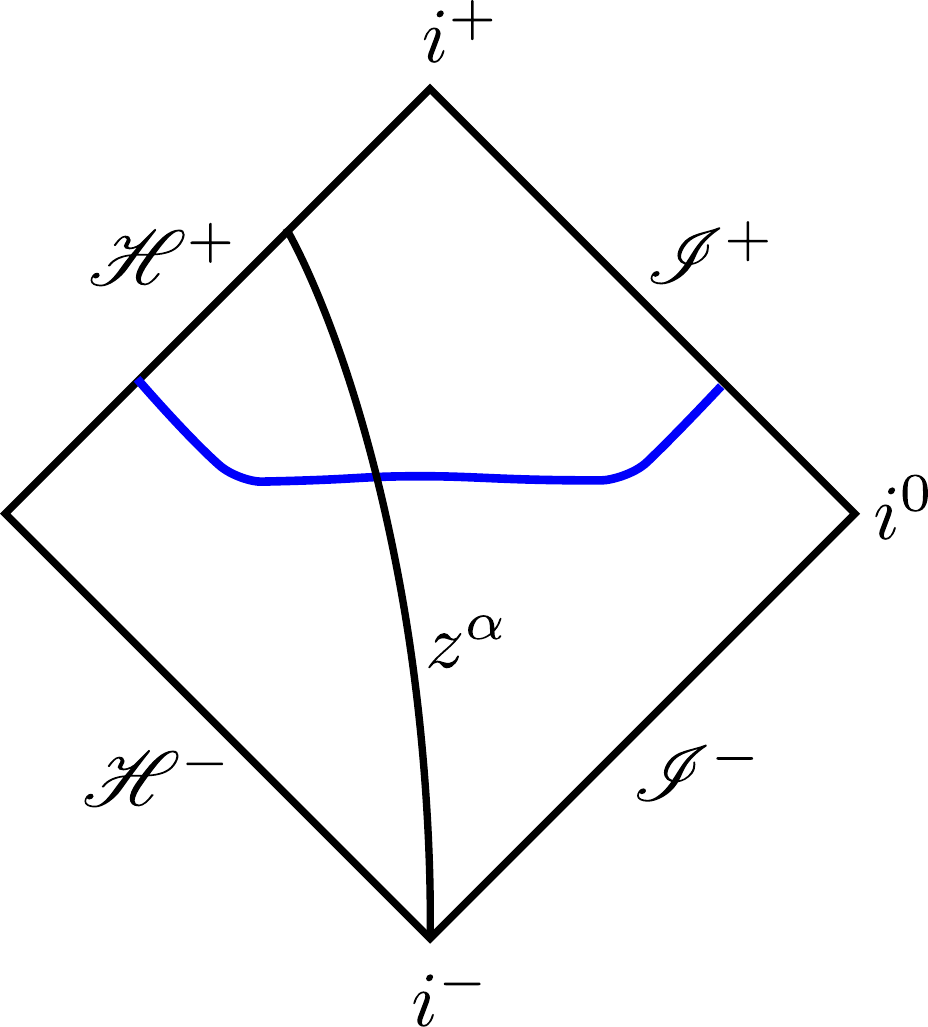}
\caption{Penrose diagram of Schwarzschild spacetime illustrating a slice (blue curve) of constant hyperboloidal time $\w=t-k(r^*)$. $\w$ transitions from advanced time $v=t+r^*$ near the future horizon, to Schwarzschild time $t$ in a region including the particle's worldline $z^\alpha$, to retarded time $u=t-r^*$ near future null infinity.}
\label{fig_penrose_diagram_hyperboloidal_time_3}
\end{figure}

In Sec.~\ref{sec_expanded_field}, we turn to the two-timescale expansion of the field equations~\eqref{EFE1-schematic} and \eqref{EFE2-schematic}. In place of the general expansion~\eqref{h multiscale}, we assume an expansion of the form\footnote{In actuality, order-$\e$ corrections appear on the right-hand side of this expansion. Here we have implicitly absorbed those corrections into the higher-$n$ mode amplitudes.} 
\beq\label{multiscale h}
\bar h^n_{\alpha\beta} = \sum_{i\ell m}\frac{a_{i\ell}}{r}\R^{n}_{i\ell m}(\tilde\w,r)e^{- i m\phi_p}Y^{i\ell m}_{\alpha\beta},
\eeq
and the analogs for $\bar h^{n\res}_{\alpha\beta}$ and $\bar h^{n\P}_{\alpha\beta}$. Here in addition to the multiscale expansion, we have performed an expansion in tensor spherical harmonics; $i=1,\ldots,10$ label the 10 linearly independent harmonics $Y^{i\ell m}_{\alpha\beta}$, given explicitly in Appendix~\ref{sec_tensor_harmonics}, and $a_{i\ell}$ is a convenient numerical factor. More importantly, we have also exploited the freedom that arises in extending the slow and fast time variables away from the worldline. Rather than simply using $\tilde t = \e t$, we have introduced a slow time $\tilde \w:=\e \w(t,r)$. This allows us to naturally account for retardation: the slow evolution of the orbit does not propagate out from $z^\alpha$ instantaneously along slices of constant $t$, but instead along null curves. In our analysis we hence consider a hyperboloidal slow time $\w$ that is equal to $t$ in a neighbourhood of the particle but becomes null as $r\to2M$ or $r\to\infty$, as illustrated in Fig.~\ref{fig_penrose_diagram_hyperboloidal_time_3}. Analogously, as our fast time, rather than $\phi_p(t,\e)=\int^t\Omega(z,\e) dz$, we use $\phi_p(\w,\e)=\int^\w \Omega(z,\e) dz$.

When we substitute the expansion~\eqref{multiscale h} into Eqs.~\eqref{EFE1-schematic} and \eqref{EFE2-schematic}, derivatives with respect to $t$ and $r$ act on the slow- and fast-time dependence, as shown in Eqs.~\eqref{chain_rule_t_1} and \eqref{chain_rule_r_1}. In this counting, a derivative with respect to slow time is suppressed by a factor of $\e$. Factoring out the spherical harmonics and fast-time phase factors, we obtain a sequence of frequency-domain equations for the mode amplitudes $\R^n_{i\ell m}$: 
\begin{align}
E^0_{ij\ell m}\R^{1\res}_{j\ell m} &= - E^0_{ij\ell m}\R^{1\P}_{j\ell m},\label{EFE1 two-timescale schematic}\\
E^0_{ij\ell m}\R^{2\res}_{j\ell m} &= 2\delta^2 G^0_{i\ell m} - E^0_{ij\ell m}\R^{2\P}_{j\ell m} - E^1_{ij\ell m}\R^1_{j\ell m},\label{EFE2 two-timescale schematic}
\end{align}
where $E^0_{ij\ell m}$ and $E^1_{ij\ell m}$ are given in Eq.~\eqref{Enijlm}, and the repeated basis label $j$ is summed over. $E^0_{ij\ell m}$ is a radial operator in which $\partial_t$ has been replaced with $-i\omega_m$, where $\omega_m:=m\Omega_0$. $E^1_{ij\ell m}$ contains the first subleading effect of the time derivatives in $E_{\alpha\beta}$; it is linear in $\partial_{\tilde \w}-im\Omega_1$. The left-hand side of Eqs.~\eqref{EFE1 two-timescale schematic} and \eqref{EFE2 two-timescale schematic} has precisely the same form as if the metric perturbations had been expanded in hyperboloidal-time Fourier modes $e^{-i\omega_m\w}$. Such use of hyperboloidal slicing in the frequency domain has been considered before in, e.g., Refs.~\cite{Zenginoglu:11,PanossoMacedo:2019npm}, though not directly for the linearized Einstein equation. In the context of our two-timescale expansion, this slicing dramatically improves the behavior of the source term $E^1_{ij\ell m}\R^1_{j\ell m}$ as $r\to2M$ or $r\to\infty$.

In Sec.~\ref{sec_combined_expansions} we derive our waveform-generation scheme. The solutions to Eqs.~\eqref{EFE1 two-timescale schematic} and \eqref{EFE2 two-timescale schematic} (for the full fields $\R^{n\res}_{i\ell m}+\R^{n\P}_{i\ell m}$) have the schematic form
\begin{align}
\R^1_{i\ell m} &= \hat\R^{1}_{i\ell m}(r_0,r) +  \bar x_{i\ell m}(M_1,S_1,r), \label{h1-form}\\
\R^2_{i\ell m} &= \hat\R^{2}_{i\ell m}(r_0,r_1,M_1,S_1,r) + \bar x_{i\ell m}(M_2,S_2,r),\label{h2-form}
\end{align}
where $\hat\R^{1}_{i\ell m} = \R^{\rm pp}_{i\ell m}$ is the usual linear perturbation due to a point particle on a circular orbit, and $M_n$ and $S_n$ are the coefficients in
\begin{align}
\delta M &= \e M_1(\tilde \w) + \e^2 M_2(\tilde \w) + \O(\e^3),\\
\delta S &= \e S_1(\tilde \w) + \e^2 S_2(\tilde \w) + \O(\e^3).
\end{align}
All the slow-time dependence is encoded in the dependence on $r_n(\tilde \w)$, $M_n(\tilde \w)$, and $S_n(\tilde \w)$. $\bar x_{i\ell m}(M_n,S_n,r)$ is the same function for all $n$, just with different arguments. It represents a linear perturbation toward a slowly evolving Kerr metric with mass and spin parameters $M+\e^n M_n(\tilde \w)$ and $\e^n S_n(\tilde \w)$; this has no fast-time dependence, containing only $\ell=0$ and $\ell=1,m=0$ contributions.

Once one has obtained the solutions~\eqref{h1-form} and \eqref{h2-form}, one can calculate the self-forces appearing in Eqs.~\eqref{r0dot-schematic}--\eqref{Omega1-schematic}. These forces depend only on the mode amplitudes $\R^n_{i\ell m}$; because $Y^{i\ell m}_{\alpha\beta}e^{- i m\phi_p}\propto e^{im(\phi - \phi_p)}$, the phases drop out of the metric perturbation on the worldline, where $\phi=\phi_p$, and therefore out of the force. The forces then have the schematic forms $f^\alpha_{1,\rm diss} = f^\alpha_{1,\rm diss}(r_0)$, $f^\alpha_{1,\rm cons} = f^\alpha_{1,\rm cons}(r_0,M_1,S_1)$, and $f^\alpha_{2,\rm diss} = f^\alpha_{2,\rm diss}(r_0,r_1, M_1,S_1)$.  Substituting these dependences into Eqs.~\eqref{r0dot-schematic}--\eqref{Omega1-schematic}, and noting that $\tilde\w=\tilde t$ on the orbit, we see that the orbital evolution equations take the form
\begin{align}
\Omega_0 &= \Omega_0(r_0), &&\frac{dr_0}{d\tilde s} = \dot r_0(r_0),\label{adiabatic}\\
\Omega_1 &= \Omega_1(r_0,r_1,M_1,S_1),\!  &&\frac{dr_1}{d\tilde s} = \dot r_1(r_0,r_1, M_1,S_1).\label{post-adiabatic}
\end{align}
We also have that
\begin{align}
\frac{dM_1}{d\tilde \w} = \dot E_H(r_0), \quad \frac{dS_1}{d\tilde \w} = \dot L_H(r_0),\label{Mdot and Sdot}
\end{align}
where $\dot E_H$ and $\dot L_H$ are the leading-order gravitational-wave fluxes of energy and angular momentum into the black hole; these depend only on the first-order mode amplitudes $R^{\rm pp}_{i\ell m}$ and therefore only on $r_0$. In Sec.~\ref{balance laws}, we re-derive Eq.~\eqref{Mdot and Sdot} from first principles, directly from our two-timescale field equations; at the same time, we re-derive the standard balance laws relating the loss of orbital energy and angular momentum, $\dot{\cal E}_0(r_0)$ and $\dot{\cal L}_0(r_0)$, to the total fluxes out of the system, $\dot E_H(r_0)+\dot E_\infty(r_0)$ and $\dot L_H(r_0)+\dot L_\infty(r_0)$.

\begin{figure}[t]
\centering
\includegraphics[width=\columnwidth]{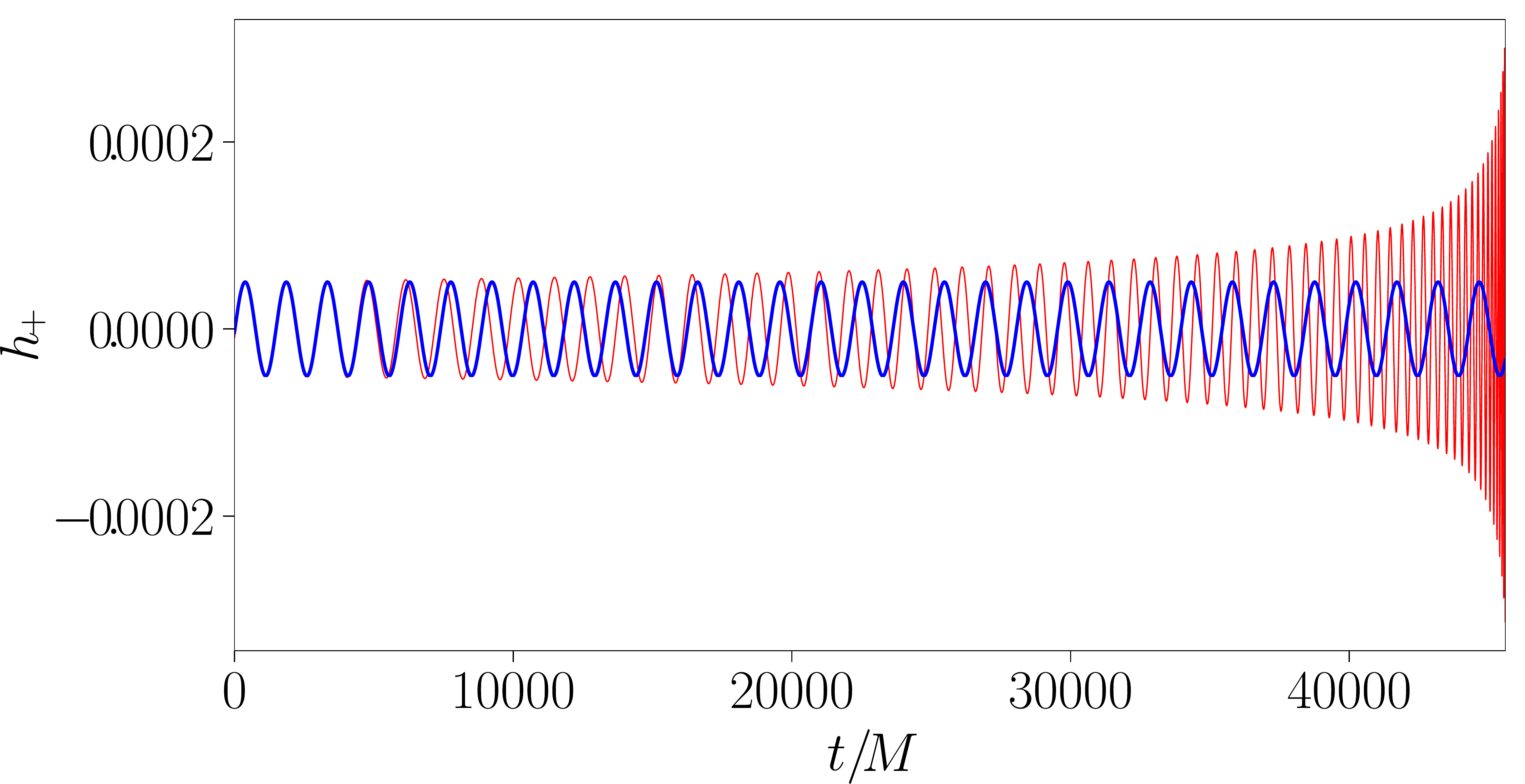}
\caption[]{The $\ell=2$, $m=\pm 2$ mode of the `$+$' polarization of the waveform at infinity in units of $\mu$. The thin red curve shows an adiabatic waveform, $h_+ =\displaystyle\lim_{r\to\infty} \sum a_{i\ell}\R^1_{i\ell m}(\e u, r)r^{-2}Y^{i\ell m}_{\theta\theta}e^{-im\int^u_0 \Omega_0(\e z)dz}$, where the sum is over $i=7,10$, $\ell=2$, $m=\pm2$. We have used an exaggerated mass ratio $\e=0.1$ to make the evolution clearly visible. For comparison, the thick blue curve shows the waveform produced by a point mass on a circular geodesic with the same initial frequency as the adiabatic orbit, $h_+ =\displaystyle\lim_{r\to\infty} \sum a_{i\ell}\R^1_{i\ell m}(0, r)r^{-2}Y^{i\ell m}_{\theta\theta}e^{-im\Omega_0(0)u}$.}
\label{fig_waveform}
\end{figure}

Equation~\eqref{adiabatic} represents an adiabatic evolution scheme. One can use $\frac{dr_0}{d\tilde\w} = \dot r_0(r_0)$ to evolve the leading-order orbital radius $r_0(\tilde s)$, recover the leading-order frequency $\Omega_0(r_0)$, and from it recover the adiabatic-order phase $\phi_p(s,\e) = \int^s \Omega_0(\e z) dz$. Since one has already computed the amplitudes $\R^{\rm pp}_{i\ell m}(r_0,r)$ as input for $\dot r_0(r_0)$, one then has the waveform  $\sum R^{\rm pp}_{i\ell m}e^{-im\int\Omega_0 ds}Y^{i\ell m}_{\alpha\beta}$ as an output (since it only evolves slowly, $\bar x_{i\ell m}$ is not required in the leading-order waveform). Fig.~\ref{fig_waveform} shows an adiabatic waveform generated with this scheme, using the  computational methods described in Ref.~\cite{Akcay:2010dx}. 

Equations~\eqref{adiabatic}--\eqref{Mdot and Sdot} together represent a post-adiabatic evolution scheme.  In this case one has four parameters to evolve: $r_0$, $r_1$, $M_1$, and $S_1$. From their evolution, one extracts the frequency evolution $\Omega_0(r_0)+\e\Omega_1(r_0,r_1,M_1,S_1)$; from the frequency, the post-adiabatic phase evolution $\phi_p(s,\e) = \int^s [\Omega_0(\e z)+\e\Omega_1(\e z)] dz$; and from the phase, the waveform. Explicitly, the two polarizations of the leading-order waveform are given by~\cite{Martel-Poisson:05} 
\begin{align}
h_{+} &= \lim_{r\to\infty}\sum \frac{a_{i\ell}}{r^2}\R^{\rm pp}_{i\ell m}(\tilde u,r)e^{-im\phi_p(u,\e)}Y^{i\ell m}_{\theta\theta},\\
h_{\times} &= \lim_{r\to\infty}\sum\frac{a_{i\ell}}{r^2 \sin^2\!\theta}\R^{\rm pp}_{i\ell m}(\tilde u,r) e^{-im\phi_p(u,\e)}Y^{i\ell m}_{\theta\phi}\!,
\end{align}
where the sums run over $i=7,10,\,\ell\geq 2,\, m\neq0$. Because the waveform's amplitude is not required to be highly accurate for matched filtering, it is unlikely that we would need to include the second-order mode amplitudes in this waveform.

\begin{figure*}[t]
\centering
\includegraphics[width=.76\textwidth]{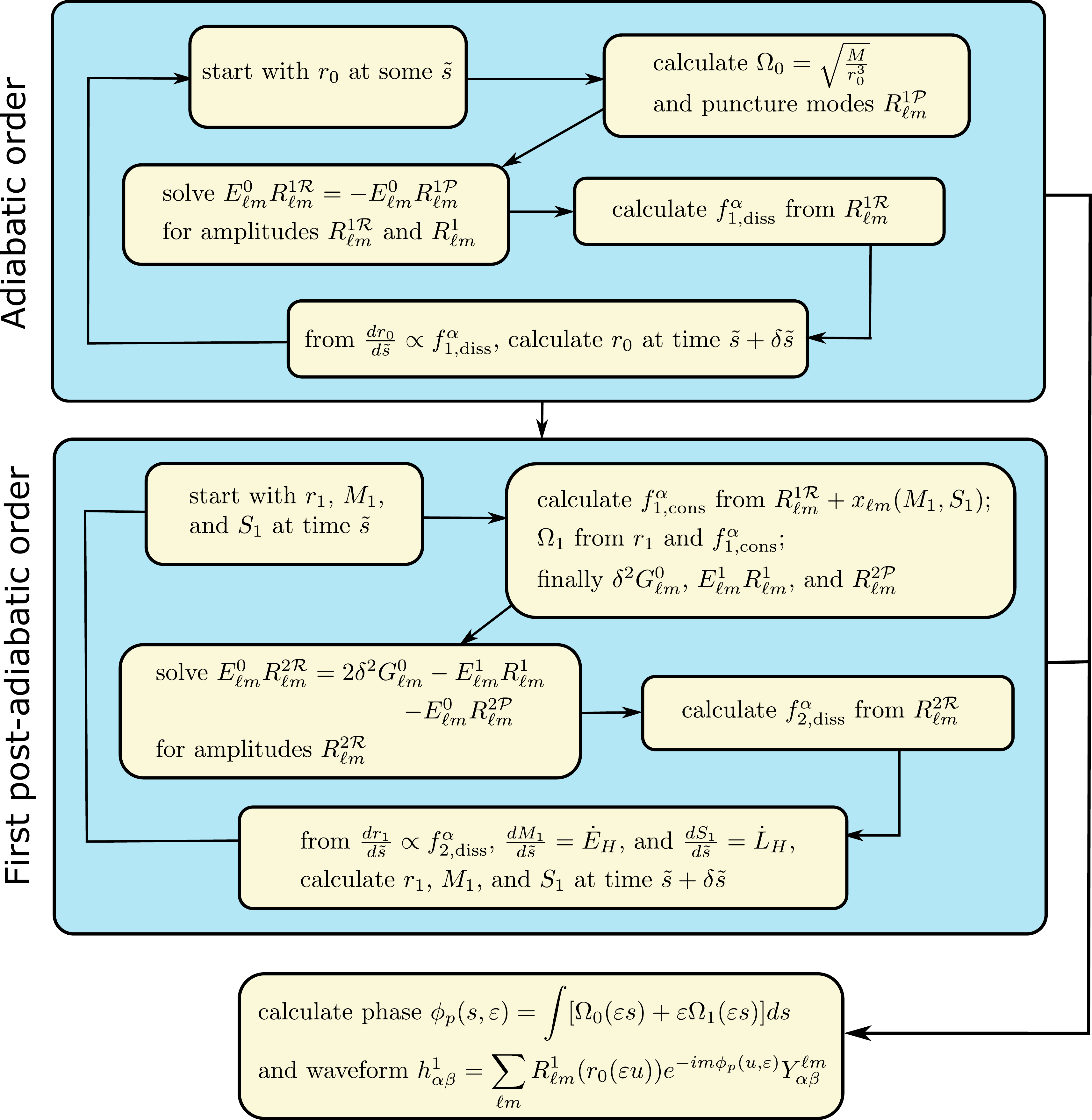}
\caption{Our procedure for computing the gravitational waveform to adiabatic and first post-adiabatic orders. The computation moves along a sequence of orbital radii, described as functions of slow time $\tilde s=\e s$ by $r_p=r_0(\tilde s)+\e r_1(\tilde s) +\O(\e^2)$. At each value of slow time, one solves (discrete) frequency-domain field equations for the metric-perturbation mode amplitudes. From the mode amplitudes, one computes the self-force, which determines the evolution to the next value of slow time as well as the post-adiabatic correction $\Omega_1$ to the orbital frequency. Once the mode amplitudes and frequency evolution are known, one can construct the full time-domain waveform. For simplicity we suppress details such as the tensor-harmonic basis labels and the numerical factor $a_{i\ell}$.}
\label{fig_flowchart}
\end{figure*}

%
%

Figure~\ref{fig_flowchart} shows the full sequence of steps required to generate adiabatic and post-adiabatic waveforms. The heart of this evolution scheme lies in solving the field equations~\eqref{EFE1 two-timescale schematic} and \eqref{EFE2 two-timescale schematic} at fixed values of slow time---i.e., for given points in the $(r_0,r_1,M_1,S_1)$ parameter space. The evolution then progresses through that parameter space. One could alter this evolution scheme in various ways. For example, one can instead work in the $(\Omega_0,\Omega_1,M_1,S_1)$ parameter space by simply rearranging Eqs.~\eqref{Omega0-schematic}--\eqref{Omega1-schematic} to obtain $r_0=r_0(\Omega_1)$ and $r_1=r_1(\Omega_0,\Omega_1,f_1^r)$. 
More significantly, one could adjust the evolution scheme to generate the waveform as a function of the full, nonperturbative frequency $\Omega$, reducing $\tilde\w$ to an auxiliary variable. This approach, which we describe in Appendix~\ref{fixed Omega}, hews slightly closer to our starting point in the self-consistent expansion, as it treats more of the particle's trajectory nonperturbatively. It also provides a convenient way to associate a local state of the binary to an asymptotic waveform: the two are naturally identified when their frequencies are the same.

But the core of the scheme, solving first- and second-order frequency-domain field equations, is largely independent of which of these approaches is taken. In all cases, the calculations at each point in the parameter space require (i) a practical method of solving the field equations, subject to given boundary conditions, (ii) a computation of the source terms in the field equations, and (iii) a specification of physical boundary conditions. We will present these remaining requirements in a sequence of followup papers~\cite{worldtube-paper, coupling-paper,source-paper,two-timescale-2,two-timescale-3}.


In this paper we use a mostly positive metric signature, $(-,+,+,+)$, and geometrical units with $G=c=1$. 
Indices are raised and lowered with the background metric $g_{\alpha\beta}$, and $\nabla$ and a semicolon both denote the covariant derivative compatible with $g_{\alpha\beta}$. $(t,r,\theta,\phi)$ denote Schwarzschild coordinates, and $\theta^A=(\theta,\phi)$. Unless otherwise stated, $g_{\alpha\beta}$ denotes the Schwarzschild metric, ${\rm diag}\left(-f,f^{-1},r^2,r^2\sin^2\theta\right)$, where $f:=1-2M/r$.

%
\section{Self-force theory in the self-consistent framework}\label{sec_field_equations}
%

In this section we review our starting point: self-force theory through second order in $\e$. We follow the self-consistent formalism of Refs.~\cite{Pound:10a,Pound:2015tma}, which provides the most direct line to a two-timescale approximation. However, as alluded to in the introduction, we extend previous descriptions to accurately incorporate the evolution of the central black hole.  Sec.~\ref{review of self-consistent expansion} reviews the standard description in the literature, and Sec.~\ref{inaccuracies} discusses what prevents that description from admitting a two-timescale expansion. Sec.~\ref{evolving background} describes how to overcome the problem using an evolving background spacetime, and Sec.~\ref{evolving perturbations} simplifies that formulation by working with small, slowly evolving perturbations rather than an evolving background.  The end result, Eqs.~\eqref{puncture_scheme_1 V2}--\eqref{puncture_scheme_2 V2}, is a self-consistent formalism that can be straightforwardly expanded in two-timescale form.

\subsection{Review of previous descriptions}\label{review of self-consistent expansion}

The self-consistent formalism begins by expanding ${\sf g}_{\alpha\beta}$ in the limit $\e\to0$ while holding $z^\mu(t,\e)$ fixed:
\begin{align}
{\sf g}_{\alpha\beta} &= g_{\alpha\beta}(x^\mu) + \sum_{n\geq1}\e^n h^n_{\alpha\beta}(x^\mu;z^\mu).\label{self-consistent expansion v0}
\end{align}
Here $g_{\alpha\beta}(x^\mu)$ is the background metric (e.g., the metric of the central Kerr black hole in an EMRI), $x^\mu$ are a set of background coordinates (e.g., Boyer-Lindquist coordinates on the Kerr background), and $h^n_{\alpha\beta}(x^\mu;z^\mu)$ is the $n$th-order perturbation due to the small object.\footnote{$\log\e$ terms also generically appear in this expansion~\cite{Pound:2012dk}. For visual simplicity, we hide these inside the coefficients $h^n_{\alpha\beta}$.}  The expansion~\eqref{self-consistent expansion v0} is taken to be accurate except in a small region around the small object, called the {\em body zone} or inner region, where the object itself is the dominant source of gravity. In the body zone, a second approximation is used, called a scaled or {\em inner expansion}. The two are linked using the method of matched asymptotic expansions. We find $h^n_{\alpha\beta}$ by solving the vacuum field equations {\em outside} the small object, subject to the {\em matching condition} that near the body zone, the metric~\eqref{self-consistent expansion v0} suitably agrees with, or ``matches'', the  inner expansion. 

Because $z^\mu$ depends on $\e$, the perturbations $h^n_{\alpha\beta}(x^\mu;z^\mu)$ do as well. This means we need a special formulation of the field equations. To see this, substitute the metric~\eqref{self-consistent expansion v0} into the vacuum Einstein equations $G_{\alpha\beta}[{\sf g}]=0$, and move nonlinear terms to the right-hand side, to obtain 
\begin{align}\label{expanded G = 0}
G_{\alpha\beta}[g] + \e\delta G_{\alpha\beta}[h^1]&+\e^2\delta G_{\alpha\beta}[h^2]\nonumber\\&= -\e^2\delta^2 G_{\alpha\beta}[h^1] + \O(\e^3), 
\end{align}
where $\delta G_{\alpha\beta}$ is the linearized Einstein tensor, and $\delta^2 G_{\alpha\beta}$ is the piece of $G_{\alpha\beta}[g+h]$ that is quadratic in $h_{\alpha\beta}$. Explicitly, in vacuum,
\beq\label{d2G}
\delta^2 G_{\alpha\beta} = \delta^2 R_{\alpha\beta} - \frac{1}{2}g_{\alpha\beta}g^{\mu\nu}\delta^2 R_{\mu\nu},
\eeq
where
\begin{align}
\delta^2 R_{\alpha\beta}[h] &=\frac{1}{2}\bar{h}^{\mu\nu}{}_{;\mu}\left(2 h_{\nu(\alpha;\beta)}- h_{\alpha\beta;\mu}\right)
+ \frac{1}{4} h^{\mu\nu}{}_{;\alpha} h_{\mu\nu;\beta}
\nonumber\\
&\quad  +\frac{1}{2}  h^{\mu}{}_{\beta}{}^{;\nu} h_{\mu\alpha;\nu} -\frac{1}{2}  h^{\mu}{}_{\beta}{}^{;\nu}  h_{\nu\alpha;\mu}
\nonumber\\
&\quad
-\frac{1}{2} h^{\mu\nu} \left( 2  h_{\mu(\alpha;\beta)\nu} - h_{\alpha\beta;\mu\nu}- h_{\mu\nu;\alpha\beta} \right)\!.
\label{second_order_ricci}
\end{align}
Because of the $\e$ dependence in  $h^n_{\alpha\beta}(x^\mu;z^\mu)$, we cannot naively equate coefficients of powers of $\e$ in Eq.~\eqref{expanded G = 0}. By virtue of the Bianchi identity, the Einstein equations constrain any material degrees of freedom in the system, and equating coefficients of powers of $\e$ in Eq.~\eqref{expanded G = 0} forces the $z^\mu$ in $h^1_{\alpha\beta}(x^\mu;z^\mu)$ to be an $\e$-independent geodesic $z^\mu_0$. The $\e$ dependence of the object's motion then appears in the higher-order perturbations $h^{n>0}_{\alpha\beta}$ in the form of deviation vectors $z_n^\mu$, which describe the small object's deviation away from $z^\mu_0$. In effect, the perturbative Einstein equations force the trajectory to be expanded in the form $z^\mu(\tau,\e)=z_0^\mu(\tau) + \e z^\mu_1(\tau) + \ldots $, with $h^1_{\mu\nu}$ only depending on $z_0^\mu$, $h^2_{\mu\nu}$ depending on $z_0^\mu$ and $z_1^\mu$, etc. This is the Gralla-Wald treatment of the motion, which (as described in the introduction) becomes inaccurate over an inspiral because the corrections $z_{n>0}^\mu$ grow large with time. 

We avoid this problem by casting the field equations in a {\em relaxed} form that does not constrain $z^\mu$. To achieve that, we impose the Lorenz gauge condition
\beq\label{gauge condition}
\nabla^\beta(\e\bar h^1_{\alpha\beta}+\e^2\bar h^2_{\alpha\beta}) = \O(\e^3).
\eeq
(Refs.~\cite{Pound:2012dk} and \cite{Pound:15b} discuss more general choices of gauge to achieve the same end.) This condition puts Eq.~\eqref{expanded G = 0} in the form of a weakly nonlinear wave equation, 
\beq
\e E_{\alpha\beta}[\bar h^1]+\e^2 E_{\alpha\beta}[\bar h^2] = 2\e^2\delta^2 G_{\alpha\beta}[h^1] + \O(\e^3),\label{relaxed EFE v0}
\eeq
where
\begin{equation}
{E}_{\alpha\beta}[\bar h] := \nabla^\mu\nabla_{\!\mu} \bar h_{\alpha\beta} + 2R^{\mu \ \nu \ }_{\ \alpha \ \beta} \bar h_{\mu\nu}.\label{Ealphabeta}
\end{equation}
Equation~\eqref{relaxed EFE v0} is the desired relaxed equation; it can be solved for arbitrary $z^\mu$. We can now equate coefficients of powers of $\e$ to obtain a solution that is valid for {\em all} $z^\mu$, yielding $G_{\alpha\beta}[g]=0$ for the background metric and 
\begin{align}
E_{\alpha\beta}[\bar h^1] &= 0,\label{EFE1 no puncture}\\
E_{\alpha\beta}[\bar h^2] &= 2\delta^2 G_{\alpha\beta}[h^1]\label{EFE2 no puncture}
\end{align}
for the perturbations.

The solutions to Eqs.~\eqref{EFE1 no puncture}--\eqref{EFE2 no puncture} are required to satisfy the matching condition. We can enforce that condition using a puncture scheme, which amounts to replacing the small object with a singular puncture in the spacetime geometry. The matching condition dictates that in a neighbourhood of $z^\mu$, the metric perturbations are required to satisfy
\begin{align} 
h^{1}_{\alpha\beta} &= h^{1\P}_{\alpha\beta}+h^{1\res}_{\alpha\beta},\label{near-z BC1}\\ 
h^{2}_{\alpha\beta} &= h^{2\P}_{\alpha\beta}+h^{2\res}_{\alpha\beta},\label{near-z BC2}
\end{align}
where $h^{n\P}_{\alpha\beta}$ are the puncture fields, which diverge on $z^\mu$, and $h^{n\res}_{\alpha\beta}$ are the residual fields, which satisfy certain regularity conditions on $z^\mu$. The puncture fields capture the dominant physical behavior of the metric near the small object, and they encode the object's multipole moments. 
If we define $z^\mu $ to be the object's center of mass, then for a generic compact object with spin $s^\mu$, the punctures have the schematic structure
\begin{align}
h^{1\P}_{\alpha\beta} &\sim \frac{\mu}{|x^\mu-z^\mu|} +\O(|x^\mu-z^\mu|^0),\label{h1P form}\\
h^{2\P}_{\alpha\beta} &\sim \frac{\mu^2+s^\mu}{|x^\mu-z^\mu|^2} + \frac{\mu h^{1\res}_{\alpha\beta}}{|x^\mu-z^\mu|} +\O(|x^\mu-z^\mu|^0),\label{h2P form}
\end{align}
given explicitly in covariant form in Ref.~\cite{Pound:2014xva} (with the exception of the $s^\mu$ term, given in local coordinates centered on $z^\mu$ in \cite{Pound:2012dk}). 
With the punctures known, we can move them to the right-hand side of the field equations and solve for the residual fields. This can be done in two equivalent ways: with a worldtube~\cite{Barack-Golbourn-Sago:07} or with a window function~\cite{Vega-Detweiler:07}. Here for simplicity we adopt the window method, in which one makes the puncture go to zero (smoothly or sharply) outside some region $\Gamma$ around $z^\mu$. The field equations then take the form
\begin{align}
E_{\alpha\beta}[\bar h^{1\res}] &= -E_{\alpha\beta}[\bar h^{1\P}],\label{puncture_scheme_1}\\
E_{\alpha\beta}[\bar h^{2\res}] &= 2\delta^2 G_{\alpha\beta}[h^1] - E_{\alpha\beta}[\bar h^{2\P}].\label{puncture_scheme_2}
\end{align}
Outside $\Gamma$, the residual field $\bar h^{n\res}_{\alpha\beta}$ reduces to the physical field $\bar h^{n}_{\alpha\beta}$. 

A solution to the wave equations~\eqref{puncture_scheme_1}--\eqref{puncture_scheme_2} only becomes a solution to the non-relaxed Einstein equations if it also satisfies the gauge constraint~\eqref{gauge condition}. By imposing this condition  on the fields near $z^\mu$, one determines that for a nonspinning, approximately spherical object, the object's center-of-mass trajectory is governed by the equation of motion~\cite{Pound:12a,Pound:2017psq}
\begin{align}
\frac{D^2 z^\mu}{d\tau^2}&= -\frac{1}{2}P^{\mu\nu}\left( g^{\ \rho}_\nu-h^{\res \rho}_\nu\right)\left( 2h^{\res}_{\beta\rho;\alpha}-h^{\res}_{\alpha\beta;\rho}\right)u^\alpha u^\beta\nonumber\\
&\quad + \O(\e^3),\label{SFhres}
\end{align}
where $P^{\mu\nu}:=g^{\mu\nu} +u^\mu u^\nu$ and  $h_{\alpha\beta}^{\res} = \e h^{1\res}_{\alpha\beta}+\e^2 h^{2\res}_{\alpha\beta}$. The punctures then move on this trajectory. The gauge condition also determines $d\mu/d\tau=\O(\e^3)$ and $Ds^\mu/d\tau=\O(\e^3)$. Note that here and throughout this paper, $\tau$ denotes proper time in $g_{\mu\nu}$, and $u^\mu:=dz^\mu/d\tau$ denotes the four-velocity normalized in $g_{\mu\nu}$.

The coupled set of equations~\eqref{puncture_scheme_1}--\eqref{SFhres} represent the self-consistent evolution scheme through second order. By sidestepping a Taylor expansion of $z^\mu$, this scheme avoids the associated, secularly growing errors that would occur in an ordinary, Gralla-Wald-type perturbative approximation. In the next section, we will  describe the remaining  secular errors that nevertheless {\em do} arise in this scheme. 

However, before doing so we remark on how the puncture formulation of self-force theory, used in the description above, relates to a point-particle description. In the field equations~\eqref{puncture_scheme_1}, $E_{\alpha\beta}[\bar h^{1\P}]$ is defined off $z^\mu$ as an ordinary function and defined on $z^\mu$ by taking the limit from off $z^\mu$. This makes the source terms regular at $z^\mu$, allowing us to enforce regularity of $\bar h^{1\res}_{\mu\nu}$. However, if we move the puncture $\bar h^{1\P}_{\alpha\beta}$ back to the left-hand side and treat the derivatives in $E_{\alpha\beta}[\bar h^{1\P}]$ distributionally, then we find that the total field $\bar h^1_{\alpha\beta}= \bar h^{1\P}_{\alpha\beta}+\bar h^{1\res}_{\alpha\beta}$ is precisely the field of a point mass, replacing Eq.~\eqref{puncture_scheme_1} with 
\beq\label{EFE1 - point mass}
E_{\alpha\beta}[\bar h^{1}] = -16\pi T^1_{\alpha\beta},
\eeq
where
\begin{equation}
T^1_{\alpha\beta}= \mu\int  u_\alpha u_\beta \frac{\delta^4\left[x^\mu-z^\mu\left(\tau\right)\right]}{\sqrt{-{\rm det}g}}d\tau\label{T1_alpha_beta}
\end{equation} 
is the stress-energy tensor of a point mass $\mu$ moving on $z^\mu$ in the background $g_{\mu\nu}$. Therefore, at first order the puncture scheme is equivalent to approximating the small object as a point particle. 

At second order in a generic gauge, the strongly singular behavior of $\delta^2 G_{\alpha\beta}[h^1]$ prevents us from straightforwardly writing a unique distributional source for the total field $\bar h^2_{\alpha\beta}=\bar h^{2\P}_{\alpha\beta}+\bar h^{2\res}_{\alpha\beta}$~\cite{Pound:2015tma} (see also Sec. 1.2.5 of Ref.~\cite{PhD_thesis}). However, forthcoming work~\cite{Upton-Pound:20} will show that there exists a canonical distributional interpretation of $\delta^2 G_{\alpha\beta}[h^1]$, under which the field equation takes the form 
\beq\label{EFE2 - point mass}
E_{\alpha\beta}[\bar h^{2}] = -16\pi T^2_{\alpha\beta} + 2\delta^2 G_{\alpha\beta}[h^1]
\eeq
with
\begin{equation}
T^2_{\alpha\beta}=-\frac{\mu}{2}\int_\gamma  u_\alpha  u_\beta Q^{\rho\sigma}h^{1\res}_{\rho\sigma}\frac{\delta^4\left[x^\mu-z^\mu\left(\tau\right)\right]}{\sqrt{-{\rm det} g}}d\tau.\label{T2_alpha_beta}
\end{equation} 
Here $Q^{\rho\sigma}:= g^{\rho\sigma}-u^\rho u^\sigma$. The sum $T_{\alpha\beta}=\e T^1_{\alpha\beta}+\e^2 T^2_{\alpha\beta}+\O(\e^3)$ can also be written as what we will call the {\em Detweiler stress-energy}
\begin{equation}
T_{\alpha\beta}=\int \mu\tilde u_\alpha \tilde u_\beta \frac{\delta^4\left[x^\mu-z^\mu\left(\tilde\tau\right)\right]}{\sqrt{-{\rm det}\tilde g}}d\tilde\tau,\label{T_alpha_beta}
\end{equation} 
where $\tilde u_\alpha:=\tilde g_{\alpha\beta}\frac{dz^\beta}{d\tilde\tau}$ is normalized in the effective metric $\tilde g_{\alpha\beta}=g_{\alpha\beta}+h^\res_{\alpha\beta}$, and $\tilde\tau$ is proper time in that metric. This stress-energy, introduced in Ref.~\cite{Detweiler:12}, is that of a point mass $\mu$ in $\tilde g_{\alpha\beta}$. Ref.~\cite{Upton-Pound:20}'s proof of its validity in Eq.~\eqref{EFE2 - point mass} is based on the class of {\em highly regular} gauges derived in Ref.~\cite{Pound:2017psq}, and there is considerable subtlety in meaningfully extending it outside that class. We will not explore these subtleties here, focusing instead on the more well-developed puncture formulation of the second-order problem. However, we will also make some use of this point-particle formulation. 

If the object has a spin $s^{\alpha}$, then the spin term in the puncture~\eqref{h2P form} is equivalent to an additional second-order stress-energy 
\beq
T^{2(\rm spin)}_{\alpha\beta} = \int u_{(\alpha} s_{\beta)}{}^\gamma  \nabla_\gamma\frac{\delta^4\left[x^\mu-z^\mu\left(\tau\right)\right]}{\sqrt{-{\rm det} g}}d\tau\label{T2 spin}
\eeq
where $s_{\alpha\beta}:=\epsilon_{\mu\alpha\beta\gamma}u^\mu s^\gamma$ and $\epsilon_{\mu\alpha\beta\gamma}$ is the Levi-Civita tensor; see, e.g., Ref.~\cite{Pound:2012dk}. We will not include this source in our two-timescale analysis, but doing so should be relatively straightforward~\cite{Akcay:2019bvk}. 

\subsection{Inaccuracies due to the black hole's evolution}\label{inaccuracies}

While accounting carefully for the evolution of $z^\mu$, our description above has ignored the evolution of another component of the system: the central black hole. We know that radiation falls into the black hole, causing it to evolve. The black hole's mass, for example, changes at an average rate~\cite{Teukolsky-Press:74, Poisson:04} $\left\langle\frac{dM_{BH}}{dt}\right\rangle\sim \e^2 M^2\partial h^1\partial h^1$. Over a radiation-reaction time $t\sim M/\e$, this accumulates to a change $\delta M \sim \e M$. Hence, on the radiation-reaction time scale, we expect the black hole mass to behave as $M_{BH} \sim M + \delta M(\e t)$, where $M$ is a constant, zeroth-order mass; the correction $\delta M$ remains $\sim\e M$ over the inspiral, and its rate of change is small as well, such that $dM_{BH}/dt\sim\e^2$. Here and below we focus on the mass, but analogous statements apply to the black hole's spin, which is expected to behave as $S_{BH} \sim aM + \delta S(\e t)$ with $\delta S\sim \e M^2$. 

These changes in the black hole must manifest themselves in the metric outside the black hole. In a two-timescale expansion, we expect that $h^1_{\mu\nu}$ will contain slowly evolving terms proportional to $\delta M$ and $\delta S$. This was the ansatz of Hinderer and Flanagan~\cite{Hinderer-Flanagan:08}, and at least in the case of a Schwarzschild background, the ansatz is borne out by a complete analysis of the Einstein field equations, sketched in Ref.~\cite{Pound-etal:19} and detailed later in this paper. 

On the other hand, if we were to perform an ordinary Taylor expansion of the metric, then the  evolving mass $M_{BH} \sim M +  \delta M(\e t) $ would become $M_{BH} \sim M + \delta M(0) + \e \frac{d\delta M}{d\tilde t}(0)t$. What was a slowly evolving first-order perturbation would hence become a linearly growing second-order perturbation. This approximation clearly goes bad on the radiation-reaction time, when $\e t\sim M$ (although it remains accurate much longer than a Gralla-Wald-type expansion, which fails on the dephasing time $t\sim M/\sqrt{\e}$ due to quadratic growth in $z^\mu_1$).

The question, then, is how these effects appear in the self-consistent formalism. Do they evolve dynamically on a long time scale, as they should in an accurate description, or do they grow linearly with time, at a forever-fixed rate, as they would in a strict Taylor series? 

Although we cannot answer this with complete certainty, we can provide strong evidence that the self-consistent evolution produces the less accurate, linear-growth description of the black hole. Consider prescribing initial data on a spatial slice $\Sigma$ and then evolving into the future of $\Sigma$ using the self-consistent equations~\eqref{EFE1 - point mass}, \eqref{EFE2 - point mass}, and \eqref{SFhres} [or \eqref{puncture_scheme_1}--\eqref{SFhres}]. Given a retarded Green's function $G_{\alpha\beta\alpha'\beta'}$ for the operator $E_{\alpha\beta}$, we can write the solution to this Cauchy-type problem as~\cite{Pound:2012dk}
\begin{align}
\bar h^1_{\alpha\beta} &= \int_\Sigma\left(\bar h^1_{\alpha'\beta'}\nabla_{\!\gamma'}G_{\alpha\beta}{}^{\alpha'\beta'}-G_{\alpha\beta}{}^{\alpha'\beta'}\nabla_{\!\gamma'}\bar h^1_{\alpha'\beta'}\right)d\Sigma^{\gamma'}\nonumber\\
							&\quad -16\pi\int_V G_{\alpha\beta}{}^{\alpha'\beta}T^1_{\alpha'\beta'}dV',\\ 
\bar h^2_{\alpha\beta} &= \int_\Sigma\left(\bar h^2_{\alpha'\beta'}\nabla_{\!\gamma'}G_{\alpha\beta}{}^{\alpha'\beta'}-G_{\alpha\beta}{}^{\alpha'\beta'}\nabla_{\!\gamma'}\bar h^2_{\alpha'\beta'}\right)d\Sigma^{\gamma'}\nonumber\\
							&\quad +\int_V G_{\alpha\beta}{}^{\alpha'\beta}\left(2\delta^2G_{\alpha'\beta'} -16\pi T^2_{\alpha'\beta'}\right)dV',\label{h2 Greens}
\end{align}
where $V$ is the future Cauchy development of $\Sigma$ and primes denote quantities at the integration point. This form splits the solution into two pieces: the integral over $V$, which is a particular solution to the wave equation~\eqref{EFE1 - point mass} or \eqref{EFE2 - point mass} with zero initial data; and the integral over $\Sigma$, which is a homogeneous solution to the wave equation with the prescribed initial data. The key fact is the following: the linearly growing perurbation proportional to $\dot{\delta M}|_\Sigma$ is a {\em homogeneous} solution to the wave equation, and for physical initial data, it always arises from the first integral in Eq.~\eqref{h2 Greens}. 

To illustrate this, we take the background spacetime to be Schwarzschild and $\Sigma$ to be the surface $t=0$. The initial data comprises the metric perturbation and its $t$ derivative at $t=0$. If the data describes a slice of time in a physical inspiral, then it must include a linear metric perturbation induced by $\delta M(0)$. If $\delta M$ were constant in time, then that perturbation could be written as $2 g_{\alpha\beta}\delta M(0)$, a homogeneous Lorenz-gauge perturbation (reviewed in Appendix~\ref{sec_low_modes}). But we must also include a nonzero $t$ derivative  $\dot{\delta M}(0)$ in the data. Imposing the Lorenz gauge condition at $t=0$ then demands that we add a $t$-$r$ component to the initial field, such that $h^{\delta M}_{\alpha\beta}(0) = 2 g_{\alpha\beta}\delta M(0) + \e p_{\alpha\beta}$, where $p_{\alpha\beta}=2\dot{\delta M}(0) (r-3M)f^{-1}\delta^t_{(\alpha}\delta^r_{\beta)}$. This form holds true even if we choose initial data with $\delta M(0)=0$, because we do not have the freedom to set $\dot{\delta M}(0)$ to zero; this is shown explicitly in Sec.~\ref{balance laws}. Finally, with $h^{\delta M}_{\alpha\beta}(0)$ as initial data, it is straightforward to show that the unique homogeneous solution to $E_{\alpha\beta}[\bar h]=0$  is $h^{\delta M}_{\alpha\beta}(t) = h^{\delta M}_{\alpha\beta}(0)+2\e g_{\alpha\beta}\dot{\delta M}(0)t$. Hence, unless the volume integral in Eq.~\eqref{h2 Greens} somehow miraculously cancels this homogeneous perturbation, the presence of a time-varying mass perturbation in the initial data automatically induces a linearly growing perturbation $h^{\delta M}_{\alpha\beta}(t)$. 

The induced perturbation $h^{\delta M}_{\alpha\beta}(t)$ is a physical one. It satisfies the Lorenz gauge condition in addition to the wave equation, and it correctly encodes the evolution of the black hole's mass. On time scales much shorter than the radiation-reaction time, it is perfectly suitable. But on the time scale of an inspiral, it becomes inaccurate. Moreover, it clearly does not admit a two-timescale expansion, meaning we cannot take the existing self-consistent expansion as our starting point for the two-timescale approximation. So we must (slightly) reformulate the self-consistent expansion. 


\subsection{Self-consistent approximation on an evolving background}\label{evolving background}

At its core, the self-consistent approximation  is based on a division of the system into its {\em field} degrees of freedom and its {\em mechanical} degrees of freedom. The field degrees of freedom are contained in the metric ${\sf g}_{\alpha\beta}$, and the mechanical degrees of freedom are in the trajectory $z^\mu$. The piece of the Einstein field equations that leaves $z^\mu$ unconstrained can be solved perturbatively by expanding ${\sf g}_{\alpha\beta}$ in a suitable perturbative series. The piece of the field equations that constrains $z^\mu$, on the other hand, can be used to obtain an approximate equation of motion for $z^\mu$ {\em without} expanding $z^\mu$ in a perturbative series.

However, our analysis above makes clear that we are putting different mechanical degrees of freedom on unequal footing: while we derive and actively enforce a uniformly accurate evolution equation for the nonperturbative $z^\mu$, we leave the evolution of the black hole parameters to be determined passively, and we end up with an approximation that is not uniformly accurate over long time scales. To obtain an accurate evolution, we must treat the black hole parameters in the same way we have treated $z^\mu$.

This motivates a generalization of the self-consistent framework. Rather than dividing the system into the metric and the trajectory, we divide it into the metric and the complete list of mechanical degrees of freedom: the effective positions and full set of multipole moments of both objects. We write the effective position of the large black hole as $Z^\mu$, and its set of multipole moments collectively as $\hat{M}$. Analogously, we write the effective position of the small object as $z^\mu$, and its multipole moments collectively as $\hat{m}$. The full set of mechanical degrees of freedom is $\hat{P}=\{Z^\mu,\hat{M},z^\mu,\hat{m})$. 

We now split the metric into a background plus a perturbation,
\begin{align}
{\sf g}_{\alpha\beta} &= \hat{g}_{\alpha\beta}(x^\mu;Z,\hat{M}) + \hat{h}_{\alpha\beta}(x^\mu,\e;\hat{P}).\label{self-consistent expansion v1}
\end{align}
Here the background metric $\hat{g}_{\alpha\beta}(x^\mu;Z,\hat{M})$ describes an evolving black hole spacetime, and $\hat{h}_{\alpha\beta}(x^\mu,\e;\hat{P})$ is the correction due to both the small object and the background's evolution. We imagine prescribing some concrete form for $\hat{g}_{\alpha\beta}(x^\mu;Z,\hat{M})$ in terms of moments with an unspecified time dependence. For concreteness, we take it to be a Kerr metric with mass $M_{\rm BH}$ and spin $S_{\rm BH}$ treated as arbitrary (order-$\mu^0$) functions of some time parameter. All its higher moments are then determined by the Kerr relationship $M_\ell + i S_\ell = M_{\rm BH}(iS_{\rm BH}/M_{\rm BH})^\ell$, where $M_\ell$ denotes the $\ell$th mass moment and $S_\ell$ the $\ell$th current moment. This choice  $\hat{g}_{\alpha\beta}$ sets $Z^\mu$ to be at the ``origin" of the Boyer-Lindquist coordinates, and the black hole's spin direction to be along the $\theta=0$ axis. In reality the large black hole's position and spin direction will be slightly perturbed by the small object's gravity, but these corrections are gauge perturbations (removable by a time-dependent translation and rotation of the coordinates) that we can incorporate into $\hat{h}_{\alpha\beta}$. We also specialize the small object to be compact, with a radius $\rho\sim \mu$, implying that its $\ell$th moments scale as $\mu\rho^\ell\sim \mu^{\ell+1}$.


For simplicity, we write the field equations formally with a stress-energy tensor rather than a puncture, as
\beq
G_{\alpha\beta}[\hat{g}] + \delta G_{\alpha\beta}[\hat{h};\hat{g}] + S_{\alpha\beta}[\hat{h};\hat{g}] = 8\pi T_{\alpha\beta}
\eeq
where $T_{\alpha\beta}=T_{\alpha\beta}[z,\hat{m},\hat{g}+\hat{h}^\res]$ is a point stress-energy associated with the small object; through order $\e^2$, it will be the sum of~\eqref{T_alpha_beta} plus \eqref{T2 spin} with $\tilde g_{\alpha\beta}$ replaced by $\hat{g}_{\alpha\beta}+\hat{h}^\res_{\alpha\beta}$. $S_{\alpha\beta}[\hat{h};\hat{g}]$ denotes the sum of all the terms in $G_{\alpha\beta}[\hat{g}+\hat{h}]$ that are nonlinear in $\hat{h}_{\alpha\beta}$. As mentioned above, a field equation of this form can be made well defined at least through order $\e^2$~\cite{Upton-Pound:20}. 

Due to the black hole's evolution, $\hat{g}_{\alpha\beta}$ is not precisely a vacuum metric, and we move its Einstein tensor to the right-hand side of the field equations, treating it as a source for the metric perturbation. We also once again impose the Lorenz gauge condition, defining the trace reversal and divergence with respect to $\hat{g}_{\alpha\beta}$, such that the field equations become 
\beq
\hat{E}_{\alpha\beta}[\hat{h}] = 2G_{\alpha\beta}[\hat{g}] +2S_{\alpha\beta}[\hat{h};\hat{g}] - 16\pi T_{\alpha\beta}.\label{EFE scr v0}
\eeq
Here $\hat{E}_{\alpha\beta}$ is given by Eq.~\eqref{Ealphabeta} with the covariant derivatives and Riemann tensor associated with $\hat{g}_{\alpha\beta}$. We write Eq.~\eqref{EFE scr v0} more compactly as
\beq
\hat{E}_{\alpha\beta}[\hat{h}] = \hat{S}_{\alpha\beta}[Z,\hat{M},z,\hat{m},\hat{h}].\label{EFE scr}
\eeq
In the true evolution, $\hat{g}_{\alpha\beta}$ is only {\em slowly} varying, meaning it is almost a vacuum metric. In that case, every term in the source  $\hat{S}_{\alpha\beta}$ is small. We have suggestively written the source as a function of $(Z,\hat{M})$ instead of $\hat{g}_{\alpha\beta}$, assuming the background metric is a prescribed function of $(Z,\hat{M})$. 


We can construct a solution to Eq.~\eqref{EFE scr} iteratively, introducing one additional multipole moment at each iteration and solving 
\beq
\hat{E}_{\alpha\beta}[\hat{h}^{(0)}] = \hat{S}_{\alpha\beta}[Z,\hat{M},z,\hat{m}=0,\hat{h}=0]
\eeq
for the zeroth iteration and
\beq
\hat{E}_{\alpha\beta}[\hat{h}^{(n)}] = \hat{S}_{\alpha\beta}[Z,\hat{M},z,\hat{m}^{(n-1)},\hat{h}^{(n-1)}]
\eeq
for subsequent iterations. Here $\hat{m}^{(n-1)}$ stands for the set of all of the small object's multipole moments up to its $(n-1)$th moment. Adopting the retarded solution at every iteration, we obtain
\begin{align}
\hat{h}^{(0)}_{\alpha\beta} &= 2\int \hat{G}_{\alpha\beta}{}^{\alpha'\beta'}G_{\alpha'\beta'}[\hat{g}]dV', \\
\hat{h}^{(1)}_{\alpha\beta} &= 2\int \hat{G}_{\alpha\beta}{}^{\alpha'\beta'}\big\{G_{\alpha'\beta'}[\hat{g}] + S_{\alpha'\beta'}[\hat{h}^{(0)};\hat{g}] \nonumber\\
&\quad- 8\pi T_{\alpha'\beta'}[z,\mu,\hat{g}+\hat{h}^{(0)}]\big\}dV',\\
\hat{h}^{(2)}_{\alpha\beta} &= 2\int \hat{G}_{\alpha\beta}{}^{\alpha'\beta'}\big\{G_{\alpha'\beta'}[\hat{g}] + S_{\alpha'\beta'}[\hat{h}^{(1)};\hat{g}] \nonumber\\
&\quad- 8\pi T_{\alpha'\beta'}[z,\mu,s,\hat{g}+\hat{h}^{(1)}]\big\}dV',
\end{align}
where $\hat{G}_{\alpha\beta}{}^{\alpha'\beta'}$ is the retarded Green's function associated with $\hat{E}_{\alpha\beta}$. In this way, $\hat{h}^{(0)}_{\alpha\beta}$ is fully determined by $Z^\mu$ and $\hat{M}$; $\hat{h}^{(1)}_{\alpha\beta}$ by $Z^\mu$, $\hat{M}$, $z^\mu$, and $\mu$; and $\hat{h}^{(2)}_{\alpha\beta}$ by $Z^\mu$, $\hat{M}$, $z^\mu$, $\mu$, and the spin $s^\mu$. In the limit $n\to\infty$, $\hat{h}^{(n)}_{\alpha\beta}$ should converge to a solution to Eq.~\eqref{EFE scr}.

The iterative solution can be found (at least formally) for any behavior of the mechanical degrees of freedom $\hat{P}$. The true behavior of $\hat{P}$ is then found by enforcing the Lorenz gauge condition, ensuring that the solution satisfies the full Einstein equation and not just the relaxed one. Here we will only conjecture that one can use the gauge condition to systematically derive uniform-in-time evolution equations for $\hat{M}$, which would complement the existing equations for $z^\alpha$ and $s^\alpha$. It will become clear from the two-timescale analysis in Sec.~\ref{balance laws} that if these self-consistent evolution equations are to admit a two-timescale solution, they must be compatible with the expected laws: the black hole's mass and spin must evolve according to the gravitational-wave flux down the horizon.

From this iterative solution, if desired, one can extract an expansion analogous to~\eqref{self-consistent expansion v0},
\beq\label{self-consistent h}
\hat{h}^{(n)}_{\alpha\beta}(x^\mu,\e;\hat{P}) =  \sum_{p=0}^n\e^n \hat{h}^{(n,p)}_{\alpha\beta}(x^\mu;\hat{P}),
\eeq
where the terms in the expansion are obtained by reading off explicit factors of $\e$ that come from the scaling of the multipole moments. For example,
\begin{align}
\hat{h}^{(1,0)}_{\alpha\beta}(x^\mu;\hat{P}) &= 2\int\! \hat{G}_{\alpha\beta}{}^{\alpha'\beta'}\big\{G_{\alpha'\beta'}[\hat{g}]\nonumber\\
		&\quad+ S_{\alpha'\beta'}[\hat{h}^{(0)};\hat{g}] \big\}dV',\\
\hat{h}^{(1,1)}_{\alpha\beta}(x^\mu;\hat{P}) &= -16\pi\!\int\! \hat{G}_{\alpha\beta}{}^{\alpha'\beta'}T^1_{\alpha\beta}[z,\hat{g}+\hat{h}^{(0)}]dV',\!
\end{align}
where $T^1_{\alpha\beta}[z,\hat{g}+\hat{h}^{(0)}]$ is given by Eq.~\eqref{T1_alpha_beta} with the four-velocity, proper time, and metric determinant all defined with respect to $\hat{g}_{\alpha\beta}+\hat{h}^{(0)}_{\alpha\beta}$. Note that in this power counting,  $\hat{h}^{(0)}_{\alpha\beta}$ is formally of order 1, and the source $S_{\alpha\beta}[\hat{h}^{(0)};\hat{g}]$ can be infinitely nonlinear in $\hat{h}^{(0)}_{\alpha\beta}$. However, once we substitute the true time dependence of the multipole moments, as determined from the gauge condition, $\hat{h}^{(0)}_{\alpha\beta}$ will be small because $G_{\alpha\beta}[\hat{g}]$ will be small. This means that in practice, one would be able to truncate $S_{\alpha\beta}[\hat{h}^{(0)};\hat{g}]$ at a finite power of $\hat{h}^{(0)}_{\alpha\beta}$.

We will not belabour this construction, as we will move onto a simpler one in the next section.


\subsection{Self-consistent approximation with evolving mass and spin perturbations}\label{evolving perturbations}

The problem becomes more tractable if we note that over the course of an inspiral, the background mass and spin change only by an amount of order $\e$. We can therefore divide $\hat M = (M_{\rm BH},S_{\rm BH})$ into constant, zeroth-order parameters $P=(M,S)$ and small, evolving corrections $\delta P = (\delta M,\delta S)$. 

Given this division of the black hole mechanical degrees of freedom, we may re-expand the metric at fixed $\{P,p\}$, where $p=\{z^\mu,\hat{m},\delta P\}$ is the complete list of perturbative degrees of freedom. 
Equation~\eqref{self-consistent expansion v1} then becomes
\begin{align}
{\sf g}_{\alpha\beta} &= g_{\alpha\beta}(x^\mu;P)  + h_{\alpha\beta}(x^\mu,\e;P,p),\label{self-consistent expansion}
\end{align}
where the background  $g_{\alpha\beta}$ is now an exact Kerr metric with parameters $P$, and the perturbation is $h_{\alpha\beta} = \hat h_{\alpha\beta}(x^\mu,\e;P,p) + \delta g_{\alpha\beta}(x^\mu,\e;P,\delta P)$ with $\delta g_{\alpha\beta}:=\hat{g}_{\alpha\beta}- g_{\alpha\beta}$.  $h_{\alpha\beta}$ contains all the information about the small mechanical parameters. Both pieces of the perturbation can be expanded with fixed mechanical parameters, as 
\begin{align}
\hat h_{\alpha\beta}(x^\mu,\e;P,p) &=\sum_{n>0}\e^n \hat h^n_{\alpha\beta}(x^\mu;P,p),\\
\delta g_{\alpha\beta}(x^\mu,\e;P,p) &=\sum_{n>0}\e^n g^n_{\alpha\beta}(x^\mu;P,\delta P),
\end{align}
where
\beq
 g^{n}_{\alpha\beta}:=\frac{1}{n!}\left[\frac{d^n}{d\e^n}\hat g_{\alpha\beta}(M+\e \delta M, S+\e\delta S)\right]_{\e=0}
\eeq
(holding $\delta M$ and $\delta S$ fixed while evaluating the derivatives). Or in total,
\beq\label{self-consistent h v2}
h_{\alpha\beta}(x^\mu,\e;P,\hat{p})=\sum_{n>0}\e^n h^n_{\alpha\beta}(x^\mu;P,p).
\eeq
At the first few orders, this reads
\begin{align}
{\sf g}_{\alpha\beta} &=  g_{\alpha\beta}(x;P) 
								+\e h^1_{\alpha\beta}(x;P,\delta P,z,\mu) \nonumber\\
								&\quad + \e^2 h^2_{\alpha\beta}(x;P,\delta P,z,\mu,s)+\O(\e^3).\label{self-consistent expansion organized}
\end{align}

Just as in the case of an evolving background, we wish to treat the evolving $\delta P$ as a source in the field equations. If we impose the Lorenz gauge condition~\eqref{gauge condition} on $h_{\alpha\beta}$ and move $\delta g_{\alpha\beta}$ to the right-hand side of the field equations, we obtain the relaxed Einstein equation
\begin{align}
\e E_{\alpha\beta}[\hat{\bar h}^1]+\e^2 E_{\alpha\beta}[\hat{\bar h}^2] &= 2\e^2\delta^2 G_{\alpha\beta}[h^1] -\e E_{\alpha\beta}[\bar{ g}^1]  \nonumber\\
&\quad -\e^2 E_{\alpha\beta}[\bar{ g}^2]+ \O(\e^3).\label{relaxed EFE v3}
\end{align}
Demanding that $\hat h^n_{\alpha\beta}$ satisfies these field equations for arbitrary $\hat{p}$, we have
\begin{align}
E_{\alpha\beta}[\hat{\bar h}^1] &= -16\pi T^1_{\alpha\beta} - E_{\alpha\beta}[\bar{ g}^1],\\
E_{\alpha\beta}[\hat{\bar h}^2] &= -16\pi T^2_{\alpha\beta} +2\delta^2 G_{\alpha\beta}[h^1]- E_{\alpha\beta}[\bar{ g}^2]. 
\end{align}
Or in a puncture scheme,
\begin{align}
E_{\alpha\beta}[\hat{\bar h}^{1\res}] &= - E_{\alpha\beta}[\bar{ g}^1] - E_{\alpha\beta}[\hat{\bar h}^{1\P}],\label{puncture_scheme_1 V2}\\
E_{\alpha\beta}[\hat{\bar h}^{2\res}] &= 2\delta^2 G_{\alpha\beta}[h^{1}] - E_{\alpha\beta}[\bar{ g}^2]- E_{\alpha\beta}[\hat{\bar h}^{2\P}].\label{puncture_scheme_2 V2} 
\end{align}
We seek the retarded solutions to these equations for arbitrary $\delta M$ and $\delta S$.

Evolution equations for $z^\mu$, $\mu$, $s^\mu$, $\delta M$, and $\delta S$ are then to be determined from the gauge condition~\eqref{gauge condition}. The local analysis near the small object, which led to Eq.~\eqref{SFhres}, $d\mu/d\tau=\O(\e^3)$, and $Ds^\mu/d\tau=\O(\e^3)$ in prior work, goes through essentially unchanged. A similar analysis near the horizon of the central black hole should yield (uniform-in-time) evolution equations for $\delta M$ and $\delta S$. However, as in the case of an evolving background, we note that in order to admit a two-timescale solution, those evolution equations must take the expected form in terms of fluxes of energy and angular momentum down the horizon. We therefore put aside any further analysis of this modified scheme, satisfying ourselves that it is now, at least in principle, compatible with a two-timescale expansion. In Sec.~\ref{balance laws} we will then use the two-timescale scheme to derive the standard flux-balance equations for $\delta M$ and $\delta S$.

At this stage, astute readers will have noticed that if we simply move $ g^n_{\mu\nu}$ back to the left-hand side of the field equations~\eqref{puncture_scheme_1 V2} and \eqref{puncture_scheme_2 V2}, then this new self-consistent framework recovers exactly the equations in the usual self-consistent scheme, ~\eqref{puncture_scheme_1} and \eqref{puncture_scheme_2}. What, then, has changed? In the old scheme, we had no way of enforcing that $h^1_{\mu\nu}$ includes a dynamically evolving $ g^1_{\mu\nu}$. Our method of solving the field equations implicitly treated $ g^1_{\mu\nu}$ as the Taylor series $ g^1_{\mu\nu}|_{t=0}+\e t \left.\frac{d g^1_{\mu\nu}}{d\tilde t}\right|_{t=0}+\O(\e^2)$, such that the evolution of $ g^1_{\mu\nu}$ covertly slipped into $h^{2}_{\mu\nu}$ in the form of a linearly growing perturbation. In the altered scheme, the perturbations $ g^n_{\mu\nu}$ instead play the role of punctures. Just as the usual punctures on $z^\mu$ enforce the correct physical behaviour near the small object, the punctures $ g^n_{\mu\nu}$ enforce the correct long-term behavior. More concretely, if we refer back to the discussion in Sec.~\ref{inaccuracies}, we see that the initial data which seeded the unwanted linear growth in $h^2_{\mu\nu}$ will not be included in the initial data for $\hat h^2_{\mu\nu}$; everything linear in $\delta M$ and its derivatives will already be accounted for in $h^1_{\mu\nu}$.

However, for simplicity in later sections, we {\em will} move $ g^n_{\mu\nu}$ back to the left-hand side of the field equations, as in Eqs.~\eqref{puncture_scheme_1}--\eqref{puncture_scheme_2}. The reason this will pose no problem is that the two-timescale framework will automatically enforce the correct long-term behavior of the solution. In that context, the quantity $x_{\mu\nu}$ that appears in the two-timescale solution will implicitly replace $ g^1_{\mu\nu}$.


%

Before proceeding to the multiscale expansion, we make one comment. Rather than starting from the self-consistent framework, one could instead build a two-timescale approximation  starting from the Gralla-Wald formalism~\cite{Gralla:08,Gralla:2012db}. As discussed above, this approach cannot remain accurate on the radiation-reaction time. However, one could still utilize its results to inform a two-timescale expansion by performing a short-time re-expansion of the two-timescale variables around each fixed value of slow time and matching the results, term by term, to the Gralla-Wald variables.\footnote{This would be made (slightly) more difficult again if Gralla's explicit second-order results in Ref.~\cite{Gralla:2012db} were used, because Gralla's gauge appears to exhibit growth in time that would not arise even in a short-time expansion of the two-timescale metric; this extra growth can be inferred from the combination of Gralla's Eqs.~(80) and (83) with his Eq. (B4)-(B6). The matching procedure would hence also require a gauge transformation.}

\section{Overview of two-timescale expansions}\label{sec_twotimescale_expansion}

Before performing the two-timescale expansion of our system of equations, we begin with a review of the method.

Consider a differential equation 
\beq\label{Dpsi=S}
D\psi(t,x^a,\e)=S(t,x^a,\e), 
\eeq
where $D$ is a linear differential operator, $x^a$ are a set of $n$ spatial coordinates, and $\e$ is a small parameter. Suppose we know that the source $S$ and solution $\psi$ are characterized by a very nearly discrete frequency spectrum, but with a slowly evolving set of discrete frequencies $\omega_k=k\omega\sim\e^0$. For simplicity, assume the timescale over which $\omega$ varies is $\sim 1/\e$, such that $\omega=\omega(\e t)$ [or more generally, $\omega=\omega(\e t,\e)$]. The system is then characterized by two timescales: a short time scale $1/\omega\sim \e^0$, and a long time scale $1/\e$. We can characterize the system's dependence on these time scales by introducing two new time variables: a {\it fast time} $\varphi(t,\e)\sim\e^0\, t$ given by $\frac{d\varphi}{dt}=\omega$; and a {\it slow time} $\tilde t(t,\e)=\e t$. $\tilde t$ changes appreciably only over the long time scale $t\sim 1/\e$, while $\varphi$ changes on the time scale $\sim 1/\omega$.

Things very quickly go wrong if we attempt to solve Eq.~\eqref{Dpsi=S} by assuming regular asymptotic expansions $\psi(t,x^a,\e)=\sum_{n=0}^\infty \e^n\psi_n(t,x^a)$ and $S(t,\e)=\sum_{n=0}^\infty \e^n S_n(t,x^a)$. The exact, slowly evolving solution will contain oscillatory functions of the fast time, of the form $e^{ik\varphi(t,\e)}$. If we expand this in a regular power series, using 
\begin{align}
\varphi &= \int_0^t \omega(\e t) \nonumber\\
		&= \int_0^t[\omega(0)+\e t \dot\omega(0)+\O(\e^2)]dt \nonumber\\
		&= \omega(0)t+\frac{1}{2}\e \dot\omega(0) t^2+\O(\e^2),\label{varphi-Taylor}
\end{align}
we obtain 
\beq
e^{ik\varphi(t,\e)}=e^{ik\omega(0)t}\left[1+\frac{ik}{2}\e\dot\omega(0)t^2+\O(\e^2)\right].
\eeq
What was a sinusoid with a slowly varying frequency and constant amplitude has become a sinusoid with a constant frequency and growing amplitude. After a time $t\sim1/\e$, the sinusoid's frequency $\omega(0)$ will differ by order $\e^0$ from the true frequency $\omega(\e t)$. After that same time, the sinusoid's amplitude will be $\sim 1/\e\gg 1$, rendering the expansion entirely invalid. For a similar reason, the $\omega_k=0$ piece of the solution also fails. A slowly varying solution of the form $\psi_0(\e t)$ becomes $\psi_0(0)+\e \dot\psi_0(0)t+\O(\e^2)$. After a time $t\sim1/\e$, every term in the expansion becomes of order $\e^0$. More dramatically, the expansion of $\varphi$ in Eq.~\eqref{varphi-Taylor} breaks down on the shorter {\em dephasing time} $t\sim 1/\sqrt{\e}$; this is especially disastrous in gravitational-wave data analysis, where accurate phasing is paramount. 

The multiscale approximation method eliminates these errors by treating $\tilde t$ and $\varphi$ as independent variables, transforming the $(n+1)$-dimensional problem~\eqref{Dpsi=S} into a more tractable $(n+2)$-dimensional problem. 

First, we assume that $\psi(t,x^a,\e)$ can be written as $\psi(t,x^a,\e)=\tilde\psi(\e t, \varphi(t,\e),x^a,\e)$,\footnote{We can relax this to the weaker assumption that $\psi(t,x^a,\e)$ is uniformly approximated, on an interval of time $\sim 1/\e$, by the asymptotic series~\eqref{multiscale_expansion}. That is, if we define $\tilde \psi_N(\tilde t,x^a,\varphi,\e)=\sum_{n=0}^N \e^n \psi_n(\tilde t,\varphi,x^a)$, then $\lim_{\e\to0}\frac{\sup|\psi(t,x^a,\e)-\psi_N(\e t,\varphi(t,\e),x^a,\e)|}{\e^N}=0$ for all integers $N\geq 0$, where the supremum is taken over $t\in[T_1,T_2/\e]$ for fixed $x^a$ and some fixed $T_{1,2}\in\mathbb{R}$.} where the function $\tilde\psi(\tilde t,x^a,\varphi,\e)$ has a regular asymptotic expansion at fixed $\tilde t$, $\varphi$, and $x^a$,
\begin{equation}
\tilde \psi\left(\tilde t,\varphi, x^a,\e\right)=\sum_{n=0}^\infty \e^n \tilde\psi_n(\tilde t,\varphi,x^a).\label{multiscale_expansion}
\end{equation}
Each coefficient $\psi_n$ is further assumed to be (i) a periodic function of $\varphi$ and (ii) a bounded function of $\tilde t$ (at fixed $x^a$). Analogously, we write $S(t,x^a,\e)=\tilde S(\e t, \varphi(t,\e),x^a,\e)$, where $\tilde S$ is the asymptotic series
\beq
\tilde S(\tilde t,\varphi,x^a,\e) = \sum_{n=0}^\infty \e^n \tilde S_{n}(\tilde t,\varphi,x^a). 
\eeq
These expansions can be compared to the self-consistent approximation, in which the expansions were carried out while holding fixed the mechanical functions rather than slow and fast times. 

When we substitute $\psi=\tilde \psi$ into Eq.~\eqref{Dpsi=S}, we can evaluate time derivatives using the chain rule, 
\begin{align}
\frac{d\psi}{d t} &= \frac{d\varphi}{dt}\frac{\partial\tilde\psi}{\partial \varphi} + \frac{d\tilde t}{dt} \frac{\partial \tilde\psi}{\partial \tilde t}\nonumber\\
&= \omega\frac{\partial \tilde\psi}{\partial \varphi} + \e \frac{\partial \tilde\psi}{\partial \tilde t}.\label{chain_rule}
\end{align}
The operator $D$ then becomes a differential operator $\tilde D$ on the $(n+2)$-dimensional manifold charted by $(\tilde t,\varphi,x^a)$, and Eq.~\eqref{Dpsi=S} becomes
\beq\label{Dpsi=S two-timescale}
\tilde D \tilde\psi = \tilde S.
\eeq
This is completely equivalent to Eq.~\eqref{Dpsi=S} when evaluated at $\tilde t = \e t$ and $\varphi=\varphi(t,\e)$. But we now consider it as a partial differential equation rather than an ordinary one, with $\tilde t$ and $\varphi$ as independent variables, and with periodicity in $\varphi$ providing an additional boundary condition. Thus, we effectively define $\tilde\psi(\tilde t,\varphi,x^a,\e)$ to be the periodic-in-$\varphi$ solution to Eq.~\eqref{Dpsi=S two-timescale}, to any desired order in $\e$. By construction, when evaluated at $\tilde t = \e t$ and $\varphi=\varphi(t,\e)$, $\tilde\psi$ will then also be a solution to the original equation~\eqref{Dpsi=S}, to any desired order in $\e$, with the desired property of approximate periodicity. 

To solve \eqref{Dpsi=S two-timescale}, we now fully expand it in powers of $\e$ at fixed $\tilde t$ and $\varphi$. In the case that $\omega$ depends on both $\tilde t$ and $\e$, this requires that we assume that it has an asymptotic expansion
\beq\label{frequency expansion}
\omega(\tilde t,\e) = \sum_{n=0}^\infty\e^n\omega_n(\tilde t).
\eeq
Given the expansion~\eqref{frequency expansion}, we can write Eq.~\eqref{chain_rule} as
\beq
\frac{d\psi}{dt} = \omega_0 \partial_\varphi \tilde\psi + \e(\partial_{\tilde t}+\omega_1 \partial_{\varphi})\tilde\psi + \O(\e^2).
\eeq

Due to their periodicity, the coefficients can be expanded in a discrete Fourier series 
\beq
\tilde\psi_n(\tilde t,\varphi,x^a) = \sum_{k=-\infty}^{+\infty}\tilde\psi_{nk}(\tilde t,x^a)e^{-ik\varphi}.
\eeq
Analogously,
\beq
\tilde S_n(\tilde t,\varphi,x^a) =  \sum_{k=-\infty}^{+\infty} \tilde S_{nk}(\tilde t,x^a)e^{-ik\varphi}. 
\eeq
This means the $\varphi$ derivatives can be explicitly evaluated, giving us
\begin{align}
\frac{ d\psi_n}{d t} = \sum_k\left[-ik\omega_0 \psi_{nk} + \e \tilde\partial_{\tilde t}\psi_{nk}+\O(\e^2)\right] e^{-ik\varphi},
\end{align}
where $\tilde\partial_{\tilde t} := \partial_{\tilde t}-ik\omega_1$. 

This treatment of derivatives leads to the operator $\tilde D$ becoming $\tilde D_{0,k}+\e \tilde D_{1,k}+\O(\e^2)$, and the original differential equation~\eqref{Dpsi=S} becoming
\beq
\sum_{n,j,k}\e^{n+j} [\tilde D_{j,k}\tilde\psi_{n,k}(\tilde t,x^a)]e^{-ik\varphi} = \sum_{n,k}\e^n \tilde S_{n,k}(\tilde t,x^a)e^{-ik\varphi}.
\eeq
If we restrict this equation to the submanifold $\tilde t=\e t$ and $\varphi=\varphi(t,\e)$, then we are not guaranteed that we can equate coefficients of explicit powers of $\e$ or Fourier coefficients in this equation. However, since we define $\tilde\psi$ to be the solution to Eq.~\eqref{Dpsi=S two-timescale} for {\em all} $\tilde t$ and $\varphi$, we {\em are} given that guarantee. Hence, we have
\beq\label{Dpsi=S-decomposed}
\sum_{j+n'=n}\tilde D_{j,k}\tilde\psi_{n',k}(\tilde t,x^a) = \tilde S_{n,k}(\tilde t,x^a).
\eeq
This is a set of partial differential equations on the new manifold charted by $(\tilde t,x^a)$. Solving these equations then determines the amplitudes' evolution with slow time. Restricting back to the submanifold defined by $\tilde t=\e t$ and $\varphi=\varphi(t,\e)$, we obtain the desired quasiperiodic solution
\beq
\psi(t,x^a,\e) = \sum_{n,k} \e^n\tilde\psi_{n,k}(\e t,x^a)e^{-ik\int^t\omega(\e s,\e)ds}.
\eeq 

The description in this section is of the textbook method of multiple scales. In our particular problem, an alternative approach is to perform an expansion at fixed $\omega$ instead of fixed $\tilde t$. The end result is then an expansion of the form $\psi(t,x^a,\e) = \sum_{n,k} \e^n\tilde\psi_{n,k}(\omega(\e t,\e),x^a)e^{-ik\int^t\omega(\e s)ds}$. This is one step less removed from the self-consistent method, as it treats the system's parameters nonperturbatively. We explore this approach in Appendix~\ref{fixed Omega}.

Ultimately, the validity of expansions such as these should be established a posteriori by showing that (i) all the equations~\eqref{Dpsi=S-decomposed} can be solved, and (ii) the coefficients $\tilde\psi_{n,k}$ are uniformly order 1 on any interval $t\in [T_1,T_2/\e]$; i.e., $\lim_{\e\to0}\sup|\psi_{n,k}(\e t)| < \infty$. (ii) rules out behavior like $\tilde\psi_{n,k} = 1/\tilde t$, which would be very large when $t\ll 1/\e$.

\section{Two-timescale expansion of the orbital dynamics}\label{sec_expanded_worldline}

We now apply the two-timescale prescription to the worldline $z^\alpha$ and its equation of motion~\eqref{SFhres}. Our presentation closely follows Sec.~IVA of Ref.~\cite{Pound:2015wva}. 

\subsection{Parametrization of quasicircular orbits}

Without loss of generality, we place the orbit in the equatorial plane, parametrizing it as
\begin{equation}\label{quasicircular}
z^\mu(t,\e) = \left(t, r_p(t,\e),\pi/2,\phi_p(t,\e)\right).
\end{equation}

To specialize to quasicircular orbits, we assume that the orbital radius $r_p$ and orbital frequency $\Omega:=\frac{d\phi_p}{dt}$ evolve slowly, on the radiation-reaction timescale $\Omega/\dot \Omega \sim 1/\e$, with no oscillations on the orbital timescale $1/\Omega\sim \e^0$. More explicitly, we assume that they can be written as slowly varying functions
\begin{align}
r_p(t,\e)= r_0(\e t)+\e\, r_1(\e t)+\O(\e^2),\label{rp-expansion}\\
\Omega(t,\e)= \Omega_0(\e t)+\e\, \Omega_1(\e t)+\O(\e^2). \label{Omegap-expansion}
\end{align}
Here we have introduced the natural slow time variable
\begin{equation}
\tilde t(t,\e)=\e t.\label{choice_of_slowtime_worldline}
\end{equation}

Our natural fast time variable is the orbital phase,
\begin{equation}
\phi_p(t,\e) = \int_0^t \Omega(s, \e) ds +\phi_p(0,\e).\label{choice_of_fasttime_worldline}
\end{equation}
As one might expect, for quasicircular inspirals the orbital dynamics is independent of this fast time; we will show this in later sections. However, the metric of the full, evolving system will have a periodic dependence on this phase, and it will play a critical role in the two-timescale expansion of the field equations.

\subsection{Equation of motion}\label{sec_expanded_eom}

We now substitute our expansion of the worldline into the equation of motion~\eqref{SFhres}. This will lead to a sequence of  equations for each $r_n$ and $\Omega_n$. 

Written in terms of the non-affine parameter $t$, Eq.~(\ref{SFhres}) reads
\begin{equation}\label{eq-mot}
\frac{d^2 z^\mu}{dt^2}+U^{-1}\frac{dU}{dt}\frac{dz^\mu}{dt}+\Gamma^\mu_{\beta\gamma}\frac{d z^\beta}{dt} \frac{dz^\gamma}{dt}=U^{-2} f^\mu,
\end{equation}
where $U:=u^t$, and where $f^\mu$ is the self force per unit mass, given by the right-hand side of Eq.~\eqref{SFhres}.

To expand Eq.~\eqref{eq-mot}, we will require an expansion of $U$. The normalization condition $u^\mu u^\nu g_{\mu\nu}=U^2 \frac{dz^\mu}{dt} \frac{dz^\nu}{dt} g_{\mu\nu}=-1 $ gives us $U^{-2}=-g_{\mu\nu}\frac{dz^\mu}{dt}\frac{dz^\nu}{dt}$. Using $\frac{dz^\mu}{dt} = (1,dr_p/dt,0,\Omega)$, Eqs.~(\ref{rp-expansion}) and (\ref{Omegap-expansion}), and $d/dt=\e d/d\tilde t$, we obtain the expansion
\begin{equation} 
U^{-2}=1 - 3M/ r_0- 2\e   r_0^2\Omega_0\Omega_1 +\O(\e^2).\label{U-expansion}
\end{equation}
This then yields
\begin{equation}
U = U_0+\e  U_1+\O(\e^2),\label{tildeUTwoTimesacaleExpansion}
\end{equation}
where 
\begin{align}
U_0 &= \left(1-3M/ r_0\right)^{-1/2},\label{U0}\\ 
U_1 &= r_0^2\Omega_0\Omega_1 \left(1-3M/ r_0\right)^{-3/2}.\label{U1}
\end{align}

We also require an expansion of $f^\mu$. As we mentioned above, the orbital dynamics is independent of the fast time $\phi_p$. We can hence write
\beq\label{multiscale f}
f^\mu(t,\e) = \e \tilde f^\mu_1(\e t) + \e^2 \tilde f^\mu_2(\e t) +\O(\e^3).
\eeq
The explicit terms in this expansion are given by Eq.~\eqref{tilde fn} below. $\tilde f^\mu_1(\tilde t)$ is the standard first-order self-force on a circular orbit of radius $r_0(\tilde t)$, but with the inclusion of terms proportional to $M_1(\tilde t)$ and $S_1(\tilde t)$. Similarly, $\tilde f^\mu_2(\tilde t)$ is specified by $r_{n\leq 1}(\tilde t)$, $M_{n\leq2}(\tilde t)$, and $S_{n\leq 2}(\tilde t)$. 

As mentioned in the introduction, dissipative  and conservative pieces of the force contribute to the long-term phase evolution at different orders. In the quasicircular case, the division into dissipative and conservative pieces is straightforward. The dissipative piece is the one that is antisymmetric under the time reversal $(t,\phi)\to (-t,-\phi)$ (at fixed $\tilde t$):
\beq
\tilde f^\mu_{n,\rm diss} = (\tilde f^t_n, 0,0,\tilde f^\phi_n).
\eeq
Analogously, the conservative piece is the one that is symmetric under that time reversal:
\beq
\tilde f^\mu_{n, \rm cons} = (0,\tilde f^r_n,0,0).
\eeq

Finally, we substitute the expansions of $z^\mu$, $U$, and $f^\mu$ into the equation of motion~\eqref{eq-mot}. The result is a sequence of equations for $\Omega_n$ and $r_n$, which take the form of (i) algebraic equations for $\Omega_n$ in terms of $r_n$ and (ii) differential equations for the slow time evolution of $r_n(\tilde t)$.

At order $\e^0$ the only nontrivial piece of Eq.~\eqref{eq-mot} is the $r$ component, from which we derive the relation
\begin{equation}\label{Omega0}
\Omega_0 = \sqrt{M/r_0^3}.
\end{equation}
This is just the standard expression for the orbital frequency of a circular geodesic orbit of radius $r_0$ in Schwarzschild spacetime.

At linear order in $\e$, from the $t$ component of Eq.~\eqref{eq-mot} we derive that
\begin{equation}
\frac{d r_0}{d\tilde t} = \frac{2( r_0-3M)^2( r_0-2M)}{M( r_0-6M)}\tilde f^t_1(\tilde t).\label{r0dot}
\end{equation}
This tells us that at leading order, the slow evolution of the radius is completely determined by $\tilde f^\mu_{1,\rm diss}$. We see that the evolution diverges when the orbit reaches the innermost stable circular orbit (ISCO), $r_0=6M$,  signalling the breakdown of our quasicircular ansatz. Physically, this corresponds to the fact that as the small object approaches the ISCO, it transitions into a plunging orbit. The transition occurs over a new slow timescale~\cite{Ori:2000zn}, and a complete treatment of the inspiral would utilize an alternative approximation in the transition region. However, the transition to plunge should have a minimal impact on LISA data analysis, since it adds relatively little to the integrated SNR. So we expect that for the purpose of matched filtering, we can simply abort the waveform model at some time prior to the transition.

Next, still at linear order in $\e$, from the  $r$ component of Eq.~\eqref{eq-mot} we find that
\begin{equation}
\Omega_1 = -\frac{1}{2 f_0 r_0  \Omega_0}\left(U_0^{-2}\tilde f^r_1+\frac{3M r_1 f_0}{ r_0^3}\right),\label{Omega1}
\end{equation}
where $f_0:=1-2M/r_0$. This equation tells us how the conservative piece of $\tilde f^\mu_1$ affects the orbital frequency at a given orbital radius.

Now moving to quadratic order in $\e$, from the $t$ component of Eq.~\eqref{eq-mot}, we obtain an equation for the slow evolution of $r_1$ and $\Omega_1$, given by
\begin{align}
\frac{2M}{ r_0^2 f_0}\frac{d r_1}{d\tilde t} &+ r_0^2U_0^2\Omega_0 \frac{d\Omega_1}{d\tilde t} \nonumber\\
&= U_0^{-2} \tilde f^t_2 + \frac{8\left( r_0-3M\right)^2\left( r_0-M\right)}{ r_0^2\left( r_0-2M\right)\left( r_0-6M\right)}  r_1 \tilde f^t_1\nonumber\\
&\quad -\left(\frac{ r_0^3\left( r_0-9M\right) f_0 }{M\left( r_0-6M\right)}+2 r_0^2\right)  \Omega_0\Omega_1 \tilde f^t_1.
\end{align}
Substituting our result~\eqref{Omega1} for $\Omega_1$ then yields an equation for $r_1(\tilde t)$ alone:
\begin{align}
\frac{dr_1}{d\tilde t} &= \frac{2r_0 f_0 (r_0-3M)^2}{M(r_0-6M)}\tilde f_2^t + \frac{r_0^3(r_0-3 M)}{M (r_0-6 M)}\frac{d\tilde f_1^r}{d\tilde t} \nonumber\\
&\quad + \frac{2 r_0^2(r_0-3 M)^2 \left(r_0^2-6 M r_0+6 M^2\right)}{M^2 (r_0-6 M)^2}\tilde f_1^r \tilde f_1^t\nonumber\\
&\quad +  \frac{4(r_0-3M)(r_0^2 - 10 M r_0 +18 M^2)}{M (r_0-6M)^2}r_1 \tilde f_1^t.\label{r1dot}
\end{align}
We see here that the slow evolution of $r_1$ and $\Omega_1$ is driven by $\tilde f^\mu_{2,\rm diss}$, the rate of change $\partial_{\tilde t}\tilde f^\mu_{1,\rm cons}$, and products of $\tilde f^\mu_{1,\rm diss}$ with first-order time-symmetric quantities.

We could next write a formula relating $\Omega_2$ to $r_2$ and $\tilde f^r_2$; this relation is contained in the order-$\e^2$ term in the $r$ component of Eq.~\eqref{eq-mot}. However, without an equation for $dr_2/d\tilde t$, this relation would not contribute to a calculation of the gravitational waveform. Our primary requirement for that calculation is an accurate evolution of the waveform phase. As we sketched in Fig.~\ref{fig_flowchart}, the waveform phase is obtained directly from the orbital phase $\phi_p$, which in turn is obtained from the orbital frequency via Eq.~\eqref{choice_of_fasttime_worldline}. Note that by changing the integration variable in that equation, we can also write $\phi_p$ as an expansion of the form~\eqref{phasing},
\begin{align}
\phi_p(t,\e) &= \frac{1}{\e}\int_0^{\e t}[\Omega_0(\tilde s)+\e\, \Omega_1(\tilde s)+\O(\e^2)]d\tilde s \nonumber\\
&\quad+\phi_p(0,\e) \\
							&:= \frac{1}{\e}\!\left[\phi_0(\e t)+\e\,\phi_1(\e t)+\O(\e^2)\right]\!+\phi_p(0,\e).
\end{align}
This equation, in combination with the above analysis, provides a simple demonstration of what is required for adiabatic and post-adiabatic accuracy:
\begin{enumerate}
\item The adiabatic approximation only requires $\Omega_0(\tilde t)$. According to Eqs.~\eqref{Omega0} and \eqref{r0dot}, the only necessary input is $\tilde f^\mu_{1,\rm diss}$.
\item The post-adiabatic approximation requires $\Omega_1(\tilde t)$ (in addition to $\Omega_0$). According to Eqs.~\eqref{Omega1} and \eqref{r1dot}, the necessary input is $\tilde f^\mu_{2,\rm diss}$ and both $\tilde f^\mu_{1,\rm diss}$ and $\tilde f^\mu_{1,\rm cons}$.
\end{enumerate}

In the next section, we will see how our results here link to the two-timescale expansion of the field equations. However, before proceeding we make one comment. Previous treatments of eternal circular orbits have typically expanded the orbit in powers of $\e$ at fixed frequency \cite{Pound:2014koa}. This corresponds to setting $\Omega_0=\Omega$ and $\Omega_n = 0$ for $n>0$. Equation~\eqref{Omega1} then becomes a formula for $r_1$ in terms of $\tilde f^r_1$:
\begin{equation}\label{Omega1Again}
 r_1 = -\frac{ r_0^3 \tilde f^r_1 }{3M U_0^2 f_0}.
\end{equation}
However, in the two-timescale expansion presented in this section, the relationship~\eqref{Omega1Again} can only be freely enforced at a single value of $\tilde t$, say $\tilde t_0$; $\Omega_n(\tilde t_0) = 0$ is simply a choice of initial condition for the evolution. After that initial time, $\Omega_1$ changes with time in a uniquely determined way, as dictated by Eqs.~\eqref{Omega1} and \eqref{r1dot}. An alternative approach, as noted several times above, would be to perform all our expansions in powers of $\e$ while holding $\Omega$, rather than $\tilde t$, fixed. Equation~\eqref{Omega1Again} would then automatically hold true throughout the evolution, with all functions on the right of the equation becoming functions of $\Omega$. We detail this approach in Appendix~\ref{fixed Omega}.

\section{Two-timescale expansion of the field equations}\label{sec_expanded_field}

We now turn to the expansion of the metric perturbation and its field equations. We first define our choice of slow and fast time variables off the worldline; our definitions are tailored to hyperboloidal slicings, but they are sufficiently general to encompass $t$ slicing and null slicings. We then derive the two-timescale expansion of the field equations with this generic slicing. We close with discussions of transformations between modes on different slicings and boundary conditions for the expanded fields.

\subsection{Choice of slow and fast times off the worldline}\label{sec_choice_of_slowtime}

To perform a two-timescale expansion for the metric perturbation, we need a suitable choice of fast and slow variables. One  option is to remain with the choice in Eq.~(\ref{choice_of_slowtime_worldline}). But as shown in Ref.~\cite{Pound:2015wva}, an asymptotically null slicing improves the behavior of the second-order source and eliminates infrared divergences in the retarded integral of that source. The advantage of such a judicious choice was also highlighted by Mino and Price \cite{Mino:2008rr}. Since similar problems arise at the horizon, this suggests taking our basic time variable to be~\cite{Zenginoglu:11}
\footnote{
For the height function $k(r)$, 
to avoid confusion with the metric perturbation
we use a notation different from that of Ref.~\cite{Zenginoglu:11}, in which $h(r)$ is used.}
\begin{equation}
\w := t-k(r^*).\label{tildeWDefinition}
\end{equation}
Here, for suitable choices of height function $k(r^*)$, surfaces of constant $\w$ foliate  the spacetime with horizon-penetrating hyperboloidal slices. Figure~\ref{fig_penrose_diagram_hyperboloidal_time_3} shows an example of a constant-$\w$ slice for a particular choice of height function.

In the present section, for the purposes of expanding the field and the field equations, we remain agnostic about our choice of time variable. We instead denote by $\w$ the generic time variable as defined in Eq.~(\ref{tildeWDefinition}), leaving $k$ unspecified. This means our versions of the field equations can be specialized to $t$ slicing (with $k=0$), null slicing (with $k=\pm r^*$), or any choice of hyperboloidal slicing.

Given our choice of basic time variable, we define our slow time to be $\tilde \w = \e\w$. As for the fast variable, we use the azimuthal phase $\phi_p$, as given in Eq.~(\ref{choice_of_fasttime_worldline}), but extended off the worldline such that it is constant on each slice of constant $\w$:
\beq
\phi_p(\w,\e)=\int^\w_0 \Omega(z,\e)dz + \phi_p(0,\e).
\eeq

The next subsection further motivates these choices of slow and fast time based on the form of the source terms in the field equations.

\subsection{Tensor-harmonic decomposition}

We now turn to the field equations (\ref{puncture_scheme_1})--\eqref{puncture_scheme_2}. Before performing the two-timescale expansion, we first slightly modify the field equations and decompose them into a basis of tensor spherical harmonics.

Our first step is to add a gauge-damping term to the field equations, following Barack and Lousto~\cite{Barack:2005nr,Barack:2007tm}. We do this by replacing $E_{\alpha\beta}$ with
\beq
\breve{E}_{\alpha\beta}[\bar h]: = E_{\alpha\beta}[\bar h] + \frac{4M}{r^2} t_{(\alpha}\breve Z_{\beta)}[\bar h], 
\eeq
where $t_{\alpha}=-\delta^t_\alpha$, $Z_{\alpha}[\bar h]:=\nabla^\beta \bar h_{\alpha\beta}$, and $\breve{Z}_\alpha$'s components in Schwarzschild coordinates are $\breve{Z}_\alpha = (Z_r,2Z_r,Z_\theta,Z_\phi)$. The field equations then become
\begin{align}
\breve{E}_{\alpha\beta}[\bar h^{1\res}] &= -\breve{E}_{\alpha\beta}[\bar h^{1\P}]  =: \breve{S}^{1\,\rm eff}_{\alpha\beta}, \label{gauge-damped EFE1}\\
\breve{E}_{\alpha\beta}[\bar h^{2\res}] &= 2\delta^2 G_{\alpha\beta}[h^1] - \breve{E}_{\alpha\beta}[\bar h^{2\P}] =: \breve{S}^{2\,\rm eff}_{\alpha\beta}.\label{gauge-damped EFE2}
\end{align} 
The solutions to these equations slightly differ from the solutions to the old ones because the individual fields do not satisfy $Z_{\alpha}[\bar h^n] = 0$. However, when the fields are summed, $\e\bar h^1_{\alpha\beta}+\e^2 \bar h^2_{\alpha\beta}$ {\em does} remain a solution to the original relaxed Einstein equation~\eqref{relaxed EFE v0} [or~\eqref{relaxed EFE v3}] because the fields {\em do} satisfy $\e Z_{\alpha}[\bar h^1] + \e^2 Z_{\alpha}[\bar h^2] = \O(\e^3)$. This step of altering the field equations is entirely nonessential for our analysis; we perform it solely because, as pointed out by Barack and Lousto, it partially decouples the field equations after the tensor-harmonic decomposition, as we review later in this section.

We next decompose the fields into tensor spherical harmonic modes, using the Barack-Lousto-Sago basis of harmonics~\cite{Barack:2005nr,Barack:2007tm}: 
\begin{equation}
\bar h^{n\res}_{\alpha\beta}=\sum_{i\ell m} \frac{a_{i\ell}}{r}\bar h^{n\res}_{i\ell m}(\w,r,\e)Y_{\alpha\beta}^{i\ell m}(r,\theta^A),\label{modeDecomposition}
\end{equation}
and analogously for $\bar h^{n\P}_{\alpha\beta}$, where $i=1,\ldots,10$, $\ell\geq0$, and $m=-\ell,\ldots,\ell$. The harmonics $Y^{i\ell m}_{\alpha\beta}$, given explicitly in Appendix~\ref{sec_tensor_harmonics}, provide an orthogonal basis for symmetric rank-2 tensors, satisfying the orthogonality property~(\ref{tensorHarmonicsOrthonormal}). $a_{i\ell}$  is a convenient numerical factor defined in Eq.~(\ref{a_il}). Following Barack-Lousto-Sago, we have also pulled out a factor of $1/r$ to simplify the field equations. The orthogonality property implies
\beq
\bar h^{n}_{i\ell m}=\frac{r}{a_{i\ell}\kappa_i}\oint d\Omega\, \eta^{\alpha\mu}\eta^{\beta\nu} \bar h^{n}_{\alpha\beta} Y^{i\ell m*}_{\mu\nu},
\eeq
with $\oint d\Omega := \int^{2\pi}_0\!d\phi\!\int^\pi_0 \!d\theta\sin\theta$,  $\kappa_i$ given by Eq.~(\ref{kappai}), and $\eta^{\alpha\beta}$ given by Eq.~\eqref{eta}. The coefficients $h^n_{i\ell m}$, if needed, are related to $\bar h^n_{i\ell m}$ by the interchange $i=3\leftrightarrow i=6$, with all other coefficients unchanged.

We similarly expand the source terms $\breve{S}^{n\,\rm eff}_{\mu\nu}$ as
\begin{align}
\breve{S}^{n\,\rm eff }_{\mu\nu} &= \sum_{i\ell m}\frac{-4a_{i\ell}}{rf}S^{n\,\rm eff}_{i\ell m}(\w,r)Y^{i\ell m}_{\mu\nu}(r,\theta^A),\label{multiscale-S}
\end{align}
where we have again introduced a factor, $\frac{-4a_{i\ell}}{rf}$, to simplify later expressions. From the orthogonality condition (\ref{tensorHarmonicsOrthonormal}), the coefficients are
\begin{align} 
S^{n\,\rm eff}_{i\ell m}(\w, r) &=\frac{-r f}{4 a_{i\ell}\kappa_i}\oint d\Omega\, \eta^{\alpha\mu}\eta^{\beta\nu} \breve{S}^{n\,\rm eff}_{\alpha\beta} Y^{i\ell m*}_{\mu\nu},\label{Seff_ilm}
\end{align}
Here and below we define mode coefficients without breves to lessen the notational load.

With these harmonic expansions, Eqs.~(\ref{gauge-damped EFE1}) and \eqref{gauge-damped EFE2} each separate into a set of ten coupled partial differential equations for the coefficients $\bar h^n_{i\ell m}$~\cite{Barack:2007tm}, which read 
\begin{equation}\label{decomposed EFE}
E_{ij\ell m}\bar h^{n\res}_{j\ell m} =  S^{n\,\rm eff}_{i\ell m}.
\end{equation}
Recall that the repeated basis label $j$ is summed over. The decomposed wave operator is given by 
\beq\label{Eijlm}
E_{ij\ell m}\bar h_{j\ell m} := \Box^{2d}_{sc}\bar h_{i\ell m} + {\cal M}^{ij}\bar h_{j\ell m},
\eeq
where $\Box_{\rm sc}^{2d}$ is the scalar-field wave operator $\partial_{uv}+V_\ell$, or
\begin{equation}
\Box^{2d}_{sc} = \frac{1}{4}\left(\partial_t^2-\partial_{r^*}^2\right)+V_\ell, \label{box_h_defined_0}
\end{equation}
with potential 
\begin{equation} 
V_\ell(r)=\frac{f}{4}\left( \frac{2M}{r^3}+\frac{\ell(\ell+1)}{r^2} \right),\label{Vl}
\end{equation}
and ${\cal M}^{ij}$ with $i,j=1,\ldots,10$ are a set of matrices comprised of first-order differential operators that couple between the various $\bar h_{j\ell m}$'s. Note that the coupling is only between different $j$'s; there is no coupling between modes of different $\ell$ and $m$. Also note that the only effect of our added gauge-damping term is to alter these coupling matrices. The explicit form of the coupling matrices can be found in Appendix A of \cite{Barack:2007tm}. 

In addition to these wave equations, we have the four gauge constraints, $\nabla^\alpha\bar h_{\alpha\beta}=0$, which separate into four conditions for $\bar h_{i\ell m}$. These also appear in Appendix A of  \cite{Barack:2007tm}. We delay discussion of them until after presenting the two-timescale expansion of the wave equations.

To more easily justify our two-timescale ansatz, rather than using the first-order field equation with an effective source, we use the equivalent form
\begin{equation}\label{decomposed EFE}
E_{ij\ell m}\bar h^1_{j\ell m} = -16\pi T^1_{i\ell m},
\end{equation}
where the modes of the stress-energy tensor are defined in analogy with Eq.~\eqref{Seff_ilm}. We can evaluate the integral in the point-particle stress-energy~\eqref{T1_alpha_beta} by changing to $t$ as the integration variable, yielding
\begin{align}
T^1_{\alpha\beta} &= \frac{\mu u_\alpha(\e t,\e) u_\beta(\e t,\e)}{U(\e t,\e) r_p^2(\e t,\e)}\nonumber\\
								&\quad \times\delta[r-r_p(\e t,\e)]\delta(\theta-\pi/2)\delta[\phi-\phi_p(t,\e)].
\end{align}
The harmonic modes are then
\begin{align}
T^1_{i\ell m} &= \frac{-r\mu f}{4 a_{i\ell}\kappa_i}\oint d\Omega\, \eta^{\alpha\mu}\eta^{\beta\nu}\frac{u_\alpha u_\beta}{U r_p^2}\delta(r-r_p)\delta(\theta-\pi/2)\nonumber\\
					&\quad\times\delta(\phi-\phi_p)Y^{i\ell m*}_{\mu\nu}(r_p,\theta,\phi) \nonumber\\
					&= t^1_{i\ell m} e^{-im\phi_p}\delta(r-r_p),\label{Tilm}
\end{align}
where
\beq\label{tilm}
t^1_{i\ell m} := \frac{-r_p\mu f_p}{4 a_{i\ell}\kappa_i}\eta^{\alpha\mu}\eta^{\beta\nu}\frac{u_\alpha u_\beta}{U r_p^2}Y^{i\ell m*}_{\mu\nu}(r_p,\pi/2,0).
\eeq
Here $\eta^{\alpha\beta}$ is evaluated at $r=r_p$ and $\theta=\pi/2$, and all of $r_p$, $f_p:=1-2M/r_p$, $U$, and $u_\alpha$ are evaluated at $(\e t,\e)$. Note, however, that since $\w$ reduces to $t$ in a neighborhood of the worldline, we have $\w(t,r_p)=t$. Since the radial delta function forces evaluation at $r=r_p$, we can hence freely treat all functions of $t$ in $T_{i\ell m}$, including $\phi_p$, as functions of $\w$. Each mode $T_{i\ell m}$ is therefore proportional to the fast-time phase factor $e^{-im\phi_p(\w,\e)}$. This motivates adopting $\phi_p(\w,\e)$ as our fast time in $\bar h^1_{\alpha\beta}$, as outlined above.

This feature extends to the second-order source as well. The source term $\delta^2 G_{i\ell m}$ can be written in terms of the first-order field modes as 
\beq
\delta^2 G_{i\ell m}=\sum_{\substack{i_1\ell_1m_1 \\ i_2\ell_2 m_2}}{\cal G}^{i\ell m}_{i_1\ell_1 m_1 i_2\ell_2 m_2}\bar h^1_{i_1\ell_1m_1}\bar h^1_{i_2\ell_2m_2},\label{coupling_formula}
\eeq
where ${\cal G}^{i\ell m}_{i_1\ell_1 m_1 i_2\ell_2 m_2}$ is a bilinear differential operator that acts linearly on $\bar h^1_{i_1\ell_1m_1}$ and, separately, linearly on $\bar h^1_{i_2\ell_2m_2}$. The explicit form of Eq.~\eqref{coupling_formula} will be presented in Ref.~\cite{coupling-paper}. Its only essential characteristic here is that it enforces $m=m_1+m_2$; this condition arises in the usual way from the coupling of two spherical harmonics, via $\delta^2 G_{i\ell m}\propto\int d\phi e^{-im\phi}e^{im_1\phi}e^{im_2\phi}d\phi\propto\delta_{m,m_1+m_2}$. Since $\bar h^1_{i_1\ell_1m_1}\propto e^{-im_1\phi_p(\w,\e)}$ and $\bar h^1_{i_2\ell_2m_2}\propto e^{-im_2\phi_p(\w,\e)}$, it follows that $\delta^2G_{i\ell m}\propto e^{-im\phi_p(\w,\e)}$. Although we do not explicitly calculate the modes of the puncture $\bar h^{2\P}_{i\ell m}$ in this paper, they too inherit the fast-time dependence of the orbit. Hence, we are again motivated to adopt $\phi_p(\w,\e)$ as our fast time in $\bar h^2_{\alpha\beta}$. 

In summary, our mode-decomposed field equations are
\begin{subequations}\label{mode-decomposed EFE v2}
\begin{align}
E_{ij\ell m}\bar h^{1}_{j\ell m} &= -16\pi\, t^1_{i\ell m}e^{-im\phi_p(\w,\e)}\delta(r-r_p),\label{mode-composed EFE1 v2}\\
E_{ij\ell m}\bar h^{2\res}_{j\ell m} &= 2\delta^2 G_{i\ell m} - E_{ij\ell m}\bar h^{2\P}_{j\ell m},\label{mode-composed EFE1 v2}
\end{align}
\end{subequations}
where $r_p = r_p(\e\w,\e)$, $t^1_{i\ell m}=t^1_{i\ell m}(\e\w,\e)$ is given by Eq.~\eqref{tilm}, and $\delta^2 G_{i\ell m}$ is given by Eq.~\eqref{coupling_formula}. The first-order source is a function of $\tilde \w$, $\phi_p$, and $\e$, periodic in $\phi_p$ (with period $2\pi/m$). If we assume $\bar h^{1}_{i\ell m}$ inherits those properties, then the second-order source $\delta^2 G_{i\ell m}$ does as well.

\subsection{Two-timescale expansion}\label{sec_expanded_field_eqns}

Prompted by the form of the source, and following the outline in Sec.~\ref{sec_twotimescale_expansion}, we assume that $\bar h^n_{i\ell m}(\w,r,\e)$ is approximated (uniformly in $\w$ at fixed $r$) by a function $\tilde{\bar h}^n_{i\ell m}(\e\w,\phi_p(\w,\e),r,\e)$ that is periodic in $\phi_p$. Expanding that function in powers of $\e$ gives us\footnote{These expressions assume smooth functions of $\e$. However, in reality, $\ln \e$ terms appear in $\tilde{\bar h}^2_{i\ell m}(\tilde \w,\phi_p,r, \e)$. Such logarithms arise from two effects: the mass $\mu$ deforms the light cones on which solutions propagate, potentially introducing $\ln(r/\mu)$ terms into the solution~\cite{Pound:2012dk}; and the waves in $h^1_{\alpha\beta}$ introduce curvature that falls off slowly at large distance, which in turn leads to hereditary effects that introduce $\ln(\e r/M)$ terms into the solution~\cite{Pound:2015wva}. Here we blithely absorb these logarithms into the `$\e$-independent' coefficients. Note that this is simply a choice to streamline the notation; we do not discard logarithms from the solutions.}
\begin{align}
\tilde{\bar h}^n_{i\ell m}(\tilde \w,\phi_p,r, \e) &= \tilde{\bar h}^n_{i\ell m}(\tilde \w,\phi_p,r,0)\nonumber\\
																	&\quad + \e \partial_\e\tilde{\bar h}^n_{i\ell m}(\tilde \w,r,\phi_p,0) +\O(\e^2).\label{hnilm expansion}
\end{align}
We then move the first subleading term from $\tilde{\bar h}^1_{i\ell m}$ into $\tilde{\bar h}^2_{i\ell m}$, defining new first- and second-order fields
\begin{align}
\tilde{\bar h}^1_{i\ell m}(\tilde \w,\phi_p,r) &:=\tilde{\bar h}^1_{i\ell m}(\tilde \w,\phi_p,r,0),\label{h1New}\\
\tilde{\bar h}^2_{i\ell m}(\tilde \w,\phi_p,r) &:=\tilde{\bar h}^2_{i\ell m}(\tilde \w,\phi_p,r,0)\nonumber\\
																	&\quad + \partial_\e \tilde{\bar h}^1_{i\ell m}(\tilde \w,\phi_p,r,0).\label{h2New}
\end{align}
Given the $2\pi/m$ periodicity of the source, we also adopt the ansatz
\beq\label{hnilm-Fourier}
\tilde{\bar h}^n_{i\ell m}(\tilde \w, \phi_p, r)  = \R^{n}_{i\ell m}(\tilde \w, r)e^{-im\phi_p}.
\eeq

Altogether, this means the two-timescale expansion of the total (trace reversed) metric perturbation is
\beq\label{multiscale h}
\bar h_{\mu\nu} = \sum_{n i\ell m}\frac{\e^n a_{i\ell}}{r} \R^{n}_{i\ell m}(\e\w,r)e^{-im\phi_p(\w,\e)}Y^{i\ell m}_{\mu\nu}(r,\theta^A).\!
\eeq
We assume analogous expansions of the puncture and residual fields.

Next we must expand $E_{ij\ell m}$ and the sources $T_{i\ell m}$ and $\delta^2G_{i\ell m}$. We begin with $E_{ij\ell m}$, applying the chain rule for derivatives as described in Sec.~\ref{sec_twotimescale_expansion}. In Eq.~\eqref{box_h_defined_0}, we expressed $\Box^{2d}_{\rm sc}$ in terms of $t$ derivatives at fixed $r$ and $r^*$ derivatives at fixed $t$. When acting on a function of $(\e\w, \phi_p(\w,\e), r)$, a $t$ derivative at fixed $r$ becomes
\begin{align}
\left(\frac{\partial}{\partial t}\right)_{\! r} &= \frac{\partial\phi_p}{\partial t}\frac{\partial}{\partial\phi_p} + \frac{\partial\tilde \w}{\partial t}\frac{\partial}{\partial\tilde \w}\nonumber\\
							&= \Omega\frac{\partial}{\partial \phi_p} + \e\frac{\partial}{\partial\tilde \w}\nonumber\\
							&= \Omega_0 \partial_{\phi_p} + \e\left(\partial_{\tilde \w} + \Omega_1\partial_{\phi_p}\right)+\O(\e^2).\label{chain_rule_t_1}
\end{align}
Similarly, $r^*$ derivatives act according to
\begin{align}
\left(\frac{\partial}{\partial r^*}\right)_{\! t} &= \left(\frac{\partial}{\partial r^*}\right)_{\! \tilde \w,\phi_p} + \frac{\partial \phi_p}{\partial r^*}\frac{\partial}{\partial \phi_p} 
																		+ \frac{\partial\tilde \w}{\partial r^*}\frac{\partial}{\partial\tilde \w}\nonumber\\
 																	&= \left(\frac{\partial}{\partial r^*}\right)_{\! \tilde \w,\phi_p} - H\Omega_0\partial_{\phi_p} \nonumber\\
 																	&\quad - \e H\left(\Omega_1\partial_{\phi_p}+\partial_{\tilde \w}\right)+\O(\e^2),\label{chain_rule_r_1}
\end{align}
where we have defined
\beq
H(r^*):=\frac{dk}{dr^*}
\eeq
and used $\partial \phi_p /\partial r^* = \left(\partial\phi_p /\partial \w\right)\left(\partial \w/\partial r^*\right) = - H \Omega$ and $\partial\tilde \w/\partial r^* = -\e H$. 

From Eq.~(\ref{chain_rule_t_1}), a $t$ derivative acting on a Fourier series of the form~\eqref{hnilm-Fourier} becomes
\begin{align}
\left(\frac{\partial}{\partial t}\right)_{\! r} \to -i\omega_m + \e\tilde\partial_{\tilde \w}+\O(\e^2).\label{tDerivativeTildeh}
\end{align}
where $\omega_m:=m\Omega_0$ and
\beq
\tilde \partial_{\tilde \w} :=\partial_{\tilde \w}-im\Omega_1.
\eeq
Similarly, from Eq.~(\ref{chain_rule_r_1}), an $r^*$ derivative becomes
\begin{align}
\left(\frac{\partial}{\partial r^*}\right)_{\! t} \to \partial_{r^*} + i\omega_m H-\e H \tilde\partial_{\tilde \w}+\O(\e^2),\label{rDerivativeTildeh}
\end{align}
where the radial derivative is taken at fixed $\tilde \w$ and $\phi_p$.

We are now ready to expand $E_{ij\ell m}$. Substituting Eqs.~(\ref{tDerivativeTildeh})~and~(\ref{rDerivativeTildeh}) into (\ref{box_h_defined_0}), we split the term $\Box^{2d}_{\rm sc}$ into 
$\Box^{2d}_{\rm sc}=\Box_0+\e\Box_1+\O(\e^2)$, where
\begin{align}
\Box_0 &= -\frac{1}{4}\Big[\partial_{r^*}^2 +i\omega_m \left(2H \partial_{r^*}+H'\right)\nonumber\\
			&\quad +\left(1-H^2\right)\omega_m^2-4V_\ell(r)\Big],\label{tildeBox0}\\
\Box_1 &= \frac{1}{4}\Big[\left(2 H\partial_{r^*}+ H'\right)\tilde \partial_{\tilde \w}\nonumber\\
			&\quad -\left(1-H^2\right)\left(2i\omega_m\tilde\partial_{\tilde \w}+i\dot\omega_m\right)\Big].\label{tildeBox1}
\end{align}
Here
\beq
H' :=\frac{dH}{dr^*} \quad \text{and}\quad  \dot\omega_m:=m\dot\Omega_0(\tilde \w).
\eeq
In the same way the coupling matrices are expanded as $\mathcal M^{ij}=\mathcal M^{ij}_0+\e\mathcal M^{ij}_1+\O(\e^2)$. The leading-order matrices ${\cal M}^{ij}_0$ are obtained from $\mathcal M^{ij}$ in Appendix A of   \cite{Barack:2007tm} with the replacements $\bar h_i\to R_i$, $\partial_r\to\partial_r+i f^{-1}\omega_m H$ and $\partial_v\to\displaystyle\frac{1}{2}\left[f\partial_r-i\left(1-H\right)\omega_m\right]$. We give the explicit forms of ${\cal M}^{ij}_0$ and ${\cal M}^{ij}_1$ in Eqs.~\eqref{Mij0} and \eqref{Mij1}.

Combining these results we obtain
\beq\label{Eijlm expansion}
E_{ij\ell m} \to E^0_{ij\ell m} + \e E^1_{ij\ell m} +\O(\e^2),
\eeq
where
\beq\label{Enijlm}
E^n_{ij\ell m} = \delta_{ij}\Box_n +{\cal M}^{ij}_n. 
\eeq
For $H=0$ (i.e., $t$ slicing), $E^0_{ij\ell m}$ is precisely the operator that appears in the standard frequency-domain Lorenz-gauge linearized field equations for a metric perturbation $\bar h_{\mu\nu}=\sum_{i\ell m}\frac{a_{i\ell}}{r}R_{i\ell m}(r)Y^{i\ell m}_{\mu\nu}e^{-i\omega_m t}$, as in Refs.~\cite{Akcay:2010dx,Wardell:2015ada}. For $H\neq0$, $E^0_{ij\ell m}$ is precisely the operator that would appear in the frequency-domain Lorenz-gauge linearized field equations for a metric perturbation $\bar h_{\mu\nu}=\sum_{i\ell m}\frac{a_{i\ell}}{r}R_{i\ell m}(r)Y^{i\ell m}_{\mu\nu}e^{-i\omega_m \w}$.
 
We next expand the source terms in the field equations. The point-particle stress-energy modes~\eqref{Tilm} are expanded as
\begin{align} 
T^1_{i\ell m} &= \left[\tilde T^{1}_{i\ell m}(\tilde \w,r)+\e \tilde T^{2}_{i\ell m}(\tilde \w,r)+\O(\e^2)\right]  e^{-im\phi_p},\!\! \label{tilde_T1_ilm_expanded}
\end{align}
where
\begin{align}
\tilde T^{1}_{i\ell m} &= \tilde t^1_{i\ell m}(\tilde \w) \delta[r-r_0(\tilde \w)],\label{T0ilm}\\
\tilde T^{2}_{i\ell m} &= \tilde t^2_{i\ell m}(\tilde \w)\delta[r-r_0(\tilde \w)] \nonumber\\
								&\quad +  \tilde t^1_{i\ell m}(\tilde \w)r_1(\tilde \w) \delta'[r-r_0(\tilde \w)].\label{T1ilm}
\end{align}
$\tilde t^1_{i\ell m}(\tilde \w):=t^1_{i\ell m}(\tilde \w,0)$ and $\tilde t^2_{i\ell m}(\tilde \w):=\frac{\partial t^1_{i\ell m}}{\partial\e}(\tilde \w, 0)$ are easily written in terms of $r_0$, $r_1$, $dr_0/d\tilde \w$, and $\Omega_1$ using the explicit expression~\eqref{tilm} and the expansions~\eqref{rp-expansion}, \eqref{Omegap-expansion}, and \eqref{tildeUTwoTimesacaleExpansion}. 

The leading term~\eqref{T0ilm} is identical to the stress-energy mode of a particle on a circular geodesic of radius $r_0$. Its exact form, also given in \cite{Barack:2007tm, Akcay:2010dx}, is 
\begin{equation}\label{t0ilm}
\tilde t^1_{i\ell m} = -\frac{1}{4}{\cal E}_0 \alpha_{i\ell m}\begin{cases}Y^*_{\ell m }(\pi/2,0 )& i=1,\ldots, 7, \\ 
																						\partial_{\theta} Y^*_{\ell m}(\pi/2, 0) & i=8,9,10. 
																					\end{cases} \!\!
\end{equation}
Here ${\cal E}_0 = \mu f_0U_0$, with $f_0= 1-2M/r_0$ is the leading-order orbital energy, and the $\alpha_{i\ell m}$'s are given by (suppressing $\ell m$ labels)
\begin{subequations}\label{alphas}
\begin{align}
\alpha_1 &=  f^2_0/r_0, &\alpha_{2,5,9} &=0, \\
\alpha_3 &= f_0/r_0, &\alpha_4 &= 2 i f_0  m \Omega_0, \\
\alpha_6 &= r_0 \Omega_0^2,  &\alpha_7 &= r_0 \Omega_0^2 [\ell(\ell+1)-2m^2], \\
\alpha_8 &= 2 f_0 \Omega_0,  &\alpha_{10} &=2 i m r_0\Omega_0^2, 
\end{align}
\end{subequations}
with $r_0=r_0\left(\tilde \w\right)$. For this source the $ i = 2,5,9 $ equations are sourceless. 

The subleading term~\eqref{T1ilm} in the stress-energy is also straightforwardly calculated. However, rather than use it directly, we will incorporate it into the field equations through a redefinition of the puncture modes $\tilde h^{2\P}_{i\ell m}$. The two-timescale expansion of the puncture will be presented in the followup paper~\cite{source-paper}.

Finally, we perform the multiscale expansion of $\delta^2 G_{i\ell m}$,
\beq\label{d2G expansion}
\delta^2G_{i\ell m} = \delta^2 G^0_{i\ell m}(\e\w,r)e^{-im\phi_p(\w,\e)}+\O(\e). 
\eeq
The leading-order term is 
\beq
\delta^2 G^0_{i\ell m}=\sum_{\substack{i_1\ell_1m_1 \\ i_2\ell_2 m_2}}{\cal G}^{(0)i\ell m}_{i_1\ell_1 m_1 i_2\ell_2 m_2}\R^{1}_{i_1\ell_1m_1}\R^{1}_{i_2\ell_2m_2},\label{coupling_formulaV2}
\eeq
where ${\cal G}^{(0)i\ell m}_{i_1\ell_1 m_1 i_2\ell_2 m_2}$ is given by ${\cal G}^{i\ell m}_{i_1\ell_1 m_1 i_2\ell_2 m_2}$ with derivatives that act on $\bar h^{1}_{i_1\ell_1m_1}$ replaced by $\partial_t\to-i\omega_{m_1}(\tilde \w)$ and $\partial_r\to \partial_{r} + i \omega_{m_1}(\tilde \w)f^{-1} H$, and analogously for derivatives of  $\bar h^{1}_{i_2\ell_2m_2}$.

With all the necessary expansions in hand, we now obtain the two-timescale form of the field equations. Substituting Eqs.~\eqref{hnilm expansion}, \eqref{hnilm-Fourier}, \eqref{Eijlm expansion}, \eqref{tilde_T1_ilm_expanded}, and \eqref{d2G expansion} into the field equations~\eqref{mode-decomposed EFE v2}, moving subleading terms from the first-order equation into the second-order equation, and treating $\tilde \w$ and $\phi_p$ as independent variables, we find the following equations for the mode amplitudes $\R^{n}_{i\ell m}$:
\begin{align}
E^0_{ij\ell m}\R^{1}_{j\ell m} &=-16\pi\, \tilde t^1_{i\ell m}\delta(r-r_0), \label{tildeEFE1}\\
E^0_{ij\ell m}\R^{2\res}_{j\ell m} &= 2\delta^2 G^0_{i\ell m} - E^0_{ij\ell m}\R^{2\P}_{j\ell m} \nonumber\\
																			&\quad - E^1_{ij\ell m}\R^{1}_{j\ell m},\label{tildeEFE2}
\end{align}
where $E^n_{ij\ell m}$ is given by Eq.~\eqref{Enijlm}, $\tilde t^1_{i\ell m}$ by Eq.~\eqref{t0ilm}, $\delta^2 G^0_{i\ell m}$ by Eq.~(\ref{coupling_formulaV2}), and as mentioned above, we have absorbed the effect of $\tilde T^2_{i\ell m}$ into a redefinition of $\R^{2\P}_{i\ell m}$.

\subsection{Expanded gauge condition}\label{sec_gauge_condition}

To satisfy the Einstein equations, the solutions to the wave equations also have to satisfy the gauge condition $Z_\mu:=\nabla^\nu\bar h_{\mu\nu}=0$.

We expand this condition in the same way we did the wave equation. After substituting the decomposition (\ref{multiscale h}), we obtain an expansion (suppressing $\ell,m$ labels)
\beq
\e Z^0_{kj}\R^1_{j}+\e^2 (Z^0_{kj}\R^2_j+Z^1_{kj}\R^1_j)=\O(\e^3),
\eeq
$k=1,2,3,4$, where
\begin{subequations}\label{eq:gauge1-w0}
\begin{align}
Z^0_{1j}\R_{j} &=i \omega_m\left(\R_{1} + f \R_{3}+H \R_{2}\right)\nonumber\\
  							&\quad +\frac{f}{r} \left( r  \R_{2,r} + \R_2 - \R_4 \right),\label{Z01}\\
Z^0_{2j}\R_j &= i \omega_m\left(\R_{2}+H \R_{1}-H f \R_{3}\right) + f\Big(\R_{1,r}- f \R_{3,r}\nonumber\\
							&\quad  +\frac{1}{r} \big[\R_{1} -\R_{5} - f\R_{3} - 2f\R_{6} \big]\Big),\label{Z02}\\
Z^0_{3j}\R_j  &= i \omega_m\left(\R_{4}+H \R_{5}\right)\nonumber\\
							&\quad + \frac{f}{r} \left[ r  \R_{5,r} + 2\R_{5} +\ell(\ell+1)\R_{6} -\R_{7} \right],\label{Z03}\\
Z^0_{4j}\R_j &= i \omega_m\left( \R_{8}+H \R_{9} \right)\nonumber\\
							&\quad +\frac{f}{r} \left(r\R_{9,r} + 2\R_{9} -\R_{10} \right),\label{Z04}
\end{align}
\end{subequations}
and
\begin{subequations} \label{eq:gauge1-w1}
\begin{align}
Z^1_{1j}\R_j &= -\tilde\partial_{\tilde \w}(\R_{1}+f\R_{3} + H\R_{2}),  \\
Z^1_{2j}\R_j &= -\tilde\partial_{\tilde \w}(\R_{2}+H\R_{1} - f H\R_{3}),  \\
Z^1_{3j}\R_j &= -\tilde\partial_{\tilde \w}(\R_{4}+H\R_{5}), \\
Z^1_{4j}\R_j &= -\tilde\partial_{\tilde \w}(\R_{8}+H\R_{9}).
\end{align}
\end{subequations}

Hence, at first order, the conditions are 
\begin{equation}
Z^0_{kj}\R^1_{j}=0.\label{eq:gauges-w0}
\end{equation}
At second order,  the conditions are
\begin{equation}
Z^0_{kj}\R^2_{j}=-Z^1_{kj}\R^1_{j}.\label{eq:gauges-w1}
\end{equation}
It is easily checked that Eqs.~(\ref{eq:gauges-w0}) are identical to the gauge constraints in \cite{Barack:2007tm} with $\partial_t\to -i\omega_m$ and $\partial_r\to\partial_r+i\omega_m f^{-1}H$.

In Sec.~\ref{sec_combined_expansions} we discuss how these gauge conditions combine with the field equations~\eqref{tildeEFE1}--\eqref{tildeEFE2} and the equations of motion~\eqref{Omega0},  \eqref{r0dot}, \eqref{Omega1}, and \eqref{r1dot} to determine (i) the mode amplitudes at fixed values of slow time, and (ii) the slow evolution of the system.

\subsection{Hierarchical structure}\label{sec_hierarchical_structure}
%

The wave equations~(\ref{tildeEFE1}) and (\ref{tildeEFE2}) each comprise 10 coupled ODEs for each value of $\ell$ and $m$. However, several properties reduce the level of coupling.

First, as is true in any gauge in Schwarzschild spacetime, the seven equations for the even-parity modes ($i=1,\ldots,7$) decouple from the three equations for the odd-parity modes ($i=8,9,10$): we have ${{\cal M}}^{ij}=0$ for any $i=1,\ldots,7$ with $j=8,9,10$, and for any $i=8,9,10$ with $j=1,\ldots,7$. (Note, however, that even- and odd-parity first-order modes do couple in the second-order source.)

Second, thanks to the gauge-damping terms added to the wave equations in Eqs.~\eqref{gauge-damped EFE1} and \eqref{gauge-damped EFE2}, the wave equations~(\ref{tildeEFE1}) and \eqref{tildeEFE2} partially decouple into a hierarchical structure. The equations for the $i=1,3,5,6,7$ modes decouple from the $i=2,4$ modes, and the equations for the $i=9,10$ modes decouple from the $i=8$ mode. Because of this partial decoupling, one can first solve for the $i=9,10$ modes and then solve for the $i=8$ mode (in the odd sector), and first solve for the $i=1,3,5,6,7$ modes and then solve for the $i=2,4$ modes (in the even sector). 

Third, as again is always true, the field equations are overdetermined: we have 14 equations (10 wave equations and 4 gauge conditions) for our 10 variables. (For $\ell=0$ this reduces to 6 equations for 4 variables, and for $\ell=1$ it reduces to 10 equations for 8 variables.) This excess of equations allows us to use the gauge condition to solve for some modes, rather than using the wave equations. For example, if $m\neq 0$, one can use the gauge condition~(\ref{eq:gauges-w0}) or~(\ref{eq:gauges-w1}) to compute these modes algebraically from the other seven modes, specifically utilizing the form of~\eqref{Z02} to obtain the $i=2$ mode, \eqref{Z04} to obtain the $i=8$ mode, and either~\eqref{Z01} or \eqref{Z03} to obtain the $i=4$ mode. 

If $m=0$, the frequency $\omega_m$ vanishes, making that procedure impossible. However, one can still utilize the gauge condition to further decouple the equations. For example, in the even sector, one can use the gauge conditions to solve for the $i=6,7$ modes in terms of the $i=1,3,5$ modes, and the $i=4$ in terms of the $i=2$. Substituting those results into the wave equations yields a coupled set of wave equations for the $1,3,5$ modes. The solutions for the $i=4,6,7$ modes can then be substituted into the wave equation for the $i=2$ mode.

For $\ell=0$ or 1, some modes do not appear, but the decoupling procedure is otherwise the same. 

There are also two additional properties that reduce the number of modes that need to be calculated. Because $\bar h_{\alpha\beta}$ is real, the amplitudes satisfy $\R^n_{i\ell,-m} = (-1)^m \R^{n*}_{i\ell m}$, meaning we need only calculate modes with $m\geq0$. This property follows from the identity $Y^{i\ell m*}_{\alpha\beta} = (-1)^m Y^{i\ell, -m}_{\alpha\beta}$. Because the system is symmetric under reflection through the equatorial plane (${\cal R}:\theta\mapsto\pi-\theta$), we also have that the even-parity modes vanish for odd values of $\ell+m$, and that odd-parity modes vanish for even values of $\ell+m$. This follows from the property ${\cal R}Y^{i\ell m}_{\alpha\beta}=-(-1)^{\ell+m}Y^{i\ell m}_{\alpha\beta}$ for even-parity harmonics and ${\cal R}Y^{i\ell m}_{\alpha\beta}=(-1)^{\ell+m}Y^{i\ell m}_{\alpha\beta}$ for odd-parity harmonics.

Putting all of these properties together, in Table~\ref{tabHierarchy} we summarize how each mode can be calculated. 

Once one has obtained all the modes, as a consistency check one can verify that they satisfy whichever wave equations or gauge conditions were not used in obtaining them. Since the wave equations and gauge conditions are only consistent with each other if the source in the field equations is conserved, this consistency check  serves as a check on the source.


\begingroup
\squeezetable
\begin{table} 
\renewcommand{\arraystretch}{1.2}
\setlength\tabcolsep{.02\columnwidth}
\begin{ruledtabular}
\begin{tabular}{ p{.34\columnwidth}  p{.2\columnwidth}  p{.14\columnwidth}  p{.225\columnwidth} }

   Sector													&  Coupled ODES 		&  Gauge condition	& Hierarchical ODEs\\

   \midrule

	$\ell\geq 2$, $\ell+m$  odd, $m\neq0$	&  $i=9,10$	 	     	& $i=8$					& 					 \\

	$\ell\geq 2$, $\ell+m$ even, $m\neq0$	&  $i=1,3,5,6,7$		& $i=2,4$					&         			\\

    $\ell=1$, $m=\pm1$								&  $i=1,3,5,6$  			&  $i=2,4$				& 					 \\

    $\ell\geq 2$ odd, $m=0$ 						&  $i=9,10$ 				& 								& $i=8$ 		\\

    $\ell\geq 2$ even, $m=0$ 						&  $i=1,3,5$				& $i=4,6,7$				& $i=2$		\\  

    $\ell=1$, $m=0$ 									&  $i=9$  					&		 						& $i=8$ 		\\

	$\ell=0$												&  $i=1,3$  				&  $i=6$  					& $i=2$ 	
    \end{tabular}
\end{ruledtabular}
\caption[Hierarchical structure of the field equations]
{Summary of how the solutions for nonvanishing modes are obtained. The first column displays the different sectors into which the modes are divided. The second column lists the modes $\R^n_{i\ell m}$ that are calculated by directly solving a coupled subset of the wave equations~(\ref{tildeEFE1}) or (\ref{tildeEFE2}). The third column lists the modes that can be extracted algebraically from the coupled modes using the gauge condition~(\ref{eq:gauges-w0}) or (\ref{eq:gauges-w1}). The fourth column lists the modes that must be extracted from the coupled modes using the remaining, hierarchically decoupled subset of the wave equations~(\ref{tildeEFE1}) or (\ref{tildeEFE2}). All unlisted modes identically vanish.} 
\label{tabHierarchy}
\end{table}
\endgroup

\subsection{Transformations between slicings}\label{sec_slicing_transformations}
%

If $|r^*|\gtrsim M/\e$, then $t$ and $\w$ can differ by a large amount, of order $\gtrsim 1/\e$. Hence, at large distances, the two-timescale expansion of the metric in one slicing can differ significantly from the two-timescale expansion of it in another slicing. This is related to the fact that the two-timescale expansion breaks down at large distances, as described in Ref.~\cite{Pound:2015wva}. This breakdown is minor in hyperboloidal slicing but dramatic in $t$ slicing.

However, within the region $|r^*|\lesssim M/\e$, the choice of slicing should be free. Given the results in one slicing, we should be able to recover the fields in another. In fact, we can straightforwardly determine the relationship between them. To eliminate ambiguities in the resulting equations, we introduce labels in square brackets, such as $[t]$ and $[\w]$, to  denote quantities defined with respect to a given a slicing. For example, $\R^{[\w]n}_{i\ell m}(\tilde \w,r)$ will denote a coefficient in the two-timescale expansion based on $\w$ slicing. $\R^{[\w]n}_{i\ell m}(\tilde t,r)$ denotes that same function evaluated at $\tilde t$; this differs from $\R^{[t]n}_{i\ell m}(\tilde t,r)$ because if $k\neq0$ then $\R^{[\w]n}_{i\ell m}$ and $\R^{[t]n}_{i\ell m}$ are different functions of $r$ (and for $n>1$, also different functions of their first argument, as we see below).

The transformations between slicings follow from $\tilde \w = \tilde t - \e k(r^*)$ and $\phi_p(\w,\e) = \phi_p(t,\e)-\Omega(\tilde t,\e)k(r^*)+\frac{1}{2}\e\dot\Omega(\tilde t) k^2(r^*)+\O(\e^2)$. Using these expansions to re-expand $\R^n_{i\ell m}(\tilde \w,r)e^{-im\phi_p(\w,\e)}$ at fixed $\tilde t$ and $\phi_p(t,\e)$, we find
\begin{align}
\R^{[\w]n}_{i\ell m}e^{-im\phi_p(\w,\e)} &= e^{im\Omega_0(\tilde t)k}\bigg\{\! 1-\e k\bigg[\tilde\partial_{\tilde t} +\frac{1}{2}im\dot\Omega_0(\tilde t)k\bigg]\nonumber\\
&\quad +\O(\e^2)\!\bigg\}\R^{[\w]n}_{i\ell m}(\tilde t,r) e^{-im\phi_p(t,\e)}.\label{slicing re-expansion}
\end{align}

For a sufficiently well-behaved metric, the reexpansion of $\sum_n \e^n \R^{[\w]n}_{i\ell m}(\tilde \w,\e) e^{-im\phi_p(\w,\e)}$ must agree with $\sum \e^n \R^{[t]n}_{i\ell m}(\tilde t,\e) e^{-im\phi_p(t,\e)}$ at each order in $\e$ at fixed $\tilde t$ and $\phi_p(t,\e)$. This equality gives us
\begin{align}
\R^{[t]1}_{i\ell m} &= e^{im\Omega_0(\tilde t)k(r^*)}\R^{[\w]1}_{i\ell m}(\tilde t,r),\label{slicing transformation1}\\
\R^{[t]2}_{i\ell m} &= e^{im\Omega_0(\tilde t)k(r^*)}\bigg\{\R^{[\w]2}_{i\ell m}(\tilde t,r)-\e k(r^*)\bigg[\tilde\partial_{\tilde t}\nonumber\\
&\quad+\frac{1}{2}im\dot\Omega_0(\tilde t)k(r^*)\bigg]\R^{[\w]1}_{i\ell m}(\tilde t,r)\bigg\}.\label{slicing transformation2}
\end{align}
If we first calculate the metric perturbations in hyperboloidal slicing, these relationships allow us to transform those results into $t$ slicing. The inverse relationships are also easily derived.

These transformations also affect the second-order source terms $\delta^2 G^0_{i\ell m}$ and $E^1_{ij\ell m}\R^1_{j\ell m}$. In particular, they alter the asymptotic behavior of $E^1_{ij\ell m}\R^1_{j\ell m}$. We will return to this issue in Sec.~\ref{sec_source_asymptotics}.

%
\subsection{Boundary conditions}\label{sec_BCs}
%

For any given boundary conditions we can solve the frequency-domain field equations~\eqref{tildeEFE1}--\eqref{tildeEFE2} using the method of variation of parameters, as done at first order in Refs.~\cite{Akcay:2013wfa,Wardell:2015ada}. We will provide additional details about the application of this method at second order, with hyperboloidal slicing, in Ref.~\cite{worldtube-paper}.

The boundary conditions themselves are also seemingly obvious: our solutions should satisfy retarded boundary conditions, with no radiation coming into the system from infinity and none coming out of the black hole's past horizon. However, determining the form of these boundary conditions in the two-timescale expansion is nontrivial because, as alluded to above, the two-timescale expansion breaks down at large $r$ and near the horizon. Here we sketch the form of the boundary conditions for (i) homogeneous solutions, which are required to obtain inhomogeneous solutions in the method of variation of parameters, (ii) first-order inhomogeneous solutions, and (iii) second-order inhomogeneous solutions.

For homogeneous solutions, we construct a complete basis. For nonstationary ($m\neq0$) modes, half of the basis solutions represent regular, purely outgoing waves at infinity, behaving like $\sim e^{-i\omega_m u}$ as $r\to\infty$; the other half represent regular, purely ingoing waves at the horizon, behaving like $\sim e^{-i\omega_m v }$ as $r\to2M$. For stationary ($m=0$) modes, half of the basis solutions are regular at infinity, and half are regular at the future horizon.

Here by regularity at the horizon we mean regularity of each component in ingoing Eddington-Finkelstein coordinates $(v,r,\theta^A)$. By expressing the Eddington-Finkelstein components of a generic (symmetric) tensor $A_{\mu\nu}$ in terms of its tensor-harmonic coefficients $A_{i\ell m}$,  we find the following: $A_{\mu\nu}$ is smooth at the future horizon if and only if
\begin{subequations}\label{horizon regularity}
\begin{align}
\text{all }A_{i\ell m}(v,r)& \text{ are smooth at } r=2M,\\
A_{2\ell m}(v,r) &= A_{1\ell m}(v,r)+\O(f^2),\label{i=2 horizon regularity}\\
A_{i\ell m}(v,r) &= A_{i+1,\ell m}(v,r)+\O(f) \nonumber\\
							&\quad \text{ for } i=4,8. \label{i=4,8 horizon regularity}
\end{align}
\end{subequations}
Here the mode amplitudes are defined without overall factors, by $A_{i\ell m}:=\frac{1}{\kappa_i}\oint d\Omega\, \eta^{\alpha\mu}\eta^{\beta\nu}A_{\alpha\beta} Y^{i\ell m*}_{\mu\nu}$; this implies that the conditions~\eqref{horizon regularity} are slightly modified for the source modes defined by Eq.~\eqref{Seff_ilm}, for example. Since the conditions describe the behavior of the mode coefficients as functions of $v$ and $r$, not as functions of $t$, and $r$, they also do not quite apply for $\R^{[t]n}_{i\ell m}$, but they do apply to the combined quantities $\R^{[t]n}_{i\ell m}e^{-i\omega_m t}=\left(\R^{[t]n}_{i\ell m}e^{i\omega_m r^*}\right)e^{-i\omega_m v}$. Given smooth mode coefficients, the Lorenz-gauge field equations tend to enforce the remaining two conditions in~\eqref{horizon regularity}.

For each $\ell m$ mode there are $2d$ independent homogeneous solutions, where $d$ is the number of coupled equations to be solved for the given $\ell m$ mode, as per the second column in Table~\ref{tabHierarchy}. We write these basis solutions as $R^{k-}_{i\ell m}$ and $R^{k+}_{i\ell m}$, with $k=1,\ldots,d$. For the nonstationary modes in $[t]$ slicing, these satisfy the boundary conditions
\begin{align}
 R^{[t]k+}_{i\ell m}(r)&=e^{i\omega_m r^*}\sum\limits^\infty_{n=0} a_{i\ell m}^{k,n}r^{-n}, \label{eq:BCout}\\
 R^{[t]k-}_{i\ell m}(r)&=e^{-i\omega_m r^*}\sum\limits^\infty_{n=0} b_{i\ell m}^{k,n}\left(r-2M\right)^n,\label{eq:BCin}
 \end{align}
for any value of $r$ near infinity or near $r=2M$, respectively. Here both $\omega_m$ and the coefficients $a_{i\ell m}^{k,n}$ and $b_{i\ell m}^{k,n}$ implicitly depend on slow time. For the stationary modes ($m=0$), the solutions satisfy
\begin{align}
 R^{[t]k+}_{i\ell0}(r)&= \sum\limits^\infty_{n=\ell}  \left(a_{i\ell m}^{k,n}+\bar a_{i\ell m}^{k,n}\ln r\right)r^{-n}, \label{eq:BCoutStatic}\\
 R^{[t]k-}_{i\ell0}(r)&= \sum\limits^\infty_{n=0} b_{i\ell m}^{k,n}\left( r-2M\right)^n,\label{eq:BCinStatic}
\end{align}
near the respective boundaries; in this case because the field equations do not depend on $\Omega_0$, the coefficients $a_{i\ell m}^{k,n}$, $\bar a_{i\ell m}^{k,n}$, and $b_{i\ell m}^{k,n}$ do not depend on slow time. In both cases the coefficients are different for each $i\ell m$ and are determined from recurrence relations derived by substituting the  ansatzes into the field equations. These can be found in Appendix A of Ref.~\cite{Akcay:2010dx}. That reference also describes how to choose the leading coefficients, $a_{i\ell m}^{k,0}$, $\bar a_{i\ell m}^{k,0}$, and $b_{i\ell m}^{k,0}$, to generate the $2d$ independent solutions. For the stationary modes, the homogeneous solutions are known analytically~\cite{Osburn-etal:14}. For the nonstationary modes, the homogeneous solutions can be obtained numerically, enforcing the boundary condition~\eqref{eq:BCout} or \eqref{eq:BCin} exactly at some radius $r_{\rm in}$ near the horizon or $r_{\rm out}$ near infinity.

The boundary conditions on $\w$ slices are easily obtained from those on $t$ slices using the transformation~\eqref{slicing transformation1}. That transformation implies $R^{[\w]k+}_{i\ell m} = R^{[t]k+}_{i\ell m}e^{-i\omega_m(\tilde t)k(r^*)}$ for $m\neq0$, and $R^{[\w]k+}_{i\ell m} = R^{[t]k+}_{i\ell m}$ for $m=0$. If the boundaries are placed sufficiently near $r=2M$ or $r\to\infty$, where $\w$ is sufficiently close to $v$ or $u$, then for the $m\neq0$ modes this amounts to using the same boundary conditions as in (\ref{eq:BCout})~and~(\ref{eq:BCin}), but without the plane-wave factor $e^{\pm i\omega_m r^*}$ in front of the sum. In that case, all modes, nonstationary and stationary, behave as simple power series near the boundaries (modulo the logarithms at large $r$ in the stationary modes).

Now moving onto the first-order inhomogeneous solutions, we note that these solutions are homogeneous at all points away from $r_0$. Hence, at the future horizon they should be a linear combination of the basis solutions $R^{k-}_{i\ell m}$, and at large $r$ they should be a linear combination of the basis solutions $R^{k+}_{i\ell m}$.  The analysis in Ref.~\cite{Pound:2015wva} showed that this is true despite the breakdown of the two-timescale expansion at large $r$, and a similar analysis, to be presented elsewhere, shows the same at the horizon. In other words, at first order the correct boundary conditions in the two-timescale expansion are the standard ones imposed in the literature on the first-order solution in the frequency domain. 

At second order, the situation is more complicated. The sources are now non-compact, extending all the way to the boundaries. This means that the solutions are not homogeneous anywhere. One might guess that integrating the source against a frequency-domain retarded Green's function would still yield the correct result, but Ref.~\cite{Pound:2015wva} showed that this is not the case. For some modes, the retarded integral fails to converge. Even when it does converge, one can show that it does not necessarily give the physically correct result; the companion paper~\cite{worldtube-paper} derives the necessary and sufficient conditions under which it does give the correct solution.

Ref.~\cite{Pound:2015wva} showed, in a scalar toy model, how to obtain physical boundary conditions by using a retarded time-domain post-Minkowski solution at large $r$. This solution is obtained analytically, using a time-domain retarded Green's function in the asymptotic region, up to some desired post-Minkowskian order. It is then re-expanded in the two-timescale form to determine the physical boundary conditions for the two-timescale solution. A similar iterative, analytical, time-domain method can be used to generate physical boundary conditions near the future horizon. These methods were used in the calculation in Ref.~\cite{Pound-etal:19}. Their full details will be presented in followup papers~\cite{two-timescale-2}~\cite{two-timescale-3}.

Near future null infinity and the future horizon, the analytical time-domain solutions take the form
\begin{align}
\bar h^{1\pm}_{i\ell m}(\w,r,\e) &= \bar h^{1H\pm}_{i\ell m}(\w,r,\e),\label{BC1 form}\\
\bar h^{2\pm}_{i\ell m}(\w,r,\e) &= \bar h^{2P\pm}_{i\ell m}(\w,r,\e) + \bar h^{2H\pm}_{i\ell m}(\w,r,\e),\label{BC2 form}
\end{align}
where `$+$' is the solution near future null infinity and `$-$' is the solution near the future horizon. The terms $\bar h^{nH\pm}_{i\ell m}$ are homogeneous solutions satisfying retarded boundary conditions: $\bar h^{nH+}_{i\ell m}$ is an outgoing, regular wave at infinity, and $h^{2H-}_{i\ell m}$ is an ingoing, regular wave at the future horizon. These homogeneous solutions are not determined internally within the analytical time-domain calculations; instead they are determined by matching the time-domain solution to the two-timescale solution. 

$\bar h^{2P\pm}_{i\ell m}$ is an analytically known particular solution to the second-order field equations~\eqref{EFE2 no puncture}, fully determined by the first-order modes $\bar h^{1H\pm}_{i\ell m}$. $\bar h^{2P+}_{i\ell m}$ may or may not be regular at future null infinity, and $\bar h^{2P-}_{i\ell m}$ may or may not be regular at the future horizon, but they are (by construction) physical, causal solutions. Any irregularity arises from the choice of gauge. For example, outgoing waves at nonlinear orders in the Lorenz gauge contain logarithms at large $r$, as is well known from post-Minkowski theory in the harmonic gauge~\cite{Blanchet-Damour:86,Blanchet-Damour:92}. As another example, at second order every single mode is impacted (through the source $\delta^2 G_{i\ell m}$) by the known pathologies of the first-order $\ell=0$ mode in the Lorenz gauge, reviewed in Appendix~\ref{sec_low_modes}. 

To match the time-domain solutions at the boundaries to the two-timescale solution in the bulk of the spacetime, Eqs.~\eqref{BC1 form}--\eqref{BC2 form} are re-expanded for $\e\ll 1$ at fixed $(\tilde\w,\phi_p(\w,\e),r)$, as in Eq.~\eqref{hnilm expansion}. This re-expansion can be highly nontrivial due to the complexity of the time-domain solution (as described in Ref.~\cite{Pound:2015wva}), but once it is complete, the results must agree term by term with the two-timescale expansion~\eqref{multiscale h}. This provides the matching conditions 
\begin{align}
\R^1_{i\ell m}e^{-im\phi_p} &= \tilde{\bar h}^{1H\pm}_{i\ell m}(\tilde\w,\phi_p, r,0),\label{matching 1}\\
\R^2_{i\ell m}e^{-im\phi_p} &= \tilde{\bar h}^{2P\pm}_{i\ell m}(\tilde\w,\phi_p, r,0) +\partial_\e\tilde{\bar h}^{1H\pm}_{i\ell m}(\tilde\w,\phi_p, r,0) \nonumber\\
											&\quad + \tilde{\bar h}^{2H\pm}_{i\ell m}(\tilde\w,\phi_p, r,0),\label{matching 2}
\end{align}
where the `$+$` refers to the matching condition at large $r$, and the `$-$' to the matching condition near the horizon. Equation~\eqref{matching 1} is used to match `outward': the output of the leading-order two-timescale solution, $\R^1_{i\ell m}$, fixes the homogeneous solutions $\bar h^{1H\pm}_{i\ell m}$. That in turn fixes the particular solutions $\bar h^{2P\pm}_{i\ell m}$. Equation~\eqref{matching 2} is then used to match `inward', providing boundary conditions for the second-order two-timescale solution $\R^2_{i\ell m}$. To enforce those conditions, one can treat the first two terms on the right as a puncture $\R^{2\P\pm}_{i\ell m}$, which is utilized in the same manner as the puncture on $z^\mu$. The residual field $\R^{2\res\pm}_{i\ell m} = \R^{2}_{i\ell m} - \R^{2\P\pm}_{i\ell m}$ must then be the homogeneous solution $\tilde{\bar h}^{2H\pm}_{i\ell m}(\tilde\w,\phi_p, r,0)$. This implies that, just like the first-order field $\R^1_{i\ell m}$, $\R^{2\res\pm}_{i\ell m}$ must reduce to a linear combination of the basis solutions $R^{k+}_{i\ell m}$ at large $r$ and of the basis solutions $R^{k-}_{i\ell m}$ near $r=2M$. Numerically solving for the residual field then fixes the homogeneous solutions $\bar h^{2H\pm}_{i\ell m}$, fully determining the full, physical fields~\eqref{BC2 form} at the future horizon and future null infinity.



%
\subsection{Asymptotics of the second-order source}\label{sec_source_asymptotics}
%

Finally, we consider the behavior of the second-order source at large $r$ and near the horizon. Our analysis will verify the expectation that the source is significantly better behaved with hyperboloidal slicing. 

We focus our attention on the term $E^1_{ij\ell m}\R^1_{i\ell m}$ in the source; we will briefly discuss $\delta^2 G^0_{i\ell m}$ at the end of the section.

First we look at the large-$r$ behavior. In $t$ slicing, the first-order field behaves as $\R^{[t]1}_{i\ell m}\sim e^{i\omega_m(\tilde t) r*}$, from Eq.~\eqref{eq:BCout}. When a slow time derivative acts on this field, it yields $\partial_{\tilde t}\R^{[t]1}_{i\ell m}\sim i\dot\omega_m r^* e^{i\omega_m r^*}$. According to Eqs.~\eqref{tildeBox1} and \eqref{Mij1}, the terms in $E^1_{ij\ell m}\R^1_{i\ell m}$ then behave as
\begin{align}
\Box_1 \R^{[t]1}_{i\ell m} &= \left[\frac{\omega_m \dot\omega_m}{2} r^*+\O(r^0)\right] \R^{[t]1}_{i\ell m},\\
{\cal M}^{ij}_1 \R^{[t]1}_{j\ell m} &= \left[\frac{i\dot\omega_m r^*}{r^2}+\O(r^{-2})\right]C^{ij} \R^{[t]1}_{j\ell m},
\end{align}
where $C^{ij}$ is an $r$-independent coupling matrix. Therefore
\begin{subequations}
\begin{align}
E^1_{ij\ell m}\R^{[t]1}_{j\ell m} &= \left[\frac{\omega_m \dot\omega_m}{2} r^*+\O(r^0)\right]\R^{[t]1}_{i\ell m},\\
&\sim r^* e^{i\omega_m r^*}.\label{E1 t slicing large r}
\end{align}
\end{subequations}
We provide a further discussion of this behavior toward the end of the section.

Next consider the large-$r$ behavior in $\w$ slicing. Assume $r$ is sufficiently large that $\w=u$. From the discussion in the previous section, $\R^{[u]1}_{i\ell m}$ is a simple power series in $1/r$ at large $r$, with coefficients that depend on $\tilde u$. According to Eqs.~\eqref{tildeBox1} and \eqref{Mij1}, the terms in $E^1_{ij\ell m}\R^1_{i\ell m}$ then behave as
\begin{align}
\Box_1 \R^{[u]1}_{i\ell m}&\sim \frac{1}{r^2}\R^{[u]1}_{i\ell m},\\
{\cal M}^{ij}_1 \R^{[u]1}_{j\ell m}&\sim \frac{1}{r^2}C^{ij} \R^{[u]1}_{j\ell m},
\end{align}
and so
\begin{subequations}
\begin{align}
E^1_{ij\ell m}\R^{[u]1}_{j\ell m} &\sim \frac{1}{r^2}.
\end{align}
\end{subequations}
This is an improvement over Eq.~\eqref{E1 t slicing large r} by {\em three orders} in $1/r$.

Now we look at the behavior near the horizon. In $t$ slicing, the first-order field behaves as $\R^{[t]1}_{i\ell m}\sim e^{-i\omega_m(\tilde t) r*}$, from Eq.~\eqref{eq:BCin}. When a slow time derivative acts on this field, it yields $\partial_{\tilde t}\R^{[t]1}_{i\ell m}\sim i\dot\omega_m r^* e^{-i\omega_m r^*}$. According to Eqs.~\eqref{tildeBox1} and \eqref{Mij1}, the terms in $E^1_{ij\ell m}\R^1_{i\ell m}$ then behave as
\begin{align}
\Box_1 \R^{[t]1}_{i\ell m}&=\left[-\frac{\omega_m \dot\omega_m}{2} r^*+\O(f^0)\right] \R^{[t]1}_{i\ell m},\\
{\cal M}^{ij}_1 \R^{[t]1}_{j\ell m}&= \left[-\frac{i\dot\omega_m r^*}{r^2}+\O(f^0)\right]C^{ij} \R^{[t]1}_{j\ell m},
\end{align}
and so
\begin{subequations}
\begin{align}
E^1_{ij\ell m}\R^{[t]1}_{j\ell m} &= \left[-\frac{\omega_m \dot\omega_m}{2} r^*+\O(f^0)\right]D^{ij}\R^{[t]1}_{j\ell m},\\
&\sim r^* e^{-i\omega_m r^*},\label{E1 t slicing near H}
\end{align}
\end{subequations}
where we have used $f=(r-2M)/r$ to count orders, and $D^{ij}$ is an $\O(f^0)$ coupling matrix.

Next consider the near-horizon behavior in $\w$ slicing. Assume $r$ is sufficiently near $2M$ that $\w=v$. From the discussion in the previous section, $\R^{[v]1}_{i\ell m}$ is a simple power series in $(r-2M)$ near the horizon, with coefficients that depend on $\tilde v$. According to Eqs.~\eqref{tildeBox1} and \eqref{Mij1}, the terms in $E^1_{ij\ell m}\R^1_{i\ell m}$ then behave as
\begin{align}
\Box_1 \R^{[v]1}_{i\ell m}\sim f,\\
{\cal M}^{ij}_1 \R^{[v]1}_{j\ell m}\sim f.
\end{align}
In the first equation, we have used $\partial_{r^*}=f\partial_r$. In the second, we have used the regularity conditions~\eqref{horizon regularity}. Hence,
\begin{align}
E^1_{ij\ell m}\R^{[v]1}_{j\ell m} \sim f.
\end{align}
This is an improvement over Eq.~\eqref{E1 t slicing near H} from a logarithmic divergence to something that vanishes at the horizon.

What is the physical cause of these wildly differing behaviors between the different slicings? In $t$ slicing, our choice of slow and fast time has left rapid oscillations in $r$ in the wave zones. Because the frequencies of those oscillations vary with slow time, varying the slow time a small amount can actually change the field by a large amount, leading to  slow time derivatives having large effects. On the other hand, in $\w$ slicing, our choice of slow and fast time in this case has correctly factored out rapid oscillations not only in $t$ at fixed $r$, but also in $r$ at fixed $t$. The amplitudes $\R^{[\w]1}_{i\ell m}$ are hence genuinely slowly varying functions in spacetime.

On the other hand, different slicings have only a trivial effect on $\delta^2 G^0_{i\ell m}$. The reason is that this source term contains  no slow time derivatives. As a consequence, $\delta^2 G^0_{i\ell m}$ transforms in the same way as $\R^1_{i\ell m}$:
\beq
\delta^2 G^{[t]0}_{i\ell m} = \delta^2 G^{[\w]0}_{i\ell m}(\tilde t,r)e^{im\Omega_0(\tilde t)k(r^*)}.
\eeq

To assess the behavior at large $r$, we can use the facts that (i) $h^{1}_{\alpha\beta}\sim \frac{e^{-i m\phi_p(u,\e)}}{r}$, (ii) $\delta^2 G_{\alpha\beta}\sim \partial h^1 \partial h^1+h^1\partial^2 h^1$, and (iii) for first-order modes with mode numbers $m_1$ and $m_2$, the mode number of $\delta^2 G_{i\ell m}$ is $m=m_1+m_2$. From these facts, we see that at large $r$,
\beq
\delta^2 G^{[t]0}_{i\ell m} \sim (rf)\frac{e^{i\omega_m r^*}}{r^2}
\eeq
and
\beq
\delta^2 G^{[\w]0}_{i\ell m} \sim (rf)\frac{1}{r^2}.
\eeq
The factor $(rf)$ arises from the overall factor introduced in the definition~\eqref{Seff_ilm}; the behaviors $\frac{e^{i\omega_m r^*}}{r^2}$ and $1/r^2$ are the `natural' behaviors in the two slicings. These scalings hold even for $m=0$ modes, due to the beating of first-order waves against each other (i.e., the combination of modes with $m_2=-m_1$).

At the horizon, we can appeal to the fact that for a horizon-regular first-order metric perturbation, $\delta^2 G_{\alpha\beta}$ is necessarily horizon-regular as well. The regularity conditions~\eqref{horizon regularity} then apply, up to the overall factor of  $(rf)$ in the definition~\eqref{Seff_ilm}. So
\beq
\delta^2 G^{[t]0}_{i\ell m} \sim (rf) e^{-i\omega_m r^*}
\eeq
and 
\beq
\delta^2 G^{[\w]0}_{i\ell m} \sim (rf).
\eeq

We summarize with two points. First, hyperboloidal slicing dramatically increases the falloff of the source term $E^1_{ij\ell m}\R^1_{j\ell m}$ near the boundaries. This lessens the numerical burden of solving the second-order field equation and simplifies the task of finding physical boundary conditions. We will show this in more detail in Ref.~\cite{worldtube-paper}. 
Our second point is that hyperboloidal slicing has no significant impact on the nonlinear source~$\delta^2 G^0_{i\ell m}$. For this source term, we must derive boundary conditions as described in the previous section, and as derived explicitly in the scalar model of Ref.~\cite{Pound:2015wva}. This was the procedure used to compute the second-order binding energy in Ref.~\cite{Pound-etal:19}.

%
\section{Structure of the solution: combined evolution of the field and the trajectory}\label{sec_combined_expansions}
%

Collecting the expanded equation of motion and the expanded field equations, we end up with the coupled set~\eqref{Omega0},  (\ref{r0dot}), \eqref{Omega1}, (\ref{r1dot}),  (\ref{tildeEFE1}), (\ref{tildeEFE2}), (\ref{eq:gauges-w0}), and (\ref{eq:gauges-w1}). Here we outline in some detail how they combine to provide the adiabatic and post-adiabatic waveform-generation schemes displayed in Fig.~\ref{fig_flowchart}.

\subsection{Adiabatic order}

At first order, the solution is fully determined by the values of $r_0$, $M_1$, and $S_1$ (together with the boundary conditions and the relation $\Omega_0 = \sqrt{M/r_0^3}$). However, $M_1$ and $S_1$ are not relevant for the evolution equations at this order, and unless one is interested in post-adiabatic evolution, one can freely set them to zero. In this section we outline how this structure emerges.

For $\ell\geq 2$, and for $\ell=1, m=\pm1$, we use the same solutions as in Refs.~\cite{Akcay:2010dx,Wardell:2015ada}. These solutions are uniquely determined by the source~\eqref{T0ilm} and the retarded boundary conditions described above. Hence, once boundary conditions are selected, each of these mode amplitudes is fully determined by the value of $r_0$. Following the notation in the introduction, we write the $\ell\geq2$ and $\ell=1,m=\pm1$ solutions as 
\beq\label{ell>=2 form}
\R^1_{i\ell m}(\tilde \w, r) = \R^{\rm pp}_{i\ell m}(r_0(\tilde \w),r).
\eeq

However, for $\ell=0$ and for $\ell=1,m=0$, our solutions differ from the traditional ones. For these modes, the solution is {\em not} uniquely determined by regularity conditions: to any linear perturbation of Schwarzschild spacetime, one can always freely add a linear perturbation toward a Kerr spacetime with mass $M+\e M_1$ and spin $\e S_1$. In standard linear perturbation theory, the corrections $M_1$ and $S_1$ must be constants; if they were not, they would violate the linearized Einstein equation. But in the two-timescale expansion, at first order they can have {\em arbitrary} dependence on slow time. This is because regardless of their slow time dependence, they will still satisfy the leading-order field equations, in which no slow-time derivatives appear. Their dependence on $\tilde \w$ is only determined by the {\em second}-order field equations. Hence our total first-order solution for these modes is
\begin{align}
\R^1_{i00}(\tilde \w, r) &= \R^{\rm pp}_{i00}(r_0(\tilde \w),r) + \bar x_{i00}(M_1(\tilde \w),r),\label{ell=0 form}\\
\R^1_{i10}(\tilde \w, r) &= \R^{\rm pp}_{i10}(r_0(\tilde \w),r) + \bar x_{i10}(S_1(\tilde \w),r),\label{ell=1, m=0 form}
\end{align}
for some yet-to-be-determined functions $M_1(\tilde \w)$ and $S_1(\tilde \w)$. We present $\R^{\rm pp}_{i00}$, $\R^{\rm pp}_{i10}$,  $\bar x_{i00}$, and $\bar x_{i10}$ explicitly in Appendix~\ref{sec_low_modes}.

The mode amplitudes at a given value of slow time feed into the evolution equations that drive the system to future slow times. To evolve the system, we require evolution equations for $r_0$, $M_1$, and $S_1$. The first of these is given by Eq.~(\ref{r0dot}), for which we require the two-timescale expansion of the self-force.

To assist in writing this expansion, we first write the two-timescale expansions of $z^\mu$, $\dot z^\mu = dz^\mu/dt$, $u^\mu$, and $h^\res_{\alpha\beta}$:
\begin{align}
z^\alpha &= z^\alpha_0(t,\tilde t,\phi_p) + \e z^\alpha_1(\tilde t,\phi_p) + \O(\e^2),\\
\dot z^\alpha &= \dot z^\alpha_0(\tilde t,\phi_p) + \e \dot z^\alpha_1(\tilde t,\phi_p) + \O(\e^2),\\
u^\alpha &= u^\alpha_0(\tilde t,\phi_p) + \e \dot z^\alpha_1(\tilde t,\phi_p) + \O(\e^2),\\
h^\res_{\alpha\beta} &= \sum_{n}\e^n \tilde h^{n\res}_{\alpha\beta}(\tilde \w,\phi_p, r,\theta^A),
\end{align}
where 
\begin{align}
z^\alpha_0 &= (t,r_0,\pi/2,\phi_p),\\
z^\alpha_1 &= (0,r_1,0,0),\\
\dot z^\alpha_0 &= (1,0,0,\Omega_0),\\
\dot z_1^\alpha &= (0,dr_0/d\tilde t,0,\Omega_1),\\
u^\alpha_0 &= U_0 \dot z^\alpha_0, \\
u^\alpha_1 &= U_1 \dot z_0^\alpha + U_0 \dot z_1^\alpha.
\end{align}

The quantity $\tilde h^{n\res}_{\alpha\beta}$ here is given by the trace reversal of the coefficient of $\e^n$ in Eq.~\eqref{multiscale h} (or more precisely, the analogue of that equation for the residual field), summed over $i\ell m$. It can be calculated from the solutions to Eqs.~\eqref{tildeEFE1}--\eqref{tildeEFE2} as 
\beq\label{hn res}
\tilde h^{n\res}_{\alpha\beta} = \sum_{i\ell m}\frac{a_{i\ell}}{r}R^{n\res}_{\underline i\ell m}(\tilde\w,r)e^{im[\phi-\phi_p(\w,\e)]}Y^{i\ell m}_{\alpha\beta}(r,\theta,0),
\eeq
where the trace reversal is achieved with
\beq
\underline i = \begin{cases} 6 & \text{if } i=3,\\
									3 & \text{if } i=6,\\
									i & \text{otherwise}.
			\end{cases}
\eeq
The amplitudes $R^{1\res}_{i\ell m}$ can be computed from the solution to Eq.~\eqref{tildeEFE1} as $R^{1}_{i\ell m}-R^{1\P}_{i\ell m}$, or directly, using a puncture scheme. In Eq.~\eqref{hn res}, we have pointedly pulled out the factor $e^{im\phi}$ from $Y^{i\ell m}_{\alpha\beta}$; this makes manifest that $\tilde h^{n\res}_{\alpha\beta}$ only depends on $\phi_p$ and $\phi$ in the combination $\phi-\phi_p$. When evaluated on $z^\alpha_0$, where $\phi=\phi_p$, the residual field hence becomes independent of fast time. The same statement naturally applies to all derivatives of the residual field, which confirms that the self-force is independent of $\phi_p$, as assumed in the expansion~\eqref{multiscale f}.

Substituting the above expansions into the self-force~\eqref{SFhres}, we find that the first- and second-order terms in the expansion~\eqref{multiscale f} are given by
\begin{subequations}\label{tilde fn}
\begin{align}
\tilde f^\alpha_1(\tilde t) &=\frac{1}{2}g^{\alpha\beta}\tilde h^{1\res}_{u_0 u_0,\beta}, \label{tilde f1}\\
\tilde f^\alpha_2(\tilde t) &=\frac{1}{2}g^{\alpha\beta}\tilde h^{2\res}_{u_0 u_0,\beta} + \frac{1}{2}\Bigl[ r_1(\partial_r g^{\alpha\beta}\tilde h^{1\res}_{u_0u_0,\beta} +g^{\alpha\beta}\tilde h^{1\res}_{u_0u_0,r\beta}) \nonumber\\
&\quad + 2g^{\alpha\beta}\tilde h^{1\res}_{u_0 u_1,\beta}+P_0^{\alpha\beta}(2\Gamma^\gamma\tilde h^{1\res}_{\beta\gamma}-2U_0 u_0^\gamma\partial_{\tilde t}\tilde h^{1\res}_{\beta \gamma}\nonumber\\
&\quad -\tilde h^{1\res}_{\beta}{}^\gamma\tilde h^{1\res}_{u_0u_0,\gamma})\Bigr].\label{tilde f2}
\end{align}
\end{subequations}
Here a comma denotes a derivative at fixed slow time, all fields are evaluated on $z^\alpha_0$, $\tilde h^{n\res}_{u_0 u_0,\beta\cdots\gamma}:=\left.\tilde h^{n\res}_{\mu\nu,\beta\cdots\gamma}\right|_{z_0^\alpha}u_0^\mu u_0^\nu$,  $P_0^{\alpha\beta}:=g^{\alpha\beta} + u_0^\alpha u_0^\beta$, and $\Gamma^\alpha := U_0^2(2\Gamma^\alpha_{\dot z_0 \dot z_1}-3\Omega_0^2f_0r_1\delta^\alpha_r)$ is the leading nonzero term in the two-timescale expansion of $\Gamma^\alpha_{\beta\gamma}(z^\mu)u^\beta u^\gamma$.

In the evolution equation for $r_0$,~\eqref{r0dot}, we require $\tilde f^t_1$. From Eq.~\eqref{tilde f1}, this evaluates to
\begin{subequations}\label{f1t modes}
\begin{align}
\tilde f^t_1 &= -\frac{1}{2}f^{-1}_0 \Omega_0\tilde h^{1\res}_{u_0 u_0,\phi_p},\\
				&= \frac{1}{2r_0 f_0} \sum_{i\ell m}  a_{i\ell}i\omega_m(\tilde\w) R^{1\res}_{\underline i\ell m}(\tilde\w, r_0)\nonumber\\
				&\quad \times Y^{i\ell m}_{\alpha\beta}(r_0,\pi/2,0)u^\alpha_0u^\beta_0.
\end{align}
\end{subequations}
Since this formula only involves $m\neq 0$ modes, $\bar x_{i\ell m}$ does not contribute. Therefore, the entirety of the right-hand side is fully determined by the value of $r_0$, and we can write $\tilde f^t_1=\tilde f^t_1(r_0)$. Again because it has no fast time dependence, we also need not include $\bar x_{i\ell m}$ in the leading-order waveform. 

Once the evolution of $r_0$ is determined, it determines $\Omega_0$ and hence the adiabatic phase $\int \Omega_0 dt$. Therefore this collection of results provides us with the adiabatic evolution scheme outlined in the introduction and summarized in the upper box of Fig.~\ref{fig_flowchart}. In this scheme, we can entirely neglect $M_1$ and $S_1$.

\subsection{Post-adiabatic order}

Before moving onto the second-order field equations, we can extract more from the first-order solutions. Specifically, if we know the values of $M_1$ and $S_1$, we can compute the other independent component of $\tilde f^\alpha_1$:
\begin{subequations}\label{f1r modes}
\begin{align}
\tilde f^r_1 &= \frac{1}{2}f_0\tilde h^{1\res}_{u_0 u_0,r}\\
					&= \frac{1}{2} f_0 \sum_{i\ell m}  a_{i\ell}\partial_{r_0} [ R^{1\res}_{\underline i\ell m}(\tilde\w, r_0)Y^{i\ell m}_{\alpha\beta}(r_0,\pi/2,0)]u^\alpha_0u^\beta_0.
\end{align}
\end{subequations}
This gets contributions from both $m\neq0$ and $m=0$ modes, meaning it depends on $\bar x_{i\ell m}(M_1,S_1,r)$. We can therefore write it as $\tilde f^r_1=\tilde f^r_1(r_0,M_1,S_1)$. (If we were interested in conservative effects on an orbit around a nonspinning black hole on a given slice of slow time, we would evaluate this with $S_1$ set to zero.)

Next we turn to the second-order field equations~\eqref{tildeEFE2} and \eqref{eq:gauges-w1}. In the source term $\delta^2 G^0_{i\ell m}$ in Eq.~\eqref{tildeEFE2}, the only required input is $\Omega_0$ and the first-order amplitudes. In the source term $E^0_{ij\ell m}\R^{2\P}_{j\ell m}$, we require $h^{1\res}_{\alpha\beta}$ evaluated on $z^\mu_0$, as we see from Eq.~\eqref{h2P form} (derivatives of $h^{1\res}_{\alpha\beta}$ are also required for higher-order terms in that equation). As mentioned below Eq.~\eqref{alphas}, we also transfer the effect of $\tilde T^1_{i\ell m}$ into $\R^{2\P}_{i\ell m}$. These terms are proportional to $r_1$, $\Omega_1$, and $\dot r_0$. $r_1$ we can consider freely specified until we begin to evolve it; $\Omega_1$ is then determined from $r_1$ and $\tilde f^r_1$ via Eq.~\eqref{Omega1}; and $\dot r_0$ is determined from $\tilde f^t_1$ via Eq.~\eqref{r0dot}. 

The final source term, $E^1_{ij\ell m}\R^1_{j\ell m}$, given by Eq.~\eqref{Enijlm} with \eqref{tildeBox1} and \eqref{Mij1}, requires as input the slow time derivative of the mode amplitudes $\R^1_{i\ell m}$. From the form of $\R^1_{i\ell m}$ in Eqs.~\eqref{ell>=2 form}--\eqref{ell=1, m=0 form}, this derivative is given by
\begin{align}
\partial_{\tilde \w}\R^{1}_{i\ell m} &= \dot r_0 \frac{\partial}{\partial r_0}\R^{\rm pp}_{i\ell m} + \dot M_1\frac{\partial}{\partial M_1}\bar x_{i\ell m} \nonumber\\
&\quad + \dot S_1\frac{\partial}{\partial S_1}\bar x_{i\ell m}.\label{dh1 form}
\end{align}
$\frac{\partial}{\partial M_1}\bar x_{i\ell m}$ and $\frac{\partial}{\partial S_1}\bar x_{i\ell m}$ are easily calculated analytically from the formulas for $x_{i\ell m}$ in Appendix~\ref{sec_low_modes}. In Ref.~\cite{worldtube-paper} we present a method of calculating $\frac{\partial}{\partial r_0}\R^{\rm pp}_{i\ell m}$ at a given value of $r_0$ (without taking a numerical derivative, which would require $\R^{\rm pp}_{i\ell m}$ in a range of neighboring $r_0$ values). $\dot r_0 $ is computed from Eq.~\eqref{r0dot}. We will return to $\dot M_1$ and $\dot S_1$ momentarily.

For $\ell\geq 2$, and for $\ell=1, m=\pm1$, the $\dot M_1$ and $\dot S_1$ terms in Eq.~\eqref{dh1 form} are not involved, and additional corrections to the black hole's mass and spin cannot appear. Given the form of the source described above, the solution to Eqs.~\eqref{tildeEFE2} and \eqref{eq:gauges-w1} for these modes hence takes the form
\beq\label{h2 ell>=2 form}
\R^2_{i\ell m}(\tilde \w, r) = \hat\R^{2}_{i\ell m}(r_0,r_1,M_1,S_1,r),
\eeq
where $r_0$, $r_1$, $M_1$, and $S_1$ are functions of $\tilde \w$. Note that $M_1$ and $S_1$ appear because the second-order source for any given $\ell m$ gets contributions from every first-order mode. Concretely, $\bar x_{i\ell m}(M_1,S_1,r)$ contributes to every mode of $\delta^2 G^0_{i\ell m}$ by coupling to other modes, and it contributes to $\R^{2\P}_{i\ell m}$ through its contribution to $\tilde h^{1\res}_{\alpha\beta}$ and $\tilde f^r_1$.

For the $\ell=0$ and $\ell=1,m=0$ modes, two things change: mass and spin perturbations, $\bar x_{i\ell m}(M_2,S_2,r)$, appear exactly as at first order, and $\dot M_1$ and $\dot S_1$ appear as sources. Just as we can neglect $\bar x_{i\ell m}(M_1,S_1,r)$ in an adiabatic evolution, we can neglect $\bar x_{i\ell m}(M_2,S_2,r)$ in a first post-adiabatic evolution. In fact, we can neglect the second-order $\ell=0$ and $\ell=1,m=0$ modes entirely. However, we do require $\dot M_1$ and $\dot S_1$ regardless, as they determine the evolution of $\bar x_{i\ell m}(M_1,S_1,r)$, which must be included in $\tilde f^r_1$, $\R^{2\P}_{i\ell m}$, and $\delta^2 G^0_{i\ell m}$.

$\dot M_1$ and $\dot S_1$ are determined directly from the $\ell=0$ and $\ell=1,m=0$ field equations. The wave equation~\eqref{tildeEFE2} with $\ell=0$ can be solved with an arbitrary value of $\dot M_1$, and the wave equation with $\ell=1,m=0$ can be solved with an arbitrary value of $\dot S_1$. However, the gauge condition~\eqref{eq:gauges-w1} then uniquely fixes the values of $\dot M_1$ and $\dot S_1$.  The result is unsurprising: these quantities grow at precisely the rate that gravitational-wave energy and angular momentum enter the black hole, 
\beq
\dot M_1 = \dot E_H, \quad\text{and}\quad \dot S_1 = \dot L_H, \label{fluxes}
\eeq
where the energy and angular momentum fluxes down the horizon, $\dot E_H$ and $\dot L_H$, are obtained from the $\ell\geq2$ amplitudes $\R^{\rm pp}_{i\ell m}$; see, for example Eq.~(76) of Ref.~\cite{Akcay:2010dx}. This is in agreement with traditional analyses of the slow evolution of the central black hole~\cite{Teukolsky-Press:74, Poisson:04}. But we emphasize here that the results are obtained directly from the vacuum Einstein equations, rather than from the dynamics of the horizon generators as in those traditional analyses. In fact, the results do not even require any knowledge of the metric anywhere near the horizon. (Physically, this is a consequence of the fact that the total flux through a surface of constant $r$ is the same for all $r<r_0$.) We present the complete derivation in the next section.

Once the mode amplitudes $\R^{2\res}_{i\ell m}$ are known, one can calculate $\tilde f^t_2$, the principal input for $\dot r_1$ in Eq.~\eqref{r1dot}. From Eq.~\eqref{tilde f2}, it is given by
\begin{align}
\tilde f^t_2(\tilde t) &= -\frac{1}{2}f^{-1}_0\Omega_0\tilde h^{2\res}_{u_0 u_0,\phi_p} -f_0^{-1}\Omega_0\tilde h^{1\res}_{u_0 u_1,\phi_p}\nonumber\\
&\quad + \frac{1}{2} f_0^{-2}\Omega_0 r_1\left(f_0'\tilde h^{1\res}_{u_0u_0,\phi_p} - f_0\tilde h^{1\res}_{u_0u_0,r\phi_p}\right)\nonumber\\
&\quad + \frac{1}{2}\left(-f_0^{-1}\delta^\beta_t+U_0 u_0^\beta\right)\Big(2\Gamma^\gamma h^{1\res}_{\beta\gamma}-2U_0u_0^\gamma\partial_{\tilde t}\tilde h^{1\res}_{\beta \gamma}\nonumber\\
&\quad + f_0^{-1}\Omega_0\tilde h^{1\res}_{\beta t}\tilde h^{1\res}_{u_0u_0,\phi_p}- f_0\tilde h^{1\res}_{\beta r}\tilde h^{1\res}_{u_0u_0,r}\nonumber\\
&\quad + r_0^{-2}\tilde h^{1\res}_{\beta \phi}\tilde h^{1\res}_{u_0u_0,\phi_p}\Big).\label{tilde f2t}
\end{align}
The notation here is as described below Eq.~\eqref{tilde f2}.  $f_0'=2M/r_0^2$, and derivatives are evaluated following the examples of Eqs.~\eqref{f1t modes} and \eqref{f1r modes}. We have used the facts that $\tilde h^{1\res}_{u_0u_0,\phi} = -\tilde h^{1\res}_{u_0u_0,\phi_p}$ and $\tilde h^{1\res}_{u_0u_0,\theta}=0$; the latter follows from the system's up-down symmetry. Note that because of the fast time derivative in the first term, to evaluate $\tilde f^t_2$ we only require the $m\neq0$ modes of $\tilde h^{2\res}_{\alpha\beta}$. Both in $\tilde f^t_2(\tilde t)$ and in $d\tilde f^r_1/d\tilde t$, which appears in Eq.~\eqref{r1dot}, we require the slow-time derivative of $\tilde h^{1\res}_{\alpha\beta}$; this is calculated as in Eq.~\eqref{dh1 form}.

With $\dot r_1$ determined, one can evolve $r_1$ and hence evolve $\Omega_1$ [through Eq.~\eqref{Omega1}] and the post-adiabatic phase $\int(\Omega_0+\e\Omega_1) dt$. Therefore this collection of results provides us with the post-adiabatic evolution scheme outlined in the introduction and summarized in Fig.~\ref{fig_flowchart}. In this scheme, we can entirely neglect $M_2$ and $S_2$. The leading-order waveform amplitudes $\R^{\rm pp}_{i\ell m}$ are also independent of $M_1$ and $S_1$, but $M_1$ and $S_1$ do contribute to the post-adiabatic phase in the numerous ways mentioned above. 

\section{Application: leading-order balance laws}\label{balance laws}

As a first demonstration of our scheme, in this section we derive the standard balance laws relating the system's evolution to the flux of energy and and angular momentum carried out to infinity and into the black hole. The end result has been well established for decades~\cite{Galtsov:82,Sago-etal:06,Isoyama-etal:19}, but our derivation, a slight variant of the one sketched in Ref.~\cite{Pound-etal:19}, is novel in that it proceeds directly from the two-timescale field equations. 

\subsection{Evolution of mass and orbital energy}

We begin with the evolution of the mass $M_1$ and leading-order orbital energy ${\cal E}_0 = \mu f_0U_0$. These evolution equations follow specifically from the field equations for the $i=2, \ell=0$ field $R^2_{200}$; this corresponds to the angle-averaged $t$-$r$ component of the field equations. 

Before writing down the field equations, let us motivate their connection to energy fluxes. Note that if we consider $-\frac{1}{8\pi} \delta^2 G^0_{\alpha\beta}$ as an effective stress-energy tensor that sources $\tilde h^{2}_{\alpha\beta}$, then the energy crossing outward (toward larger radii) through a surface of constant $r$ is
\begin{subequations}
\begin{align}
\Delta E_{{\cal S}_r} &= \frac{1}{8\pi}\int_{{\cal S}_r} \delta^2 G^0_{\alpha\beta} t^\beta dS^\alpha \\
				&= \frac{1}{8\pi}\int \delta^2 G^0_{\alpha\beta} t^\beta r^\alpha r^2 dt d\Omega.
\end{align}
\end{subequations}
Here $t^\alpha = \delta_t^\alpha $ and $r_\alpha = \partial_\alpha r$. The average rate of energy transfer across the surface is therefore
\begin{subequations}\label{Edot to ddG}
\begin{align}
\dot E_{{\cal S}_r} &= \frac{1}{8\pi}f r^2 \int \delta^2 G^0_{tr}d\Omega \\
						&= -\frac{r}{\sqrt{4\pi}f} \delta^2 G^0_{200},
\end{align}
\end{subequations}
where we have used Eq.~\eqref{Seff_ilm}. This demonstrates the connection between the energy flux and the $i=2,\ell=0$ mode of the source. 

We can make the connection to the specific fluxes $\dot E_H$ and $\dot E_\infty$ by appealing to the concrete form of $\delta^2 G_{200}$. At any vacuum point (i.e., at all points off the worldline), the contracted Bianchi identity reads $\nabla^\beta\delta^2G_{\alpha\beta}=0$.\footnote{This equality holds at fixed slow time for the coefficient of $e^{-im\phi_p}$ in $\nabla^\beta\delta^2G_{\alpha\beta}$; we do not need to include slow-time derivatives from $\nabla^\beta\delta G_{\alpha\beta}=0$, as one might expect from performing a two-timescale expansion of the Bianchi identity. The reason is that the coefficient of each $e^{-i{m}\phi_p}$ in the first-order field is identical to what the coefficient of $e^{-i\omega_{m} s}$ would be in ordinary perturbation theory. It follows that if we omit all slow time derivatives, the coefficient of each $e^{-im\phi_p}$ in $\nabla^\beta\delta^2G_{\alpha\beta}$ is also identical to what the coefficient of $e^{-i\omega_m s}$ would be in ordinary perturbation theory, and therefore it satisfies the same identities.} The mode decomposition of this is identical to the mode decomposition of $\nabla^\beta\bar h_{\alpha\beta}=0$, and the analog of Eq.~\eqref{Z01}, specialized to $\ell=0$, is 
\beq\label{Bianchi i=2}
r\partial_r(f^{-1}\delta^2 G^0_{200})+f^{-1}\delta^2 G^0_{200}=0.
\eeq
This applies for all $r\neq r_0$. The extra factors of $f^{-1}$, relative to Eq.~\eqref{Z01}, arise from the differing factors in the decompositions \eqref{modeDecomposition} and \eqref{multiscale-S} of $\bar h_{\alpha\beta}$ and $\delta^2 G^0_{\alpha\beta}$. 
The solution to Eq.~\eqref{Bianchi i=2} is 
\beq
\delta^2 G^0_{200} = \frac{f s^-_{200}}{r}\theta(r_0-r) + \frac{f s^+_{200}}{r}\theta(r-r_0)
\eeq
for some constants $s^\pm_{200}$. Together with Eq.~\eqref{Edot to ddG}, this result implies that $\dot E_{{\cal S}_r} = - s^+_{200}/\sqrt{4\pi}$ for all $r>r_0$, and $\dot E_{{\cal S}_r} = - s^-_{200}/\sqrt{4\pi}$ for all $r<r_0$. Therefore, for all radii smaller than $r_0$, we have $\dot E_{{\cal S}_{r}}=-\dot E_H$, and for all radii larger than $r_0$, we have $\dot E_{{\cal S}_{r}}=\dot E_\infty$. (The minus sign appears in front of $\dot E_H$ because we define $\dot E_H$ to be an inward flux, toward smaller radii.) Combining these results, we obtain the following simple expression for $\delta^2 G^0_{200}$:
\beq\label{ddG200 to flux}
\delta^2 G^0_{200} = \frac{\sqrt{4\pi}f}{r}\left[\dot E_{H}\theta(r_0-r) - \dot E_{\infty}\theta(r-r_0)\right].
\eeq
A more detailed analysis is required, starting from the form of the puncture field, to establish that no term proportional to $\delta(r-r_0)$ appears. For the purpose of our simple demonstration, we omit those details here.

Because $\delta^2 G^0_{200}$ is manifestly integrable, we can simplify the field equations by using the second-order Detweiler stress-energy~\eqref{T_alpha_beta} instead of a puncture. That stress-energy is proportional to $\mu u_\alpha u_\beta$, meaning the $t$-$r$ component is proportional to $\mu u_r \sim \e^2$. Explicitly, the $t$-$r$ component has the simple form
\beq
T_{tr} = \mu \frac{U_0 \dot r_0}{r_0^2} \delta(r-r_0)\delta(\theta-\pi/2)\delta(\phi-\phi_p) +\O(\e^3); 
\eeq
corrections due to $h^\res_{\mu\nu}$ and $r_1$ appear only at third order. From this we can straightforwardly read off $T_{200} = \e^2 \tilde t^2_{200}\delta(r-r_0)$, following the notation in Eq.~\eqref{tilde_T1_ilm_expanded}, where
\beq
\tilde t^2_{200} = - \frac{{\cal E}_0 f_0\dot r_0}{4\sqrt{\pi}r_0}.
\eeq
We have used ${\cal E}_0 = \mu f_0 U_0$ to express the source in terms of the orbital energy.

Now, there are two field equations involving $R^2_{200}$: the wave equation~\eqref{tildeEFE2} with $i=2$ and $\ell=0$, which reads
\begin{align}
-\frac{1}{4}(\partial^2_{r^*}&-2Mf/r^3) R^2_2 + {\cal M}_0^{2j}R^2_j \nonumber\\
   &= -16\pi\tilde t^2_{200}\delta(r-r_0) +2\delta^2 G^0_{200} \nonumber\\
   &\quad + \frac{M}{r^2}\left[f H \dot R^1_3+(1-H)\dot R^1_1\right],\label{200v1}
\end{align}
and the gauge condition \eqref{eq:gauges-w1} with $k=1$  and $\ell=0$, which reads
\beq\label{LR2=R1dot}
\mathscr{L}_2R^2_2 = \dot R^1_1 + f\dot R^1_3,
\eeq 
where $\mathscr{L}_2=f(\partial_r +1/r)$. Here and below we omit the $\ell m$ label on $R^n_{i\ell m}$, and we adopt a first-order monopole solution in which $R^1_{200}=0$ (implying $\dot R^1_{200}=0$). The term ${\cal M}_0^{2j}R^2_j$ in the field equation couples $R^2_2$ to $R^2_1$, $R^2_3$, and $R^2_6$, but we can eliminate that coupling by substituting the gauge condition~\eqref{eq:gauges-w1} with $k=2$, which reduces ${\cal M}_0^{2j}R^2_j$ to
\beq
{\cal M}_0^{2j}R^2_j = \frac{ff'}{2}\partial_r R^2_2 + \frac{f^2}{2r^2}R^2_2 - \frac{f'}{2}H(\dot R^1_1 - f \dot R^1_3).
\eeq
Equation \eqref{200v1} then becomes
\begin{align}\label{DR2=S2}
\mathscr{D}_2 R^2_2  = S^2_2,
\end{align}
where $\mathscr{D}_2 = \left(f^2\partial_r^2  - ff'\partial_r - \frac{ff'}{r} - \frac{2f^2}{r^2}\right)$ and 
\begin{align}
S^2_2 &= \frac{16\sqrt{\pi} {\cal E}_0 f_0\dot r_0}{r_0}\delta(r-r_0)  - \frac{4M}{r^2}\dot R^1_1 \nonumber\\
	&\quad -\frac{16\sqrt{\pi}f}{r}\left[\dot E_{H}\theta(r_0-r) - \dot E_{\infty}\theta(r-r_0)\right].
\end{align}
Note that the dependence on the height function has now vanished, implying that the evolution equations will be independent of the choice of slow-time slicing. 

In the field equation~\eqref{DR2=S2}, $\dot{\cal E}_0$ (or equivalently, $\dot r_0$) and $\dot M_1$ act as sources for $R^2_2$. We could solve this field equation for arbitrary values of those sources. The gauge condition~\eqref{LR2=R1dot} would then determine the relationships between $\dot {\cal E}_0$, $\dot M_1$, $\dot E_{H}$, and $\dot E_{\infty}$. That was the general method sketched in Ref.~\cite{Pound-etal:19}, and it is in line with the general descriptions in Sec.~\ref{sec_field_equations}. 

However, the explicit analytical solution to Eq.~\eqref{DR2=S2} is distractingly lengthy. For compactness, we take another approach. It is well known that a solution to $E_{\alpha\beta}[\bar h]=S_{\alpha\beta}$ satisfies $\nabla^\beta\bar h_{\alpha\beta}=0$ if and only if $\nabla^\beta S_{\alpha\beta}=0$; this follows from $\nabla^\beta S_{\alpha\beta}=\nabla^\beta E_{\alpha\beta}[\bar h] = \Box \nabla^\beta \bar h_{\alpha\beta}$. So any information about the system's evolution that we can glean from the gauge condition, we can equally well extract directly from the source. Inspired by that, we will derive a differential equation for the source, which will yield the desired evolution equations.

We first note the commutation relation
\beq
\mathscr{L}_2\mathscr{D}_2 - \mathscr{D}_2\mathscr{L}_2 = \frac{2f}{r^2}\mathscr{L}_2. 
\eeq
Applying $\mathscr{L}_2$ to the field equation~\eqref{DR2=S2}, using this commutation relation, and substituting Eq.~\eqref{LR2=R1dot} for $\mathscr{L}_2R^2_2$, we obtain
\beq
\left(\mathscr{D}_2 + \frac{2f}{r^2}\right)(\dot R^1_1 + f\dot R^1_3) = \mathscr{L}_2 S^2_2.
\eeq
We write each side of this equation in the form $a(r_0)\delta'(r-r_0)+b(r_0)\delta(r-r_0)+c^+(r)\theta(r-r_0)+c^-(r)\theta(r_0-r)$, allowing us to equate the coefficients of $\delta'$, $\delta$, and $\theta[\pm(r-r_0)]$. The coefficients of $\delta'$ yield 
\beq\label{i=2 delta' eqn}
[\dot R^1_1+f_0\dot R^1_3] = \frac{16\sqrt\pi{\cal E}_0\dot r_0}{r_0},
\eeq
where $[x]:={\displaystyle \lim_{r\to r_0^+}x - \lim_{r\to r_0^-}x}$. The coefficients of $\delta$ yield
\begin{align}
\!\![(\partial_r-f'f^{-1})(\dot R^1_1+f\dot R^1_3)] &= - \frac{4M}{r_0^2f_0}[\dot R^1_1]  +\frac{16\sqrt\pi{\cal E}_0\dot r_0}{r_0^2f_0}\nonumber\\
			&\quad +\frac{16\sqrt\pi}{r_0}(\dot E_H+\dot E_\infty),\label{i=2 delta eqn}
\end{align}
and the coefficients of the Heaviside functions yield
\begin{align}\label{i=2 theta eqns}
\left(\mathscr{D}_2 + \frac{2f}{r^2}\right)(\dot R^{1\pm}_1 + f \dot R^{1\pm}_3) =  \mathscr{L}_2S_2^{2\pm},
\end{align}
where the $+$ applies for $r>r_0$ and the minus for $r<r_0$.

The two equations~\eqref{i=2 theta eqns} suffice to determine $\dot M_1$ and $\dot{\cal E}_0$. Substituting the first-order monopole field $R^1_{i00}=R^{pp}_{i00}+\bar x_{i00}(M_1)$ given in Appendix~\ref{infinity-regular monopole}, and using $\dot {\cal E}_0 = (d{\cal E}_0/dr_0)\dot r_0$, we quickly obtain\footnote{The same result is also obtained if we use the monopole solution discussed in Appendix~\ref{infinity-regular monopole}. On the other hand, if instead we use the Berndtson solution $R^1_{i00}=R^{\rm Bern}_{i00}+\bar x_{i00}(M_1)$, then we still obtain Eq.~\eqref{orbital E balance}, but \eqref{BH E balance} is replaced by Eq.~\eqref{M1dot Berndtson}, corresponding to the fact that the black hole's mass is $M_{\rm BH}=M+\e(M_{\rm Bern}+M_1)$ in this solution.}
\begin{align}
\dot M_1 &= \dot E_H\label{BH E balance}
\end{align}
from the $r<r_0$ equation, and
\begin{align}
\dot {\cal E}_0 + \dot M_1 = - \dot E_\infty\label{E balance}
\end{align}
from the $r>r_0$ equation. The first of these, as discussed in previous sections, tells us that the black hole's mass grows at the rate that energy flows down the horizon. The second tells us that the system's total energy changes at the rate that energy is carried out of it. We can also combine the two equations to write
\begin{align}
\dot {\cal E}_0 = - (\dot E_\infty+\dot E_H),\label{orbital E balance}
\end{align}
which says that at leading order, all the energy leaving the region $2M<r<\infty$ comes from the particle's orbital energy. We stress that obtaining these equations for $\dot{\cal E}_0$ and $\dot M_1$ does not require evaluating Eqs.~\eqref{i=2 theta eqns} at any particular values of $r$; Eqs.~\eqref{orbital E balance} and \eqref{BH E balance} ensure that Eqs.~\eqref{i=2 theta eqns} are satisfied at {\em all} $r$. 

Although we can also solve Eqs.~\eqref{i=2 delta' eqn} and \eqref{i=2 delta eqn}, they do not contain any additional information. Eq.~\eqref{i=2 delta' eqn} is satisfied for any of the solutions $R^1_{i00}=R^{pp}_{i00}+\bar x_{i00}(M_1)$ in Appendix~\ref{monopole solutions}, regardless of the value of $\dot r_0$, while Eq.~\eqref{i=2 delta eqn} again yields Eq.~\eqref{orbital E balance}. $\bar x_{i00}(M_1)$ does not enter into either of the equations because it is smooth at $r=r_0$.


We have already emphasised that our derivation proceeds directly from the field equations, but there is one other aspect worth drawing attention to: it shows that $-\frac{1}{8\pi} \delta^2 G^0_{\alpha\beta}$ provides a meaningful notion of gravitational stress-energy throughout the spacetime, in the sense that it defines appropriate fluxes of gravitational energy (and in the next section, of gravitational angular momentum). This may appear obvious, since the Isaacson gravitational-wave stress-energy is derived from  $\delta^2 G_{\alpha\beta}$~\cite{Isaacson}. However, earlier analyses of the evolution of a black hole (e.g., Refs.~\cite{Teukolsky-Press:74, Poisson:04}) have stressed that Isaacson's derivation does not apply at the horizon, and they have instead derived the black hole's evolution from the shear of the horizon generators. The ``fluxes" that emerge from such a derivation are then directly the rates of change of the black hole parameters rather than surface integrals of a stress-energy. To our knowledge, our derivation provides the first demonstration that the relevant physical fluxes are precisely those defined from $-\frac{1}{8\pi} \delta^2 G^0_{\alpha\beta}$. Reference~\cite{Moxon-etal} will extend this result to generic bound orbits around a Kerr black hole (and to post-adiabatic order).

\subsection{Evolution of spin and orbital angular momentum}

We next derive the evolution of the spin $S_1$ and leading-order orbital angular momentum ${\cal L}_0 = \mu r^2_0\Omega_0U_0$. These evolution equations follow specifically from the field equations for the $i=9, \ell=1, m=0$ field $R^2_{910}$; this corresponds to the angle-averaged $r$-$\phi$ component of the field equations. 

The analysis is much the same as in the preceding section, and our presentation will be terse. The angular momentum carried outward across a surface of constant $r$ is
\beq
\Delta L_{{\cal S}_r} = -\frac{1}{8\pi}\int \delta^2 G^0_{\alpha\beta} \phi^\beta dS^\alpha,
\eeq
where $\phi^\beta = \delta^\beta_\phi$, implying that the average rate of angular momentum transfer is
\begin{subequations}\label{Ldot to ddG}
\begin{align}
\dot L_{{\cal S}_r} &= -\frac{1}{8\pi}fr^2\int \delta^2 G^0_{r\phi}d\Omega \\
								&= \frac{r^2}{2\sqrt{3\pi} f}\delta^2 G^0_{910}.
\end{align}
\end{subequations}
Here we have again used Eq.~\eqref{Seff_ilm}.

The contracted Bianchi identity implies
\beq
\partial_r(f^{-1}\delta^2G^0_{910})+\frac{2}{rf}\delta^2G^0_{910} = 0
\eeq
for all $r\neq r_0$; c.f. Eq.~\eqref{Z04}, specialized to $\ell=1, m=0$. The solution for $r\neq r_0$ is
\beq
\delta^2 G^0_{910} = \frac{s^-_{910}f}{r^2}\theta(r_0-r) + \frac{s^+_{910}f}{r^2}\theta(r-r_0)
\eeq
for constants $s^\pm_{910}$. Substituting this into Eq.~\eqref{Ldot to ddG}, we see that $\dot L_{{\cal S}_r} = \frac{s^\pm_{910}}{2\sqrt{3\pi}}$, where the $+$ applies for all $r>r_0$ and the minus for all $r<r_0$. Therefore, writing $\dot L_{{\cal S}_{r>r_0}}=\dot L_\infty$ and $\dot L_{{\cal S}_{r<r_0}}=-\dot L_H$, we have
\beq\label{ddG200 to flux}
\delta^2 G^0_{910} = -\frac{2\sqrt{3\pi}f}{r^2}\left[\dot L_{H}\theta(r_0-r) - \dot L_{\infty}\theta(r-r_0)\right]\!.
\eeq

We once again use the Detweiler stress-energy~\eqref{T_alpha_beta} instead of a puncture. The $r$-$\phi$ component is 
\begin{align}
T_{r\phi} &= \mu f_0^{-1}U_0 \dot r_0 \Omega_0\delta(r-r_0)\delta(\theta-\pi/2)\delta(\phi-\phi_p) \nonumber\\&\quad+\O(\e^3),
\end{align}
from which we read off $T_{910}=\e^2\tilde t^2_{910}\delta(r-r_0)$, where
\beq
\tilde t^2_{910} = -\frac{1}{4}\mu\sqrt{\frac{3}{\pi}}f_0 U_0 \dot r_0\Omega_0.
\eeq

With these sources in hand, we can write down the two field equations involving $R^2_{910}$: First, the wave equation~\eqref{tildeEFE2} with $i=9, \ell=1, m=0$, which reads
\begin{align}\label{DR9=S9}
\mathscr{D}_9 R^2_{910} = S^2_{9}, 
\end{align}
where $\mathscr{D}_9=f^2\partial_r^2+ff'\partial_r-\frac{2f}{r^3}(3r-8M)$ and
\begin{align}
S^2_9 &= 64\pi\tilde t^2_{910}\delta(r-r_0) -8\delta^2 G^0_{910} \nonumber\\
   &\quad + 2f H\partial_r \dot R^1_{910}+H'\dot R^1_9.\label{S910}
\end{align}
Second, the gauge condition \eqref{eq:gauges-w1} with $k=4,\ell=1,m=0$, which reads
\beq\label{LR9=R8dot}
\mathscr{L}_9R^2_{910} = \dot R^1_{810} + H\dot R^1_{910},
\eeq 
where $\mathscr{L}_9=f(\partial_r+2/r)$. Here we have suppressed $\ell m$ indices as in the previous subsection.

The operators $\mathscr{L}_9$ and $\mathscr{D}_9$ satisfy the commutation relation
\beq
\mathscr{L}_9\mathscr{D}_9 -\mathscr{D}_9 \mathscr{L}_9 = \frac{4}{r^3}f(r-4M)\mathscr{L}_9.
\eeq
Applying $\mathscr{L}_9$ to Eq.~\eqref{DR9=S9}, using this relation, and substituting Eq.~\eqref{LR9=R8dot}, we obtain an equation that relates $\dot{\cal L}_0$ and $\dot S_1$ to $\dot L_H$ and $\dot L_\infty$:
\beq
\left(\mathscr{D}_9+\frac{4f}{r^3}(r-4M)\right)(\dot R^1_{810} + H\dot R^1_{910}) = \mathscr{L}_9 S_9^2.
\eeq
This can again be divided into one equation for the coefficients of $\delta'$, one for the coefficients of $\delta$, and one each for the coefficients of $\theta[\pm(r-r_0)]$. As with the energy balance laws, all the information can be obtained from the two equations in the regions $r<r_0$ and $r>r_0$. The former yields
\begin{align}
\dot S_1 &= \dot L_H,\label{BH L balance}
\end{align}
and the latter yields
\begin{align}
\dot {\cal L}_0 + \dot S_1 = - \dot L_\infty.\label{L balance}
\end{align}
Substituting the first into the second, we obtain
\begin{align}
\dot {\cal L}_0 = - (\dot L_\infty+\dot L_H).\label{orbital L balance}
\end{align}
We note that once again, the results are independent of the choice of height function in our slow time slicing.

\section{Conclusion and outlook}

Since the seminal work of Hinderer and Flanagan~\cite{Hinderer-Flanagan:08}, the two-timescale approach has provided a common language for discussions of the long-term evolution of EMRIs. However, it has never actually been implemented, nor has any attempt been made (until now) to apply it to the full system of equations in an EMRI, which involve not only the small companion's equation of motion, but also the perturbed Einstein equations.

In this paper, we have taken a first step toward a complete two-timescale treatment of EMRIs. Our analysis is restricted to the narrow case of quasicircular orbits around a Schwarzschild black hole, but it suffices to highlight many of the key features of the full problem:
\begin{enumerate}
\item The gravitational field is written as a sum of modes~\eqref{multiscale h}, each with a slowly varying amplitude, which evolves on the radiation-reaction time scale, and a rapidly oscillating phase, which varies on the orbital time scale.
\item The field equations split into two sets. One set, given by Eqs.~\eqref{tildeEFE1} and \eqref{tildeEFE2}, take the form of standard frequency-domain equations for the mode amplitudes at fixed values of slow time. The other set, given by Eqs.~\eqref{r0dot}, \eqref{r1dot}, and \eqref{fluxes}, take the form of evolution equations, determining how the field evolves as a function of slow time. 
\item Slow-time derivatives of the first-order field amplitudes also appear as source terms in the second-order field equation~\eqref{tildeEFE2}. The asymptotic behavior of these sources, near infinity as well as near the future horizon, depend strongly on the choice of time foliation. 
\end{enumerate}
Our formulation in this restricted case should therefore provide a guide to the full problem. 

As part of our two-timescale expansion, we have also introduced a frequency-domain formulation of the field equations with hyperboloidal slicing. Although we have cast our field equations in the language of a two-timescale expansion, the equations~\eqref{tildeEFE1} apply equally well in an ordinary frequency-domain calculation, with a metric perturbation $\bar h_{\alpha\beta}=\sum_{i\ell m}\frac{a_{i\ell}}{r}R_{i\ell m}(r)e^{-i\omega_m s}Y^{i\ell m}_{\alpha\beta}$. Such hyperboloidal slicing in the frequency domain has been considered in the past for the Teukolsky equation~\cite{Zenginoglu:11,PanossoMacedo:2019npm}, but this is the first time, to our knowledge, that it has been used for the full linearized Einstein equation. A companion paper will present a practical method of solving these equations with various choices of hyperboloidal slicing~\cite{worldtube-paper}. That paper will also apply the same method to calculate the slow-time derivatives of the first-order field. 

Our expansions in the body of the paper were formulated in terms of series in $\e^n$ at fixed values of a slow time variable. In Appendix~\ref{fixed Omega}, we also presented an equivalent method based on expansions at fixed values of the orbital frequency, which works with a nonperturbative $\Omega$ rather than the series $\Omega_0+\e\Omega_1+\ldots$. This alternative provides slightly neater equations, and it may look more familiar to researchers working on waveform generation in other regimes of the two-body problem, such as post-Newtonian theory~\cite{Blanchet:14}.

In sequel papers, we will present several extensions and applications of this work. Still considering quasicircular orbits in Schwarzschild, we will present the complete calculation of the second-order source, including the second-order Einstein tensor and punctures at the particle~\cite{coupling-paper,source-paper}; the formulation of physical boundary conditions for the second-order field, obtained from a post-Minkowskian expansion near future null infinity and an analogous expansion near the future horizon; the calculation of quasistationary quantities such as the Detweiler redshift~\cite{Pound:2014koa}; and the post-adiabatic inspiral and emitted waveforms. Going beyond the quasicircular case, we will present a two-timescale expansion for the full problem of generic orbits in Kerr spacetime~\cite{two-timescale-0,two-timescale-1}, along with accompanying treatments near the horizon and infinity~\cite{two-timescale-2,two-timescale-3}.

This two-timescale expansion will ultimately provide a complete, flexible wave-generation framework, as outlined in the quasicircular case in Sec.~\ref{this_paper} and Fig.~\ref{fig_flowchart}. For a generic orbit in Kerr spacetime, the system is characterized by a set of adiabatic-order mechanical parameters $J_0^a=({\cal E}_0,{\cal L}_0,{\cal Q}_0,M,S)$ and post-adiabatic ones $J_1^a=({\cal E}_1,{\cal L}_1,{\cal Q}_1,M_1, S_1)$, where $S=aM$ is the central black hole's leading-order spin, ${\cal L}$ is the orbital angular momentum around the black hole's spin axis, and ${\cal Q}$ is the Carter constant. With the exception of the background quantities $M$ and $S$, each of these evolves as a function of slow time (after eliminating any fast-time dependence using near-identity averaging transformations~\cite{vandeMeent-Warburton:18}). One can pre-compute the rates of change $\frac{dJ^a_0}{d\tilde t}(J_0^b)$ and $\frac{dJ^a_1}{d\tilde t}(J_0^b,J^b_1)$, along with the waveform amplitudes $h^{\omega_{mkn}}_{\alpha\beta}(J_0^B)$, across the parameter space. One can then quickly evolve through the parameter space by integrating simple ODEs. The trajectory through parameter space then determines the post-adiabatic evolution of the frequencies $\omega_{mkn} = \omega^{(0)}_{mkn}(J_0^b)+\e\omega^{(1)}_{mkn}(J_0^b,J_1^b)$ and the waveform $\sum_{mkn}h^{\omega_{mkn}}_{\alpha\beta}(J_0^b(\e t))e^{-i\int \omega_{mkn}(J_0^b(\e t),J_1^b(\e t))dt}$. This framework is described by one of us in Ref.~\cite{Pound-Wardell:21} (which was submitted some time after the present paper). It is easily extended to include additional effects, such as the small object's spin, in a modular way; such effects can again be precomputed before simulating an evolution. In principle this framework can stand alone, but to cope with the very large parameter space, its outputs can also be used as input for effective-one-body and surrogate models or as training data for neural networks~\cite{Antonelli:2019fmq,Rifat:2019ltp,Chua:2018woh}.

%
%


\acknowledgments
We are grateful to Leor Barack, Niels Warburton, and Barry Wardell for helpful discussions and to the anonymous referee for correcting numerous typos. AP additionally thanks Eanna Flanagan and Jordan Moxon for enlightening comments on the central black hole's evolution, and he acknowledges support from a Royal Society University Research Fellowship.

\appendix

\section{Two-timescale expansion with nonperturbative mechanical degrees of freedom}\label{fixed Omega}

In Sec.~\ref{sec_field_equations}'s description of the self-consistent framework, the governing philosophy is to express the metric perturbations as functionals of the system's mechanical degrees of freedom, and then to obtain approximate evolution equations for those degrees of freedom. While the evolution equations are approximate, the mechanical degrees of freedom themselves are treated nonperturbatively (i.e., never expanded in an asymptotic series).

In the two-timescale framework developed in the body of the paper, and summarized in Fig.~\ref{fig_flowchart}, we abandon that nonperturbative treatment and treat the metric and mechanical parameters on an equal footing: all quantities are expanded in powers of $\e$ at fixed $\tilde\w$ and $\phi_p$.

In this appendix, we present a slightly different formulation of the two-timescale expansion that preserves more of the spirit of the self-consistent expansion. We do so by expanding all quantities at fixed $J^a:=(\Omega,\delta{\cal M}, \delta{\cal S})$ instead of at fixed $\tilde\w$. Here we have introduced the order-1 quantities $\delta{\cal M}:=\delta M/\e$ and $\delta{\cal S}:=\delta S/\e$. The conceptual distinction is that rather than solving fast-time field equations to find the state of the system at a given value of slow time, and then evolving to the next slow time, here we will solve equations to find the state of the system for a given set of mechanical parameters, and then evolve to new values of those parameters. However, we stress that both expansions yield solutions that are uniformly accurate on the slow time scale. They differ only in details of implementation.

In the new expansion considered here, in place of Eq.~\eqref{quasicircular}, the worldline becomes $z^\mu(t,\e) = \left(t,r_p(J^a(t,\e),\e),\pi/2,\phi_p(t,\e)\right)$, where
\begin{align}
r_p(J^a,\e) &= r_0(\Omega)+\e r_1(J^a)+\O(\e^2),\label{r_p(J)}
\end{align}
using the fact that $\delta M$ and $\delta S$ do not appear in the adiabatic-order term $r_0$. Put another way, we expand the worldline as
\beq\label{z(J)}
z^\mu = z_0^\mu(t,\Omega,\phi_p) + \e z_1^\mu(J^a) +\O(\e^2),
\eeq
where $z_0^\mu(t,\Omega,\phi_p) = (t,r_0(\Omega),\pi/2,\phi_p)$ and $z^\mu_1=(0,r_1(J^a),0,0)$. Rather than a nonperturbative trajectory $z^\mu$, as in the self-consistent expansion, here $J^a$ and $\phi_p$ are our exact mechanical degrees of freedom. 

Similarly, in place of~\eqref{multiscale h} the metric is expanded as
\beq\label{multiscale h alt}
\bar h_{\mu\nu} = \sum_{n i\ell m}\frac{\e^n a_{i\ell}}{r} \R^{n}_{i\ell m}(J^a(\tilde\w,\e),r)e^{-im\phi_p(\w,\e)}Y^{i\ell m}_{\mu\nu}(r,\theta^A).\!
\eeq
The metric's time dependence is entirely encoded in $J^a$ and $\phi_p$.

Finally, because {\em all} functions are expanded at fixed $J^a$, we also expand the functions $\dot J^a:=dJ^a/d\tilde\w$:
\beq
\dot J^a(J^b,\e) = F^a_0(\Omega) + \e F^a_1(J^b) + \O(\e^2),\label{dot J}
\eeq
again using the fact that $\delta{\cal M}$ and $\delta{\cal S}$ do not appear at adiabatic order. The terms on the right are related to (but not identical to) $\dot J^a_n=\frac{dJ^a_n}{d\tilde\w}$, the derivative of the $n$th-order term in the expansion of $J^a$ at fixed $\tilde\w$. The two quantities can be related by re-expanding the right-hand side of Eq.~\eqref{dot J} at fixed $\tilde\w$ and comparing the result to $\frac{d}{d\tilde\w}[J^a_0(\tilde\w)+\e J^a_1(\tilde\w)+\O(\e^2)]$. This yields the relationships $\dot J^a_0(\tilde\w) = F^a_0(\Omega_0(\tilde\w))$ and $\dot J^a_1(\tilde\w) = F^a_1(J_0^b(\tilde\w)) + \Omega_1\partial_{\Omega_0}F^a_0(\Omega_0(\tilde w))$.

Equation~\eqref{dot J} is precisely what governs the system's slow evolution from one state to the next. Its form, an approximate equation for the exact state variables $J^a$, is akin to the treatment of the motion in the self-consistent expansion, in which we write an approximate equation $\frac{D^2z^\alpha}{d\tau^2}=\e f_1^\alpha+\e^2 f^\alpha_2 +\O(\e^3)$ for the exact worldline $z^\alpha$. 

Given the above expansions, the equation of motion and field equations are solved just as in Secs.~\ref{sec_expanded_worldline} and \ref{sec_expanded_field}. We substitute the expansions, apply the chain rules $\left(\frac{\partial}{\partial t}\right)_r = \e \dot J^a\partial_a + \Omega\partial_{\phi_p}$ and $\left(\frac{\partial}{\partial r^*}\right)_t = \partial_{r^*} - \e H (\dot J^a\partial_a + \Omega\partial_{\phi_p})$, and then solve the resulting equations while treating $J^a$ and $\phi_p$ as independent variables. 

We will merely summarize the results here. Substituting the expansions~\eqref{z(J)}, \eqref{dot J}, $U=U_0(\Omega)+\e U_1(J^a)+\O(\e^2)$, and
\begin{align}
 f^t &= \e \tilde f^t_1(\Omega) + \e^2 \tilde f_2^t(J^a) +\O(\e^3),\\
 f^r &= \e \tilde f^r_1(J^a) +\O(\e^2),
\end{align}
into  the normalization condition $g_{\alpha\beta}u^\alpha u^\beta =-1$ and equation of motion $\frac{D^2z^\alpha}{d\tau^2}=f^\alpha$, we obtain a sequence of equations for 
$U_n$ (from the normalization condition), $r_n$ (from the radial component of the equation of motion), and $F^{\Omega}_n$ (from the time component). The solutions are
\begin{align}
U_0 = \frac{1}{\sqrt{1-3(M\Omega)^{2/3}}},\qquad U_1 = 0,
\end{align}
\begin{align}
r_0 = \frac{M}{(M\Omega)^{2/3}},\qquad r_1 = \frac{\tilde f_1^r}{3U_0^2f_0\Omega^2},\label{r0(Omega)}
\end{align}
with $f_0=1-2(M\Omega)^{2/3}$, and
\begin{align}
F^\Omega_0 &= -\frac{3f_0\Omega \tilde f^t_1}{(M\Omega)^{2/3}U_0^4[1-6(M\Omega)^{2/3}]},\label{Omegadot0}\\
F^\Omega_1 &= -\frac{3f_0\Omega \tilde f^t_2}{(M\Omega)^{2/3}U_0^4[1-6(M\Omega)^{2/3}]} \nonumber\\
&\quad - \frac{2 \dot J^a_0\partial_{a}\tilde f_1^r}{MU_0^4[(M\Omega)^{1/3}-8M\Omega+12(M\Omega)^{5/3}]}\nonumber\\
&\quad - \frac{4[1-6(M\Omega)^{2/3}+12(M\Omega)^{4/3}]\tilde f_1^r\tilde f_1^t}{M\Omega U_0^6f_0[1-6(M\Omega)^{2/3}]^2}.\label{Omegadot1}
\end{align}
Note that the expressions for $U_0$, $U_1$, $r_0$, and $r_1$ here are identical to Eqs.~\eqref{U0}, \eqref{U1}, \eqref{Omega0}, and \eqref{Omega1} with $\Omega_0=\Omega$ and $\Omega_1=0$.

Next, the expansion of the field equations proceeds just as in Sec.~\ref{sec_expanded_field}, and the results in that section apply with only small changes. $\tilde\partial_{\tilde\w} = \partial_{\tilde\w}-im\Omega_1$ becomes $\partial_{\tilde\w}=F^a_0\partial_a$ ($=\dot J_0^a\partial_a$). $\omega_m=m\Omega_0$ becomes $\omega_m=m\Omega$. $\dot\omega_m$ becomes $m F^\Omega_0$ ($=m\dot\Omega_0$). The two-timescale expansion of $T^1_{i\ell m}$ (or equivalently, of the puncture) is slightly altered by the condition $\Omega_1=0$, but we leave a concrete description to a sequel paper. 

This naturally provides us with a wave-generation scheme that differs slightly from the one in Fig.~\ref{fig_flowchart}: 
\begin{enumerate}
\item Start with initial values of $\Omega$, $\delta M$, and $\delta S$ (along with $\mu$ and $M$).
\item Calculate $r_0(\Omega)$ from Eq.~\eqref{r0(Omega)}.
\item Calculate $\tilde T^1_{i\ell m}(\Omega)$ from $r_0(\Omega)$. 
\item Solve the field equations for the mode amplitudes $R^1_{i\ell m}(J^a,r)$ [and, separately, for $\partial_a R^1_{i\ell m}(J^a,r)$, as will be described in Ref.~\cite{worldtube-paper}].
\item From $R^1_{i\ell m}(J^a,r)$ and $\partial_a R^1_{i\ell m}(J^a,r)$, calculate $\delta^2G^0_{i\ell m}(J^a)$, $\tilde f^\alpha_1(J^a)$, $F^{\delta\cal M}_0=\dot E_{H}(\Omega)$, and $F^{\delta\cal S}_0=\dot L_{H}(\Omega)$. From $\tilde f^t_1(\Omega)$ in \eqref{Omegadot0}, calculate $F^\Omega_0$. 
\item From $\delta^2G^0_{i\ell m}(J^a)$, $E^1_{i\ell m}\sim F^a_0\partial_a R^1_{i\ell m}$, and the puncture, construct the second-order source. 
\item Solve the second-order field equations for the mode amplitudes $R^2_{i\ell m}(J^a,r)$. 
\item From $R^2_{i\ell m}(J^a,r)$, calculate $\tilde f^t_2(J^a)$, and from $\tilde f^t_2(J^a)$ in \eqref{Omegadot1}, calculate $F^\Omega_1$.
\item Use the post-adiabatic approximations $d\Omega/d\tilde\w = F^\Omega_0(\Omega)+\e F^\Omega_1(J^a)$, $d({\delta\cal M})/d\tilde\w = F^{\delta\cal M}_0(\Omega)$, and $d({ \delta\cal S})/d\tilde\w = F^{\delta\cal S}_0(\Omega)$ to evolve forward to new values of $J^a$. 
\item Repeat the above steps for as long as desired. 
\item Once the evolution of the frequency is known, obtain the orbital phase from $\phi_p = \frac{1}{\e}\int \Omega(\tilde \w)d\tilde \w$ and construct the waveform from Eq.~\eqref{multiscale h alt}. 
\end{enumerate}

\begin{widetext}
\section{Barack-Lousto-Sago tensor spherical harmonics}\label{sec_tensor_harmonics}

The Barack-Lousto-Sago harmonics are given in Schwarzschild coordinates by
\begin{subequations} \label{eqIII20}
\begin{equation} \label{eqIII20(1)}
Y^{1\ell m}_{\alpha\beta}=\frac{1}{\sqrt{2}}\left(
\begin{array}{c c c c}
1 & 0 & 0 & 0 \\
0 & f^{-2} & 0 & 0 \\
0 & 0 & 0 & 0 \\
0 & 0 & 0 & 0
\end{array}\right) Y^{\ell m},
\quad\quad
Y^{2\ell m }_{\alpha\beta}=\frac{f^{-1}}{\sqrt{2}}\left(
\begin{array}{c c c c}
0 & 1 & 0 & 0 \\
1 & 0 & 0 & 0 \\
0 & 0 & 0 & 0 \\
0 & 0 & 0 & 0
\end{array}\right) Y^{\ell m},
\quad\quad
Y^{3\ell m }_{\alpha\beta}=\frac{1}{\sqrt{2}}\left(
\begin{array}{c c c c}
f & 0 & 0 & 0 \\
0 & -f^{-1} & 0 & 0 \\
0 & 0 & 0 & 0 \\
0 & 0 & 0 & 0
\end{array}\right) Y^{\ell m},
\end{equation}
\begin{equation} \label{eqIII20(4)}
Y^{4\ell m }_{\alpha\beta}=
\frac{r}{\sqrt{2\lambda_1}}\left(
\begin{array}{c c c c}
0                 & 0 & \partial_{\theta} & \partial_{\phi} \\
0                 & 0 &        0          &         0          \\
\partial_{\theta} & 0 &        0          &         0          \\
\partial_{\phi}& 0 &        0          &         0
\end{array}\right) Y^{\ell m},
\quad\quad
Y^{5\ell m }_{\alpha\beta}=
\frac{rf^{-1}}{\sqrt{2\lambda_1}}\left(
\begin{array}{c c c c}
0 &        0           &        0          &          0         \\
0 &        0           & \partial_{\theta} & \partial_{\phi} \\
0 & \partial_{\theta}  &        0          &          0         \\
0 & \partial_{\phi} &        0          &          0
\end{array}\right) Y^{\ell m},
\end{equation}
\begin{equation} \label{eqIII20(6)}
Y^{6\ell m}_{\alpha\beta}=
\frac{r^2}{\sqrt{2}}\left(
\begin{array}{c c c c}
0 & 0 & 0 & 0 \\
0 & 0 & 0 & 0 \\
0 & 0 & 1 & 0 \\
0 & 0 & 0 & s^2
\end{array}\right) Y^{\ell m},
\quad\quad
Y^{7\ell m}_{\alpha\beta}=
\frac{r^2}{\sqrt{2\lambda_2}}\left(
\begin{array}{c c c c}
0 & 0 & 0   & 0        \\
0 & 0 & 0   & 0        \\
0 & 0 & D_2 & D_1       \\
0 & 0 & D_1 & -s^2 D_2
\end{array}\right) Y^{\ell m},
\end{equation}
\end{subequations}
\begin{subequations} \label{eqIII25}
\begin{equation} \label{eqIII25(8)}
Y^{8\ell m}_{\alpha\beta}=\frac{r}{\sqrt{2\lambda_1}}\left(
\begin{array}{c c c c}
0 &        0            & s^{-1}\partial_{\phi} & -s\,\partial_{\theta} \\
0 &        0            &          0               &                       \\
s^{-1}\partial_{\phi}&          0               &     0    &    0       \\
-s\,\partial_{\theta}   &          0               &     0    &    0
\end{array}\right) Y^{\ell m},
\end{equation}
\begin{equation} \label{eqIII25(9)}
Y^{9\ell m}_{\alpha\beta}=\frac{rf^{-1}}{\sqrt{2\lambda_1}}\left(
\begin{array}{c c c c}
0 &        0                &        0                &           0         \\
0 &        0                & s^{-1}\partial_{\phi} & -s\,\partial_{\theta} \\
0 & s^{-1}\partial_{\phi}&        0                &          0          \\
0 & -s\,\partial_{\theta}   &        0                &          0
\end{array}\right) Y^{\ell m},
\end{equation}
\begin{equation} \label{eqIII25(10)}
Y^{10\ell m}_{\alpha\beta}=
\frac{r^2}{\sqrt{2\lambda_2}}\left(
\begin{array}{c c c c}
0 & 0 & 0          & 0            \\
0 & 0 & 0          & 0            \\
0 & 0 & s^{-1}D_1 & -s\,D_2       \\
0 & 0 & -s\,D_2     & -s\,D_1
\end{array}\right) Y^{\ell m},
\end{equation}%
\end{subequations}%
\end{widetext}
where $s:=\sin\theta$, $\lambda_1 :=\ell(\ell+1)$, $\lambda_2:= (\ell-1)\ell(\ell+1)(\ell+2)$, $Y^{\ell m}= Y^{\ell m}(\theta,\phi)$ are the standard scalar spherical harmonics, and
\begin{align}\label{D1D2}  
D_1 &:=2(\partial_{\theta}-\cot\theta)\partial_{\phi},\\
D_2 &:= \partial_{\theta\theta}-\cot\theta\,\partial_{\theta}-s^{-2} \partial_{\phi\phi}.
\end{align}
The radial factors involving $r$ and $f$ are introduced to make the modes $\bar h_{ilm}$ dimensionless and to ensure that if the components $\bar h_{\alpha\beta}$ are regular at the future horizon in horizon-penetrating coordinates, then each of the modes $\bar h_{ilm}$ is as well.

This basis is orthogonal with respect to a certain inner product, satisfying 
\begin{equation} 
\oint d\Omega\, \eta^{\alpha\mu}\eta^{\beta\nu} Y_{\mu\nu}^{i\ell m} Y_{\alpha\beta}^{\ast j \ell^\prime m^\prime}=\kappa_i\delta_{ij}\delta_{\ell \ell^\prime}\delta_{mm^\prime},
\label{tensorHarmonicsOrthonormal}
\end{equation}
where $d\Omega=\sin\theta d\theta d\phi$ is the surface element on the unit sphere, 
\beq\label{eta}
\eta^{\alpha\beta}:=\text{diag}\left(1,f^{2},r^{-2},r^{-2}\sin^{-2}\theta\right),
\eeq
and 
\begin{equation}\label{kappai}
\kappa_i:=\begin{cases}
f^2\quad & \text{if }i=3,        \\
1\quad  &\text{otherwise}.
\end{cases}
\end{equation}
Note that our expression for $\eta^{\alpha\beta}$ corrects a typo in the original formula in Ref.~\cite{Barack:2005nr} (also corrected in Ref.~\cite{Wardell:2015ada}).

The coefficients $a_{i\ell}$ are introduced in Eq.~(\ref{modeDecomposition}) for the purpose of simplifying the field equations. They are defined to be
\begin{equation}
a_{i\ell}=\frac{1}{\sqrt{2}}\times\begin{cases}
1 \quad &\text{ for }i=1,2,3,6,\\ 
1/\sqrt{\lambda_1} \quad &\text{ for } i=4,5,8,9,\\
1/\sqrt{\lambda_2} \quad &\text{ for }i=7,10,\label{a_il}
\end{cases}
\end{equation}
where $\lambda_1$ and $\lambda_2$ are as defined above.

\section{Field equations}\label{sec_Mij0}

We give here explicit expressions for the matrix elements ${\cal M}_n^{ij}$ appearing in the wave operators~\eqref{Enijlm} and field equations (\ref{tildeEFE1})--(\ref{tildeEFE2}). 
For brevity, we omit the labels $n,\ell,m$ and the arguments $\tilde \w, r$  of the fields $\R^n_{i\ell m}(\tilde \w,r)$. For the same reason, we use $\lambda_1=\ell(\ell+1)$, as in Appendix~\ref{sec_tensor_harmonics}, and introduce $\lambda := \lambda_2/\lambda_1 = (\ell+ 2)(\ell -1)$. The partial derivative $\partial_r$ is taken with fixed $\tilde \w$.

\begin{widetext}
With those shorthand notations, the quantities ${\cal M}_{0}^{ij}\R_j$ in Eqs.~\eqref{tildeEFE1}  are given by
\begingroup
\allowdisplaybreaks
\begin{subequations}\label{Mij0}
\begin{align}
{\cal M}_0^{1j}\R_j &= \frac{ f^2 f'}{2}\left( \partial_r \R_{3}+\frac{i\omega_m  H}{f}\R_3\right)
										+\frac{f \left(1-\frac{4M}{r}\right)}{2r^2}\left(\R_1-\R_5 -f \R_3 \right)
										-\frac{f^2}{2r^2}\left(1-\frac{6M}{r}\right)\R_6, \label{M0-1}\\
{\cal M}_0^{2j}\R_j &= \frac{f^2 f'}{2}\left(\partial_r\R_3+\frac{i\omega_m  H}{f}\R_3\right)
										+\frac{ff'}{2}\partial_r \left(  \R_2 - \R_1\right)-\frac{i\omega_m }{2}\left(1-H\right)f'\left(  \R_2 - \R_1\right)\nonumber\\
									&\quad +\frac{ f^2}{2 r^2} \left(\R_2-\R_4\right) 
										-\frac{ff'}{2r}\left(\R_1-\R_5-f\R_3-2 f \R_6\right), \label{M0-2}\\
{\cal M}_0^{3j}\R_j &= -\frac{f}{2r^2}\left[\R_1-\R_5-\left(1-\frac{4M}{r}\right)\left(\R_3 + \R_6\right)\right],\label{M0-3}\\
{\cal M}_0^{4j}\R_j &= \frac{ff'}{4}\partial_r\left( \R_4- \R_5\right)-\frac{i \omega_m  \left(1-H\right)f'}{4}\left( \R_4- \R_5\right)
										-\frac{\lambda_1}{2}\,\frac{f}{r^2}\R_2
										-\frac{ ff'}{4 r}\left(3\R_4+2\R_5-\R_7+\lambda_1\R_6\right),\label{M0-4}\\
{\cal M}_0^{5j}\R_j &= \frac{f}{r^2}\left[\left(1-\frac{9M}{2r}\right)\R_5 -\frac{\lambda_1}{2}\left(\R_1-f\R_3\right)
										+\frac{1}{2}\left(1-\frac{3M}{r}\right)\left(\lambda_1\R_6-\R_7\right)\right],\label{M0-5}\\
{\cal M}_0^{6j}\R_j &= -\frac{f}{2r^2}\left[\R_1-\R_5-\left(1-\frac{4M}{r}\right)\left(\R_3+\R_6\right)\right],\label{M0-6}\\
{\cal M}_0^{7j}\R_j &= -\frac{f}{2r^2}\left(\R_7+\lambda\,\R_5\right),\label{M0-7}\\
{\cal M}_0^{8j}\R_j &= \frac{ff'}{4}\partial_r\left( \R_8- \R_9  \right)-\frac{i\omega_m \left(1-H\right)f'}{4}\left( \R_8- \R_9  \right)
										-\frac{ff'}{4r} \left(3\R_8+2\R_9-\R_{10}\right),\label{M0-8}\\
{\cal M}_0^{9j}\R_j &=\frac{f}{r^2}\left(1-\frac{9 M}{2r}\right)\R_9 - \frac{f}{2r^2}\left(1-\frac{3M}{r}\right)\,\R_{10},\label{M0-9}\\
{\cal M}_0^{10j}\R_j &= -\frac{f}{2r^2}\left(\R_{10} +\lambda \R_9\right),\label{M0-10}
\end{align}
\end{subequations}
\endgroup
\end{widetext}
where $f'=\partial f/\partial r = 2M/r^2$.

The quantities $\mathcal{ M}^{ij}_1\R^j$ in Eqs.~\eqref{Enijlm} and (\ref{tildeEFE2})  are given by
\begin{subequations}\label{Mij1}
\begin{align}
{\cal M}^{1j}_1\R_j &= -\frac{1}{2}ff'H\tilde\partial_{\tilde \w}\R_3, \label{M1-1}\\
{\cal M}^{2j}_1\R_j &= -\frac{f'}{2}\left[ f H \tilde\partial_{\tilde \w}\R_3\right.\nonumber\\
									&\quad\left. - \left(1-H\right)\tilde\partial_{\tilde \w}\left(\R_2-\R_1\right) \right],\label{M1-2}\\
{\cal M}^{4j}_1\R_j &= \frac{f'}{4}\left(1-H\right)\tilde\partial_{\tilde \w}\left(\R_4-\R_5\right),\label{M1-4}\\
{\cal M}^{8j}_1\R_j &= \frac{f'}{4}\left(1-H\right)\tilde\partial_{\tilde \w}\left(\R_8-\R_9\right),\label{M1-8}\\
{\cal M}^{ij}_1\R_j &= 0\,\,\,\,\text{for}\,\,\,\,\,i=3,5,6,9,10.
\end{align}
\end{subequations}

\section{Analytic solutions for the $\ell=0$ and $\ell=1,m=0$ first-order modes}\label{sec_low_modes}
In this appendix we present the explicit analytical formulas for the solutions~\eqref{ell=0 form} and \eqref{ell=1, m=0 form}

\subsection{Dipole solution}

We first consider $\ell=1,m=0$, which only has an odd-parity contribution. For $\R^{\rm pp}_{i10}$ we use
\begin{subequations}\label{odd dipole}
\begin{align}
\R^{\rm pp}_{810} &= -16\sqrt{\frac{\pi}{3}}{\cal L}_0 \begin{cases}r^2/r_0^3 & \text{for } r\leq r_0, \\ 1/r & \text{for } r\geq r_0,\end{cases} \\
\R^{\rm pp}_{910} &= -\sqrt{\frac{\pi}{3}}\frac{256M^4 {\cal L}_0}{r_0^3 r^2} ,
\end{align}
\end{subequations}
where ${\cal L}_0=\mu r_0^2U_0\Omega_0$ is the zeroth-order orbital angular momentum. Note that our solution~\eqref{odd dipole} differs from the one that has been used historically. The historical solution, dating back to Zerilli's classic paper~\cite{Zerilli:70}, agrees with our $\R^{\rm pp}_{810}$ but sets $\R^{\rm pp}_{910}=0$; though this solution is often assumed to be horizon regular, it actually violates the regularity condition~\eqref{i=4,8 horizon regularity}, leading to divergent behavior of the second-order source term $\delta^2 G^0_{i\ell m}$ at the horizon. The addition of an $i=9$ mode (which is pure gauge) satisfying Eq.~\eqref{i=4,8 horizon regularity} cures this ill behavior.\footnote{This longstanding error was discovered in collaboration with Leor Barack, Niels Warburton, and Barry Wardell. We thank Leor Barack for deriving the correct, nonzero $\R^{\rm pp}_{910}$.} In components, the perturbation at large $r$ reads
\beq
h^{pp,\ell=1,m=0}_{t\phi} = -\frac{2{\cal L}_0\sin^2\theta}{r} +\O(1/r^4),
\eeq
the expected asymptotic form of the metric in a spacetime with angular momentum ${\cal L}_0$. One can straightforwardly verify, using Komar or Abbott-Deser integrals~\cite{Dolan-Barack:13} over surfaces of constant $r$ (at fixed slow time), that the spacetime $g_{\alpha\beta}+\e h^{\rm pp}_{\alpha\beta}$ has zero angular momentum inside every sphere of radius $r<r_0$, corresponding to a nonspinning central black hole, and angular momentum $\e{\cal L}_0$ inside every sphere of radius $r>r_0$.

On top of $\R^{\rm pp}_{i10}$, we can freely add a perturbation generated by the central black hole's small, evolving spin $S_1$, given by 
\begin{subequations}\label{dJ perturbation}
\begin{align}
\bar x_{810} &=  -\sqrt{\frac{\pi}{3}}\frac{16 S_1}{r},\\
\bar x_{910} &= - \sqrt{\frac{\pi}{3}}\frac{32 M S_1}{r^2}. 
\end{align}
\end{subequations}
The $i=9$ mode is again added to ensure regularity at the future horizon. $\bar x_{i\ell m}$ corresponds to a metric perturbation with $t\phi$ component $x^{\ell=1,m=0}_{t\phi} = -\frac{2 S_1\sin^2\theta}{r} +\O(1/r^4)$, and one can verify that the spacetime $g_{\alpha\beta}+\e x_{\alpha\beta}$ contains angular momentum $\e S_1$ for all $r>2M$.

\subsection{Monopole solutions}\label{monopole solutions}

Next, we consider $\ell=0$. The analysis of this mode is complicated by a well-known pathology of the Lorenz gauge: in the Lorenz gauge, there is no globally regular homogeneous solution with nonzero mass. A homogeneous solution with nonzero mass is always irregular at either the horizon or at infinity (or both)~\cite{Dolan-Barack:13,Akcay:2013wfa}. These singularities are purely a gauge artefact; outside the Lorenz gauge, it is easy to find mass perturbations that are regular at the future horizon and at infinity. For example, in an `Eddington-Finkelstein gauge', a homogeneous perturbation with mass $M_1$ has the simple, manifestly regular form
\beq\label{hEF}
x^{\rm EF}_{\alpha\beta} = \frac{2M_1}{r}\delta^v_\alpha\delta^v_\beta
\eeq
(obtained by replacing $M$ with $M+\e M_1$ in the Eddington-Finkelstein components of the exact Schwarzschild metric). However, here we restrict our attention to Lorenz-gauge solutions. We consider several solutions and their relative merits.

\subsubsection{Basis of homogeneous solutions}

All possible solutions to Eq.~\eqref{tildeEFE1} are easily constructed from the complete basis of homogeneous solutions provided by Dolan and Barack~\cite{Dolan-Barack:13}, the main members of which we denote $\left\{h^{(A)}_{\alpha\beta}, h^{(B)}_{\alpha\beta}, h^{(C)}_{\alpha\beta}, h^{(D)}_{\alpha\beta}\right\}$, following Dolan and Barack's labelling. With the metric perturbation written as $h_{\alpha\beta}={\rm diag}(h_{tt}, h_{rr}, Hr^2, Hr^2\sin^2\theta)$, these are given by
\begin{align}
h^{(A)}_{\alpha\beta} &= g_{\alpha\beta},\\
h^{(B)}_{tt} &= -\frac{2M f P(r)}{r^3},\ h_{rr}^{(B)} = \frac{2 Q(r)}{fr^3},\\
H^{(B)} &= \frac{2 f P(r)}{r^2},\\
h^{(C)}_{tt} &= -\frac{2 M^4}{r^4},\ h_{rr}^{(C)} = -\frac{2 M^3 (2 r - 3 M)}{f^2r^4},\\ 
H^{(C)} &= \frac{2 M^3}{r^3},\\
h^{(D)}_{tt} &= -\frac{2}{3r^4} \big[r W(r) + M r f P(r) \ln f \nonumber\\
					&\quad - 8 M^4 \ln(r/M)\big],\\
h_{rr}^{(D)} &= \frac{2}{3r^4f^2} \big[L(r) \ln f -r K(r) \nonumber\\
					&\quad + 8 M^3 (2 r - 3 M) \ln(r/M)\big],\\ 
H^{(D)} &= \frac{2}{3r^3}\big\{(r^3 - 8 M^3) \ln f -r [P(r) - M r]\nonumber\\
				&\quad - 8 M^3 \ln(r/M)\big\},
\end{align}
where 
\begin{align}
P &:= r^2 + 2 r M + 4 M^2,\\
Q &:= r^3 - r^2 M - 2 r M^2 + 12 M^3,\\
W &:= 3 r^3 - 7 r^2 M - r M^2 - 4 M^3,\\ 
K &:= r^3 - 5 r^2 M - 5 r M^2 + 12 M^3,\\ 
L &:= r^4 - 3 r^3 M + 16 r M^3 - 24 M^4.
\end{align}
(The $H$ here should not be confused with $H=dk/dr^*$.) There is also a fifth independent solution, consisting solely of a $t$-$r$ component,
\beq
h^{(E)}_{tr} = \frac{M^2}{f r^2}.
\eeq
This was not identified by Dolan and Barack, but it appears as part of one of the solutions they consider. We will not require explicit use of it. 

Given the metric components, the Barack-Lousto-Sago amplitudes are
\begin{subequations}\label{comps to modes}
\begin{align}
\R_{100} &= \sqrt{4\pi}  r \left(h^{\ell=0}_{tt} + f^2 h^{\ell=0}_{rr}\right),\\
\R_{200} &= \sqrt{16\pi} r f h^{\ell=0}_{tr},\\
\R_{300} &= \sqrt{16\pi} r H^{\ell=0},\\
\R_{600} &= \sqrt{4\pi} r f^{-1}\left(h^{\ell=0}_{tt} - f^2 h^{\ell=0}_{rr}\right).
\end{align}
\end{subequations}

Together, $h^{(A)}_{\alpha\beta}$ through $h^{(E)}_{\alpha\beta}$ form a complete basis. From Table~\ref{tabHierarchy}, we see that for $\ell=0$ there are two coupled field equations for $i=1,3$, plus the hierarchically decoupled field equation for $i=2$, with the $i=6$ mode recovered algebraically from the gauge condition~\eqref{eq:gauges-w0} with \eqref{Z02}. The $i=1,3$ equations should have a total of four independent homogeneous solutions; these are provided by the $(A)$--$(D)$ solutions. When these solutions are substituted into the $i=2$ field equation, one finds that the $i\neq2$ modes cancel one another in ${\cal M}^{ij}_0 R_{j\ell m}$, leaving a fully decoupled equation for $i=2$. The gauge condition~\eqref{eq:gauges-w0} with \eqref{Z01} also provides a decoupled equation for $i=2$, and it is easy to verify that the sole solution to the gauge condition automatically satisfies the field equation; this is the $(E)$ solution  

$h^{(A)}_{\alpha\beta}$ is the only one of the five solutions to contain mass, equal to $M/2$; all others are pure gauge. It is regular at the future horizon but irregular at infinity (where its components go to constants instead of decaying to zero). $h^{(B)}_{\alpha\beta}$ is regular at the future horizon but irregular at infinity. $h^{(C)}_{\alpha\beta}$ and  $h^{(E)}_{\alpha\beta}$ are regular at infinity but not at the future horizon. $h^{(D)}_{\alpha\beta}$ is irregular at both boundaries.

\subsubsection{Asymptotically regular, horizon-irregular solution}\label{infinity-regular monopole}

We first consider solutions that are asymptotically flat and in which $\bar x_{i00}(M_1,r)$ contains the full perturbation to the black hole's mass. 

An inhomogeneous solution to Eq.~\eqref{tildeEFE1} that is regular at infinity, contains no mass in the region $r<r_0$ (and therefore no correction to the black hole's mass), and is maximally regular at the horizon is given by
\begin{align}
h^{{\rm pp},\ell=0}_{\alpha\beta} &= \left(2{\cal E}_0 h^{(A)}_{\alpha\beta}+a^{(B)}_+ h^{(B)}_{\alpha\beta}+a_+^{(C)}h^{(C)}_{\alpha\beta} +a^{(D)}_+ h^{(D)}_{\alpha\beta}\right)\!\theta^+\nonumber\\
								&\quad + \left(a_-^{(B)}h^{(B)}_{\alpha\beta} + a_-^{(C)}h^{(C)}_{\alpha\beta} +a_-^{(D)}h^{(D)}_{\alpha\beta}\right)\!\theta^-,\!\label{monopole}
\end{align}
where $\theta^\pm := \theta[\pm(r-r_0)]$ and
\begin{subequations}
\begin{align}
a^{(B)}_+ &= -\frac{4}{3}{\cal E}_0,\\
a_+^{(C)} &= -\frac{{\cal E}_0}{3 Mr_0f_0} \left[M^2 (8 \ln 2-44)+r_0^2\right.\nonumber\\
			&\quad \left.+8 M (r_0-3 M) \ln(r_0/M)+20 M r_0\right],\\
a_+^{(D)} &= - {\cal E}_0,\\			
a_-^{(B)} &= -\frac{(r_0-3M)\ln(f_0) {\cal E}_0}{3r_0f_0},\\
a_-^{(C)} &= -\frac{4M(5+8\ln 2){\cal E}_0}{3r_0f_0},\\
a_-^{(D)} &= - \frac{M{\cal E}_0}{r_0f_0}.
\end{align}
\end{subequations}
The mode amplitudes are then given by Eq.~\eqref{comps to modes}. Since only $h^{(A)}_{\alpha\beta}$ contains mass, we can read off that in the spacetime $g_{\alpha\beta}+\e h^{\rm pp}_{\alpha\beta}$, a sphere of radius $r<r_0$ contains mass $M$, and a sphere of radius $r>r_0$ contains mass $M+\e{\cal E}_0$. This describes the spacetime of a particle of mass $\mu$ orbiting a black hole of mass $M$. 

The perturbation is regular at large $r$, where 
\beq
h^{{\rm pp},\ell=0}_{\alpha\beta}=\frac{2{\cal E}_0}{r}{\rm diag}(1,1,r^2,r^2\sin^2\theta) +\O(1/r^2),
\eeq
and
\begin{subequations}\label{Rpp large r}
\begin{align}
\R^{\rm pp}_{100} &= 8\sqrt{\pi} {\cal E}_0+\O(1/r),\\
\R^{\rm pp}_{200} &= 0,\\
\R^{\rm pp}_{300} &= 8\sqrt{\pi} {\cal E}_0 +\O(1/r),\\
\R^{\rm pp}_{600} & = \O(1/r).
\end{align}
\end{subequations}
But it diverges logarithmically at the horizon, where
\begin{subequations}\label{Rpp near horizon}
\begin{align}
\R^{\rm pp}_{100} &= \O(f^2),\\
\R^{\rm pp}_{200} &= 0,\\
\R^{\rm pp}_{300} & = \O(f),\\
\R^{\rm pp}_{600} &= \frac{8\sqrt{\pi}}{r_0f_0}\bigg[(r_0-3M)\ln f_0 \nonumber\\
								&\quad+M\ln\left(\frac{r}{2M}-1\right)\bigg] +\O(f\ln f).
\end{align}
\end{subequations}
One can construct other inhomogeneous solutions with the same properties by adding perturbations proportional to $h^{(E)}_{\alpha\beta}$. But it is straightforward to check that no such solution can improve upon the above solution's regularity at the horizon without sacrificing regularity at infinity.

Next, an asymptotically flat mass perturbation with maximal regularity at the horizon is given by
\begin{align}
x^{\ell=0}_{\alpha\beta}(M_1,r) &= 2 M_1 h^{(A)}_{\alpha\beta} -\frac{4}{3}M_1 h^{(B)}_{\alpha\beta}-\frac{4}{3}(5+\ln 4)M_1 h^{(C)}_{\alpha\beta} \nonumber\\
					&\quad - M_1 h^{(D)}_{\alpha\beta}.\label{x}
\end{align}
At large $r$, it behaves as in Eq.~\eqref{Rpp large r} with the replacement ${\cal E}_0\to M_1$. Near the horizon it behaves as
\begin{subequations}
\begin{align}
\bar x_{100} &= \O(f^2),\\
\bar x_{200} &= 0,\\
\bar x_{300} &= 8\sqrt{\pi} M_1\left[2 +\ln\left(\frac{r}{2M}-1\right)\right] + \O(f),\\
\bar x_{600} & = 16\sqrt{\pi} M_1 +\O(f).
\end{align}
\end{subequations}
Again,  one can easily check that this is the maximal regularity that an asymptotically flat Lorenz-gauge mass perturbation can have at the horizon.

The perturbation $h^{{\rm pp},\ell=0}_{\alpha\beta}+x^{\ell=0}_{\alpha\beta}(M_1,r)$ in Eqs.~\eqref{monopole} and \eqref{x} fits neatly into Sec.~\ref{sec_combined_expansions}'s description of the full two-timescale solution and its evolution. In particular, the black hole's total mass is $M_{\rm BH} = M+\e M_1$, and it satisfies the flux equation $dM_{\rm BH}/d\tilde\w=\e dM_1/d\tilde\w = \e\dot E_H$.

The disadvantage of this solution is obvious. Because it is singular at the horizon, it makes the second-order source term $\delta^2 G^0_{i\ell m}$ also singular there. And because the $\ell=0$ mode of $\tilde h^1_{\alpha\beta}$ contributes to every mode $\delta^2 G^0_{i\ell m}$, this irregularity spreads into every second-order mode. Of course once the solution is obtained, the singularity can be eliminated with a gauge transformation. But one must first carefully obtain the correct (singular) boundary conditions for each mode.

\subsubsection{Berndtson solution}\label{Berndtson monopole}

We next consider asymptotically flat solutions in which $x^{\ell=0}_{\alpha\beta}(M_1,r)$ does {\em not} contain the full perturbation to the black hole's mass. This implies that the inhomogeneous solution $h^{{\rm pp},\ell=0}_{\alpha\beta}$ contains part of the correction to the black hole mass, which turns out to allow it to be regular at both boundaries.

The relevant regular inhomogeneous solution was first derived by Berndtson~\cite{Berndtson:07}. Denoting it by $h^{\rm Bern}_{\alpha\beta}$ to distinguish it from the $h^{{\rm pp},\ell=0}_{\alpha\beta}$ of Eq.~\eqref{monopole}, we can write it as 
\begin{align}
h^{\rm Bern}_{\alpha\beta} &= \bigg[2M_> h^{(A)}_{\alpha\beta}+b_+^{(B)} h^{(B)}_{\alpha\beta} + b^{(C)}_+h^{(C)}_{\alpha\beta} \nonumber\\
									&\quad + b^{(D)}_+ h^{(D)}_{\alpha\beta}\bigg]\theta^+  + \bigg[2M_< h^{(A)}_{\alpha\beta} + b^{(B)}_-h^{(B)}_{\alpha\beta} \nonumber\\
								&\quad+ b^{(C)}_-h^{(C)}_{\alpha\beta} +b^{(D)}_-h^{(D)}_{\alpha\beta}\bigg]\theta^-,
\end{align}
where 
\begin{subequations}
\begin{align}
b^{(B)}_\pm &= a_\pm^{(B)} -\frac{4}{3}M_<,\\ 
b^{(C)}_\pm &= a_\pm^{(C)} -\frac{4}{3}(5+\ln 4)M_< ,\\
b^{(D)}_\pm &= a_\pm^{(D)}  -M_<.
\end{align}
\end{subequations}
Here 
\beq
M_< = -\frac{{\cal E}_0}{r_0f_0} =: M_{\rm Bern}
\eeq
is the mass contained in a sphere of radius $r<r_0$, and $M_>= (M_{\rm Bern}+{\cal E}_0)$ is the mass contained in a sphere of radius $r>r_0$. 

$h^{\rm Bern}_{\alpha\beta}$  is regular at both future null infinity and the future horizon, satisfying all the conditions of \eqref{horizon regularity}, as desired. If we exclude any nonzero $t$-$r$ components, $h^{\rm Bern}_{\alpha\beta}$ is in fact the unique Lorenz-gauge solution that is regular at both boundaries. (If we include $h^{(E)}_{\alpha\beta}$, we can construct other regular solutions, but none of them improve upon the Berndtson solution.) $h^{\rm Bern}_{\alpha\beta}$ also differs from Eq.~\eqref{monopole} in that it contains mass, $M_{\rm Bern}$, in the region $r<r_0$; this corresponds to a perturbation to the black hole's mass. Contrary to statements in Refs.~\cite{Dolan-Barack:13,Akcay:2013wfa}, this does not suggest $h^{\rm Bern}_{\alpha\beta}$ is an unphysical solution. It tells us that $h^{\rm Bern}_{\alpha\beta}$ describes the physical spacetime of a point mass $\mu$, with specific energy ${\cal E}_0$, on a circular geodesic orbit around a black hole with mass $M_{\rm BH} = M+\e M_{\rm Bern}$.

A minor drawback of the Berndtson solution is that it does not fit quite so cleanly into the evolution scheme in Sec.~\ref{sec_combined_expansions}. If we take our total first-order $\ell=0$ solution to be $h^{\rm Bern}_{\alpha\beta}+x^{\ell=0}_{\alpha\beta}(M_1)$, with $x^{\ell=0}_{\alpha\beta}(M_1)$ as in Eq.~\eqref{x}, then the total black hole mass is $M_{\rm BH}=M+\e(M_{\rm Bern}+M_1)$. Since $dM_{\rm BH}/d\tilde\w=\e\dot E_H$, this implies that the evolution equation for $M_1$ becomes 
\beq\label{M1dot Berndtson}
\frac{dM_1}{d\tilde\w} = \dot E_H - \frac{dM_{\rm Bern}}{d\tilde \w},
\eeq
where $\frac{dM_{\rm Bern}}{d\tilde \w}=\frac{dr_0}{d\tilde\w}\frac{dM_{\rm Bern}}{dr_0}$.

Of course, since $x^{\ell=0}_{\alpha\beta}$ is irregular in any case, the total solution $h^{\rm Bern}_{\alpha\beta}+x^{\ell=0}_{\alpha\beta}$ is just as irregular as the solution $h^{\rm pp,\ell=0}_{\alpha\beta}+x^{\ell=0}_{\alpha\beta}$ from the previous subsection. The only potential advantage it might have over the previous solution is if the changes in $M_1$ were numerically negligible; in that case, we could simply neglect $M_1$, and $h^{\rm Bern}_{\alpha\beta}$ would provide a well-behaved first-order field at the boundaries, leading in turn to a second-order Einstein tensor $\delta^2 G^0_{i\ell m}$ that is maximally well behaved at the boundaries. This may in fact be the case, since changes in the black hole parameters are typically extremely small over an inspiral~\cite{Hughes:2018qxz}.

A major advantage of the Berndtson solution lies in cases where we are not actually concerned with evolving the system but only with calculating some physical quantity at a fixed value of slow time. In that case we can freely omit $x_{\alpha\beta}$, and the Berndtson solution provides a first-order field that is regular at both boundaries, leading to a second-order source that is maximally well behaved at the boundaries. This was the approach taken in Ref.~\cite{Pound-etal:19}.

As a final comment on the Berndtson solution, we note that if one leaves the Lorenz gauge, it provides a simple way to construct a solution that is continuous at $r=r_0$, regular at both boundaries, has zero mass for $r<r_0$, and has mass ${\cal E}_0$ for $r>r_0$. To find this solution, start with the Berndtson solution and add an Eddington-Finkelstein solution~\eqref{hEF} with mass $M_1=-M_{\rm Bern}$.

\subsubsection{Asymptotically irregular, horizon-regular solution}\label{horizon-regular monopole}

In most Lorenz-gauge calculations, neither the solution~\eqref{monopole} nor the Berndtson solution has been used. Instead, to our knowledge, all authors have adopted a solution for $h^{\rm pp,\ell=0}_{\alpha\beta}$ that is regular at the horizon but not at infinity. Since this solutions is displayed frequently, we do not repeat it here; it can be found in Eq.~(114) of Dolan and Barack, for example. It satisfies the horizon-regularity conditions~\eqref{horizon regularity}, but at infinity it goes to a constant:
\beq\label{hpp HR at infinity}
\lim_{r\to\infty}h^{\rm pp,\ell=0}_{\alpha\beta} = 2M_{\rm Bern}\delta_\alpha^t \delta_\beta^t.
\eeq
Here $M_{\rm Bern}$ is {\em not} the mass in the solution; instead, it appears as a measure of the mismatch between intervals of coordinate time $t$ in the perturbed spacetime and intervals of proper time for asymptotic observers. 

A simple mass perturbation that is horizon-regular and minimally asymptotically irregular is given by
\begin{align}
x^{\ell=0}_{\alpha\beta}(M_1,r) &= 2 M_1 h^{(A)}_{\alpha\beta} -M_1 h^{(B)}_{\alpha\beta}.
\end{align}
At large $r$, it behaves as 
\beq\label{x HR at infinity}
\lim_{r\to\infty}x^{\ell=0}_{\alpha\beta} = -2M_1 \delta_\alpha^t \delta_\beta^t.
\eeq

Just as the solutions in the previous two sections made the second-order source ill behaved at the horizon, the solution $h^{\rm pp,\ell=0}_{\alpha\beta}+x^{\ell=0}_{\alpha\beta}$ in this section makes the source ill behaved at infinity, decaying slowly as $r\to\infty$. That slow decay is numerically burdensome to integrate over, and it complicates the procedure of deriving boundary conditions. 

This solution also requires special considerations in order to obtain useful results. In practice we are almost always interested in calculating quantities in gauges that admit the same preferred asymptotic reference frame as the background. As a consequence, in a typical first-order calculation, after computing a physical quantity in the asymptotically irregular Lorenz gauge, one performs a gauge transformation to find the value of that quantity in an asymptotically flat gauge~\cite{Barack-Pound:18}. With the transformation written in the form $x^\alpha\to x^\alpha-\e \xi^\alpha$, it requires the linearly growing gauge vector
\beq
\xi^\alpha =\delta_t^\alpha M_{\rm Bern}\, t,
\eeq
which corresponds to a rescaling of time, 
\beq
t\to (1-\e M_{\rm Bern})t
\eeq
and also consequently a rescaling of the frequencies,
\beq
\omega_m\to (1+\e M_{\rm Bern} )\omega_m.
\eeq
In the first-order context, this transformation only affects the monopole mode of the metric perturbation, and it leaves that perturbation static. But at second order, such a transformation would affect every mode and lead to spurious, growing solutions in the second-order field, violating the presumed form of the two-timescale expansion.

We can recast the linearly growing transformation in a form consistent with the two-timescale form of the field by writing it as 
\beq
\xi^t = \int M_{\rm Bern}(\e t)dt = \frac{1}{\e}\int M_{\rm Bern}(\tilde t)d\tilde t;
\eeq 
if reexpanded at fixed $t$, this reduces to the form above. It is a large, $\O(1)$ transformation of $t$, but it is better understood as a small transformation of the slow time, 
\beq
\tilde t\to \tilde t - \e \int M_{\rm Bern}(\tilde t)d\tilde t.
\eeq
Such a transformation filters through the entire two-timescale expansion, altering the frequencies $\omega_m$ as above, the field equations in which $\omega_m$ appears, and the fast time $\phi_p=\int\Omega dt$. It also requires additional care if we use $\tilde s$ instead of $\tilde t$ as our slow time.
 
Ref.~\cite{two-timescale-0} will discuss the gauge freedom in the two-timescale expansion in greater generality, including this freedom to transform the slow time.

\bibliography{refs}

\end{document}